\documentclass[preprint,aps,amsmath,nofootinbib,superscriptaddress]{revtex4}
\usepackage{graphicx}
\usepackage{bm}
\usepackage{epsfig}
\usepackage{pst-grad}

\newif\ifpdf
\ifx\pdfoutput\undefined
\pdffalse 
\else
\pdfoutput=1 
\pdftrue
\fi

\def\Slash#1{{#1\!\!\!\slash}}
\def\Aslash{A\!\!\!\!\slash}
\def\Bslash{B\!\!\!\!\slash}
\def\Dslash{D\!\!\!\!\slash}

\def\Mslash{M\!\!\!\!\!\slash}

\def\nslash{n\!\!\!\slash}
\def\bnslash{\bar n\!\!\!\slash}
\def\vslash{v\!\!\!\slash}
\def\pslash{p\!\!\!\slash}
\def\qslash{q\!\!\!\slash}
\def\tslash{t\!\!\!\slash}

\def\Aslash{A\!\!\!\slash}
\def\kslash{k\!\!\!\slash}

\def\vepslash{\varepsilon\!\!\!\slash}
\def\Delslash{\Delta\!\!\!\!\slash}
\def\OMIT#1{}

\newcommand{\nn}{\nonumber} 

\newcommand{\bn}{{\bar n}}
\newcommand{\bea}{\begin{eqnarray}}
\newcommand{\eea}{\end{eqnarray}}

\newcommand{\bnP}{\bar {\cal P}}

\newcommand{\cP}{{\cal P}}
\newcommand{\cPslash}{ {\cal P}\!\!\!\!\slash}

\newcommand{\mcdot}{\!\cdot\!}

\newcommand{\LQCD}{{\Lambda}}

\newcommand{\DvppPl}{\: iv \mcdot\! \overleftarrow D_c^\perp}
\newcommand{\DvppPr}{\: iv \mcdot\! \overrightarrow D_c^\perp}

\newcommand{\DSppP}{{{iD\!\!\!\!\hspace{0.04cm}\slash}}_c^\perp}
\newcommand{\DSppPl}{\,i\!\overleftarrow {D\!\!\!\!\hspace{0.04cm}\slash}_c^\perp}
\newcommand{\DSppPr}{\,i\!\overrightarrow {D\!\!\!\!\hspace{0.04cm}\slash}_c^\perp}

\newcommand{\DgppP}{{{D}}_c^{\perp\mu}}
\newcommand{\DgppPl}{\,\overleftarrow D{}_c^{\perp\mu}}
\newcommand{\DgppPr}{\,\overrightarrow D{}_c^{\perp\mu}}
\newcommand{\DgppPld}{\,\overleftarrow D{}_{\!c\,\alpha}^{\perp}}
\newcommand{\DgppPrd}{\,\overrightarrow D{}_{\!c\,\alpha}^{\perp}}

\newcommand{\lD}{\overleftarrow D{}_c}
\newcommand{\rD}{\overrightarrow D{}_c}

\begin{document}
\ifpdf
\DeclareGraphicsExtensions{.pdf, .jpg}
\else
\DeclareGraphicsExtensions{.eps, .jpg}
\fi


\preprint{ \vbox{\hbox{INT-PUB-02-49} \hbox{}  
}}

\title{\phantom{x}\vspace{0.5cm} 
A Complete Basis for Power Suppressed \\ Collinear-Ultrasoft Operators
\vspace{0.5cm} }

\author{Dan Pirjol}
\affiliation{Department of Physics and Astronomy, 
        The Johns Hopkins University,\\[-4pt]
        Baltimore, MD 21218\footnote{Electronic address: dpirjol@pha.jhu.edu}}

\author{Iain W. Stewart\vspace{0.4cm}}
\affiliation{Institute for Nuclear Theory,  University of Washington, Seattle, 
        WA 98195 \footnote{Electronic address: iain@phys.washington.edu}
        \vspace{0.5cm}}


\begin{abstract}
\vspace{0.5cm}
\setlength\baselineskip{18pt}

We construct operators that describe power corrections in mixed
collinear-ultrasoft processes in QCD.  We treat the ultrasoft-collinear
Lagrangian to ${\cal O}(\lambda^2)$, and heavy-to-light currents involving
collinear quarks to ${\cal O}(\lambda)$ including new three body currents.  A
complete gauge invariant basis is derived which has a full reduction in Dirac
structures and is valid for matching at any order in $\alpha_s$. The full set of
reparameterization invariance (RPI) constraints are included, and are found to
restrict the number of parameters appearing in Wilson coefficients and rule out
some classes of operators. The QCD ultrasoft-collinear Lagrangian has two ${\cal
  O}(\lambda^2)$ operators in its gauge invariant form.  For the ${\cal
  O}(\lambda)$ heavy-to-light currents there are $(4,4,14,14,21)$ subleading
(scalar, pseudo-scalar, vector, axial-vector, tensor) currents, where
$(1,1,4,4,7)$ have coefficients that are not determined by RPI. In a frame where
$v_\perp=0$ and $n\cdot v=1$ the total number of currents reduces to
$(2,2,8,8,13)$, but the number of undetermined coefficients is the same. The
role of these operators and universality of jet functions in the factorization
theorem for heavy-to-light form factors is discussed.

\end{abstract}

\maketitle


\newpage

\setlength\baselineskip{15pt}

\section{Introduction}

The soft-collinear effective theory (SCET) constructed in
\cite{bfl,bfps,cbis,bpssoft} offers a systematic description of processes
involving energetic particles. It has an expansion in a small parameter
$\lambda\sim p_\perp/Q$, where $p_\perp$ is a typical transverse momenta and $Q$
the large energy scale. Hard exclusive and inclusive processes in QCD are
usually described using the powerful techniques of QCD factorization and
light-cone expansions~\cite{known1,known2}. SCET encompasses and extends these
frameworks, and in particular allows a model independent description of effects
caused by the interplay between energetic collinear particles and soft particles
beyond leading order in the power expansion. These effects can be described in a
rigorous way based solely on QCD, but are not included in purely collinear
expansions.\OMIT{~\cite{nosoft}.}  The study of operators that describe these
mixed collinear-ultrasoft (collinear-usoft) effects is the purpose of this
paper.  For recent applications of SCET in hard scattering processes and
B-decays see Refs.~\cite{bps,bfprs,scet_apps,chay,mmps,bpspc,bcdf,bps4}.

Since our focus is on mixed collinear-usoft interactions, we consider collinear
quark fields $\xi_{n,p}$, collinear gluon fields $A_{n,p}^\mu$, usoft heavy
quark fields $h_v$, usoft light quark fields $q_{us}$, and usoft gluons
$A_{us}^\mu$. (We follow the notation in Refs.~\cite{bfps,cbis}, but for
simplicity will often suppress the momentum label $p$ on the collinear fields.)
These degrees of freedom can interact in a local manner in Lagrangians and
currents.  This is in contrast with collinear-soft couplings whose
interactions are mediated by offshell fluctuations~\cite{bpssoft}, and appear in
external operators. We comment on collinear-soft interactions at the end of the
paper.

The derivation of the leading order collinear quark and gluon Lagrangians ${\cal
L}_{\xi\xi}^{(0)}$ and ${\cal L}_{cg}^{(0)}$ can be found in
Ref.~\cite{bfps,bpssoft}, and a description of the gauge symmetries of SCET can
be found in Refs.~\cite{cbis,bpssoft}. For details on power counting we refer to
Ref.~\cite{bpspc}.  The heavy-to-light currents at large energy, $J_{hl}$, were
derived to leading order in Ref.~\cite{bfps}, including one-loop matching
for all the Wilson coefficients.  The running of these Wilson coefficients was
considered in Refs.~\cite{bfl,bfps}.

In the context of the SCET, power suppressed corrections were first considered
in Ref.~\cite{chay}, and the ${\cal O}(\lambda)$ suppressed currents $J_{hl}$
and collinear quark Lagrangians were derived. The authors showed that a
reparameterization invariance (RPI) uniquely fixes the Wilson coefficients of
their subleading currents and Lagrangian in terms of the leading order
coefficients.\footnote{A similar application of Lorentz invariance was used to
derive constraints on the form of higher-twist contributions to structure
functions in deep inelastic scattering in \cite{DIS}. For this case, invariance
under changes in the light-cone vector $\bn_\mu$ was used to derive constraints
on matrix elements $\langle p|T \bar \psi(0)\Gamma \psi(\lambda
\bn_\mu)|p\rangle$.} In Ref.~\cite{mmps} the RPI of SCET was extended to the
most general three classes (I,II,III), and the multipole expansion of the
collinear quark Lagrangian was treated to higher orders in $\lambda$ and were
shown not to receive anomalous dimensions.  In Ref.~\cite{bpspc} the presence of
additional ${\cal O}(\lambda)$ heavy-to-light currents was pointed out that were
missing in Ref.~\cite{chay}.

The study of power corrections in SCET was continued in Ref.~\cite{bcdf} and
several important results were obtained for mixed usoft-collinear operators. In
particular the mixed usoft-collinear quark Lagrangian ${\cal L}_{\xi q}$ was
first considered and was derived to ${\cal O}(\lambda,\lambda^2)$ working at
tree level, but to all orders in attachments of $\bn\mcdot A_n\sim\lambda^0$
gluon fields.  In a similar fashion heavy-to-light currents were derived to
${\cal O}(\lambda^2)$, and linear combinations of currents that are invariant
under the three types of RPI were identified. It was also shown that the
operators in ${\cal L}_{\xi q}$ are not renormalized based on an analysis of
arbitrary $N$-loop diagrams in the hard region of QCD. The mixed usoft-collinear
quark Lagrangian ${\cal L}_{\xi q}$ was extended to a gauge invariant form with
covariant derivatives in Ref.~\cite{Feldmann}.

The purpose of the present paper is to answer some open questions regarding our
knowledge of the power suppressed usoft-collinear Lagrangian and heavy-to-light
currents. This includes the number of $J_{hl}$ currents at ${\cal O}(\lambda)$,
since even at tree level the full reduction of Dirac structures has not yet been
implemented. For both $J_{hl}$ and ${\cal L}_{\xi q}$ we also construct a
complete basis which is valid for matching at any order in $\alpha_s$, and
therefore includes all operators that can be induced by radiative corrections or
operator mixing.  We work in the most general possible frame throughout
(eg.~allowing $v_\perp\ne 0$, $v\mcdot n\ne 1$), and consider all the
restrictions from RPI including the transformation of Wilson coefficients.
Finally, we include the mixed usoft-collinear pure glue Lagrangian beyond LO
(which follows from an extension of work in Refs.~\cite{bpssoft,mmps}).  The
above results are obtained by considering the full implications of RPI, and
including all possible operators allowed from collinear gauge invariance, power
counting, and the reduction of Dirac structures from the effective theory
fields.\footnote{ Note that in deriving the complete basis for $J_{hl}$ we
  restrict ourselves to ${\cal O}(\lambda)$ which is one order less than the
  order to which the tree-level matching results are known from
  Ref.~\cite{bcdf}.  We treat ${\cal L}_{\xi q}$ to ${\cal O}(\lambda^2)$, and
  give a detailed account of how the gauge invariant form in Ref.~\cite{bps4}
  was derived.  In cases where our results are restricted to those in
  Refs.~\cite{chay,bcdf} we find agreement, as discussed in more detail
  in the body of the paper. The results derived here are sufficient for the
  proof of a factorization theorem for heavy-to-light form factors to all orders
  in $\alpha_s$ and leading order in $1/Q$~\cite{bps4}.  }

For the heavy-to-light currents at ${\cal O}(\lambda)$ an important result we
find is a new type of ``three-body'' currents, which have not been previously
considered in the literature.\footnote{In the final stages of this paper,
Ref.~\cite{NH} appeared where soft-collinear light-to-light currents are
considered. Although different from the usoft-collinear heavy-to-light case
studied here, we note that 3-body currents were also found.  Further remarks are
left to a note added at the end.} In Refs.~\cite{chay,bpspc,bcdf} the attention
was restricted to SCET operators of two-body type $J=(\bar\xi \ldots W)(h_v)$,
where the two products in parenthesis are collinear gauge invariant, and the
ellipses denote combinations of collinear derivatives.  Beyond tree level but at
the same order in $\lambda$, we find that three-body structures can appear for
some of the currents, having the form ${\cal J} = (\bar\xi \ldots W) (W^\dagger
\ldots W) (h_v)$ with three collinear gauge invariant factors.  We show the RPI
can be used to determine for which currents this happens. We also show that RPI
greatly restricts the form of the three-body operators, so that they always
involve a collinear gluon field strength. The two-body operators have hard
Wilson coefficients which are functions of a single parameter $C(\omega_1)$,
while the new three-body operators have two parameter coefficients
$C(\omega_1,\omega_2)$.  Analogous three-body structures could appear in the
usoft-collinear Lagrangian ${\cal L}_{\xi q}$ at higher orders in perturbation
theory, however using constraints from symmetries of SCET we prove that this
does not occur.

Our results are relevant to the study of decay channels for $B$ mesons which
involve energetic hadrons in the final state.  For instance, the results derived
in this paper are necessary ingredients in the factorization formula for
heavy-to-light form factors proven in Ref.~\cite{bps4} (for earlier work on
factorization in heavy-to-light form factors see Ref.~\cite{early,bf}, and for
results from QCD sum rules see Refs.~\cite{QCDsumrules}). The factorization
theorem is valid to all orders in $\alpha_s$ and leading order in $1/Q$,
$Q=\{m_B,E\}$, and separates contributions from the scales $p^2\sim Q^2$,
$p^2\sim Q\LQCD$, and $p^2\sim \LQCD^2$, where $\Lambda$ is a hadronic scale. It
states that a generic form factor can be split into two types of contributions
$F = f^{\rm F}(Q) + f^{\rm NF}(Q)$ where~\cite{bps4}
\begin{eqnarray}\label{fFintro}
 f^{F}(Q) &=& N_0 \int_0^1\!\!\!\! dz\! 
    \int_0^1\!\!\!\! dx\! \int_0^\infty\!\!\!\!\! dr_+ \,
    T(z,Q,\mu_{\rm 0}) \\
 && \quad \times J(z,x,r_+,Q,\mu_{\rm 0},\mu) \phi_M(x,\mu) 
  \phi_B(r_+,\mu) \,, \nn\\
 f^{\rm NF}(Q) &=&C_k(Q,\mu)\: \zeta_k^M(Q,\mu) \,,
 \label{fNFintro}
\end{eqnarray}
$N_0=f_B f_M\, m_B/(4 E^2)$, and the two terms both scale as $1/Q^{3/2}$. This
scaling is model independent and is in agreement with that derived from QCD Sum
Rules~\cite{CZ}.  In Eq.~(\ref{fFintro}) $\phi_M$ and $\phi_B=\phi_B^{\pm}$ are
standard nonperturbative light-cone distribution amplitudes, c.f.~\cite{GN,bf}.
The hard coefficients $C_k$ and $T$ can be calculated in an expansion in
$\alpha_s(Q)$ and are simply related to the Wilson coefficients of the ${\cal
  O}(\lambda^0,\lambda^1)$ current operators $J_{hl}$.  The jet function $J$ is
dominated by momenta $p^2\simeq Q\Lambda$. If we wish to expand in
$\alpha_s(\sqrt{Q\LQCD})$ then using the techniques developed in
Ref.~\cite{bps4}, $J$ is {\em calculable} in terms of time-ordered products of
the ${\rm SCET}_{\rm I}$ operators $J_{hl}$ and ${\cal L}_{\xi q}$ that we study
here. At tree-level [ie. ${\cal O}(\alpha_s(\sqrt{Q\LQCD})^1\,\alpha_s(Q)^0)$]
one finds that $J$ contains a $\delta(z-x)$, and in ratios of form factors the
results for $f^F$ then agree with terms computed in Ref.~\cite{bf}.  The $z$
dependence first shows up at ${\cal O}(\alpha_s^2)$ as does possible dependence
on $\phi_B^-$. However, as we show in section~\ref{Bpisection} it is possible to
absorb the $\phi_B^-$ terms into a redefinition of the $\zeta_k^M$ to all orders
in perturbation theory.

The factorization formula provides a clean separation of the ``soft''
non-factorizable (NF) contributions and ``hard'' factorizable (F) terms without
double counting.  It also gives us a procedure to systematically improve the
predictions to any order in perturbation theory at leading order in $1/Q$.  The
value of $T$ and $C_k$ can depend on which heavy-to-light process we consider,
whereas $\phi_M$ and $\phi_B^{\pm}$ are universal functions.  The $\zeta_k$'s
are also universal since only a $\zeta^M(E)$ appears for decays to pseudoscalars
$M$, and a $\zeta^M_\perp(E)$ and $\zeta^M_\parallel(E)$ appear if $M$ is a
vector meson. The jet functions $J$ are common among certain classes of form
factors and also do not depend on the precise state (eg.  $\pi$ or $\eta$).  The
$f^{NF}$ terms satisfy the so-called large energy form factor
relations~\cite{bps4}, as expected from the prior loose definitions of these
terms as ``soft'' contributions~\cite{charles,bf,bfps}.\footnote{These relations
  were first derived in Ref.~\cite{charles} using LEET~\cite{dg}.  However for
  studying energetic hadrons with QCD the LEET framework is known to be
  inconsistent~\cite{Ugo,bfl}, for instance it does not bind an energetic
  quark-antiquark pair into a meson in heavy-to-light decays~\cite{pw}.}  Note
that we have not bothered to separate $p^2\sim Q\Lambda$ and $p^2\sim \Lambda^2$
fluctuations in the $\zeta_k^M$ functions, since it is not clearly beneficial
phenomenologically. The factorization theorem does tell us that $\zeta_k^M\sim
(\Lambda/Q)^{3/2}$, however it does not distinguish between factors of $m_b$ and
$E$ in this $Q^{-3/2}$. It also does not numerically favor the $f^F$ or $f^{NF}$
term, for instance it is possible that the leading $\alpha_s(\sqrt{Q\LQCD})$ in
$J$ is compensated by an analogous factor in $\zeta_k^M$.

We start in Section II by reviewing the general constraints imposed on SCET
operators following from collinear gauge invariance, spin structure reduction,
and reparameterization invariance. In Section III we study the implications of
these predictions for the subleading usoft-collinear Lagrangian ${\cal L}_{\rm
uc}$.  In Section IV we present detailed results for SCET currents.  Using the
example of the scalar current as the pedagogical example, we demonstrate the
construction of the complete basis of $O(\lambda)$ operators contributing to the
weak currents, which closes under RPI transformations.  Explicit results are
then also derived for the pseudo-scalar, vector, axial-vector, and tensor
heavy-to-light currents to ${\cal O}(\lambda)$.  In Section V we summarize the 
one-loop matching results for the currents, give explicit results for
${\cal L}_{\xi q}$ Feynman rules, and discuss the basis of currents in the
particular frame $v_\perp=0$, $n\mcdot v=1$.

\OMIT{
\input ps1con.tex
\input ps1Luc.tex
\input ps1Js.tex
\input ps1Jv.tex
\input ps1Jax.tex
\input ps1Jt.tex
\input ps1mat.tex
\input ps1Bpi.tex
\input ps1Conclusion.tex
}

\section{Operator Constraints in SCET} \label{sect_const}

In this section we briefly review the symmetries and structure of SCET which
will be important for our construction of operators. We refer to
Refs.~\cite{bfl,bfps,cbis,bpssoft,chay,mmps} for more details.

SCET includes infrared degrees of freedom corresponding to the relevant low
energy scales in the problem.  These are typically those with momentum that are
collinear $p_c^\mu\sim Q(\lambda^2,1,\lambda)$, soft $p_s^\mu\sim Q(\lambda,
\lambda, \lambda)$, or ultrasoft (usoft) $p_{us}^\mu \sim Q(\lambda^2, \lambda^2,
\lambda^2)$, where the components here are in a light-cone basis $(+,-,\perp)$. 
Each type of mode has effective theory quark and gluon fields, which are then
organized into operators with a well-defined power counting in $\lambda$. It is
convenient to introduce light-cone unit vectors $(n_\mu,\bar n_\mu)$ satisfying
$n^2 = \bar n^2 = 0$, $n\mcdot \bar n = 2$, in terms of which a vector has
components $p^\mu=(n\cdot p, \bn\cdot p, p_\perp^\mu)$. The couplings of the
fields are described by an effective Lagrangian, while the couplings to external
sources appear as additional operators or currents. Both the Lagrangian and
currents are constructed such that they include constraints from power counting,
spin symmetries, and collinear and (u)soft gauge invariance.  

The soft-collinear effective theory also contains a kinematical
reparameterization invariance symmetry.  Lorentz invariance is broken by
introducing the vectors $n$ and $\bar n$, but is restored order by order in
$\lambda$, by requiring invariance of operators under a simultaneous change in
$n$ and/or $\bn$ and compensating changes in the effective theory fields. This
reparameterization invariance (RPI) symmetry of SCET was first considered in
Ref.~\cite{chay}, and was then extended to the most general three classes
(I,II,III) of allowed transformations in Ref.~\cite{mmps}.\footnote{The nature
  of Lorentz symmetries on the light-cone are well known~\cite{lcone}. The new
  point in SCET~\cite{chay,mmps} is that for any collinear process these
  symmetries are realized in a way that leads to non-trivial restrictions both
  on operators at a given order in the power counting {\em and} between
  operators at different orders in the power counting.} The three types are
defined by the infinitesimal change they induce on the light-cone unit vectors:
type-I ($n\to n+\Delta_\perp$), type-II ($\bn\to \bn+\varepsilon_\perp$), and
type-III ($n\to (1+\alpha) n, \bn\to (1-\alpha) \bn$). Here
$\alpha\sim\varepsilon_\perp\sim \lambda^0$, while $\Delta_\perp\sim \lambda$.
It is the analog of the reparameterization invariance of heavy quark effective
theory (HQET) under changes in the heavy quark velocity $v$~\cite{LM}, where
$v^2=1$.  We will use HQET for heavy quark fields~\cite{bbook}.

The restrictions we consider for finding the most general set of power
suppressed gauge invariant operators are:
\begin{itemize}
 \item[i)] Power counting and gauge invariance which determine what basic
     building blocks are allowed at the order we are considering.
 \item[ii)] What auxiliary vectors are available (such as $n$, 
     $\bn$, $v$, $\ldots$), which can be used to construct the most general 
     set of allowed scalars/tensors/Dirac structures.
 \item[iii)] Eliminate operators which are redundant by integration by parts,
             or equations of motion.
 \item[iv)] Impose type-III reparameterization invariance. If a non-trivial
 invariant can be formed with the label operators, such as ($n\cdot v\: \bnP$),
 then include Wilson coefficients that depends on these quantities.
 \item[v)] Impose all constraints from type-I and type-II reparameterization 
     invariance.
\end{itemize}

To impose the five constraints we start by writing minimal sets of independent
operators compatible with the general principles in i), ii), and iii).  We then
require RPI invariance order by order in the $\lambda$ power counting. To do
this we found it useful to split the RPI transformations into two categories,
those that act within the order we are considering $\delta_j^{(\lambda^0)}$, with
$j=$I,II,III, and those which connect operators to one higher order
$\delta_j^{(\lambda)}$. At leading order the type-II and type-III
$\delta_j^{(\lambda^0)}$ transformations already provide non-trivial constraints
on the allowed form of operators.  In contrast the $\delta_j^{(\lambda)}$
transformations allow us to derive relations valid to all orders in $\alpha_s$
between the Wilson coefficients of operators at different orders in $\lambda$,
These relations are similar to the case of RPI in HQET~\cite{LM,rpiother}, where
we note in particular the important relations derived for coefficients of
subleading heavy-to-heavy currents in Ref.~\cite{rpiNeubert}. We start by
summarizing restrictions that follow from collinear gauge invariance and power
counting in section~\ref{section_ginvpc}, spin structure reductions in
section~\ref{section_dirac}, and RPI in section~\ref{section_rpi}.

To separate the momentum scales we follow Ref.~\cite{bfps} and use collinear
quark fields $\xi_{n,p}(x)$ (and gluon fields $A_{n,p}^\mu(x)$) which have
momentum labels $p$ for the large components of the collinear momenta, and
residual coordinates $x^\mu\sim 1/\lambda^2$~\cite{bfps}. Thus, all derivatives
on collinear fields are the same size as derivatives on usoft fields,
$\partial^\mu\sim \lambda^2$. This setup implements the multipole expansion in
momentum space. Note that our analysis of power corrections differs from
Ref.~\cite{bcdf} in two ways, the first being that in Ref.~\cite{bcdf} the
momentum scales were separated by performing the multipole expansion in position
space, which however leads to an equivalent formulation.  We do find that
concise results for the power suppressed corrections are obtained with the
momentum space version. Secondly, we derive our basis of operators and implement
all symmetry constraints working order by order in the power counting, rather
than constructing invariants and then expanding in $\lambda$.  This made it
simpler to derive a complete gauge invariant basis at the desired order while
working in a general frame.

\subsection{Power Counting and Gauge Invariance} \label{section_ginvpc}

The SCET is derived from QCD by integrating out fluctuations with $p^2\gg
Q^2\lambda^2$, where in typical processes $\lambda=(\Lambda_{\rm QCD}/Q)^k$ with
$k=1$ or $k=1/2$. Infrared fluctuations are then described by effective theory
fields. A gauge invariant power counting for fields can be fixed by demanding
that the kinetic terms in the action are order $\lambda^0$.  For the collinear
fields this gives $\xi_{n}\sim \lambda^0$ for the quarks, and $(n\mcdot
A_n,\bn\mcdot A_n, A_n^{\perp\mu})\sim (\lambda^2,\lambda^0,\lambda)$ for the
collinear gluons, $h_v\sim q\sim\lambda^3$ for usoft quarks, and $A_{us}^\mu\sim
\lambda^2$ for usoft gluons~\cite{bfl,bfps}. Derivatives on these fields count
as $\partial^\mu\sim \lambda^2$. The larger collinear momenta are picked out by
introducing label operators $\bnP\sim\lambda^0$ and $\cP^\mu_\perp \sim 
\lambda$~\cite{cbis}.  For example $\bnP\,\xi_{n,p}=(\bn\!\cdot\!  p)\, 
\xi_{n,p}$. For notational convenience we define collinear covariant derivatives
\begin{eqnarray}
  i\bn\mcdot D_c=\bnP+g\bn\mcdot A_{n}\,,\qquad
  iD_c^{\perp\mu} = \bnP_\perp^\mu + gA_{n}^{\perp\mu}\,,
\end{eqnarray}
and ultrasoft covariant derivatives
\begin{eqnarray}
  i\bn\mcdot D_{us}=i\bn\mcdot\partial+g\bn\mcdot A_{us}\,,\qquad
  iD_{us}^{\perp\mu} = i\partial_\perp^\mu + gA_{us}^{\perp\mu}\,.
\end{eqnarray}
For the $n^\mu$ components, it is only the combination
\begin{eqnarray}
 in\mcdot D = i n\mcdot \partial + g n\mcdot A_n + g n\mcdot A_{us}\,,
\end{eqnarray}
that ever appears. In general a derivative without a subscript involves the sum
of the collinear and usoft pieces, $D^\mu = D_c^\mu+D_{us}^\mu$, and it is this
combination which is RPI invariant~\cite{mmps} (implying that the anomalous
dimensions of terms that appear in the multipole expansion are related).

Integrating out the offshell fluctuations builds up a collinear Wilson line,
$W$, built out of collinear gluon fields which are not suppressed in
the power counting~\cite{cbis}
\begin{eqnarray} \label{W}
  W &=& \Big[ \sum_{\rm perms} \exp\Big( -\frac{g}{\bnP}\: \bn\cdot A_{n,q}(x) \
  \Big) \Big] \,,
\end{eqnarray} 
where the label operators only act on fields inside the square brackets.  Up to
the important fact that $W$ has been multipole expanded, it is the Fourier
transform of a standard position space Wilson line, $W(-\infty,x)$.  Factors of
$W\sim\lambda^0$ can be included in operators without changing the order in the
power counting. However, their location is restricted by collinear gauge
transformations, $U_c$, under which $W\to U_c W$~\cite{cbis}.  Since $\bnP\sim
\lambda^0$ in the power counting the hard Wilson coefficients can be 
arbitrary functions of the momentum or momenta, $\omega_i$, picked out by these
operator, $C(\omega_i,\mu)$~\cite{cbis}. These coefficients can be computed by
matching with QCD at the hard scale $\mu\simeq Q$ and running with the
renormalization group.

If we consider a general Wilson coefficient and operator $C\otimes {\cal O}$,
then the covariant derivative
\begin{eqnarray} \label{DcW}
 i\bn\cdot D_c = W\, \bnP\, W^\dagger\,,
\end{eqnarray}
so it is always possible to put all the Wilson lines in ${\cal O}$ and the
dependence on the momenta picked out by $\bnP$ into $C$. We will find it
convenient to use the notation
\begin{eqnarray} \label{cdefn}
 ( \bar\xi_n W )_{\omega_1} 
  &=& \big[ \bar\xi_n W \delta(\omega_1\!-\!n\mcdot v\bnP^\dagger) \big]\,,\nn\\
 ( W^\dagger D_{c}^{\perp\,\mu} W)_{\omega_2} 
 &=& \big[  W^\dagger D_{c}^{\perp\,\mu} W\: 
  \delta(\omega_2\!-\!n\mcdot v \bnP^\dagger) \big]\,,
\end{eqnarray}
where again the label operators do not act outside the square brackets.  The
factor of $n\mcdot v$ is included next to $\bnP$ to make it a type-III RPI
invariant.  Thus, the momentum labels $\omega_i$ do not transform under RPI.
The products of fields in Eq.~(\ref{cdefn}) are color singlets under the
collinear gauge symmetry, so the momentum labels $\omega_i$ are gauge
invariant. These products still transform under an usoft gauge
transformation $U_{us}$ as $(\bar\xi_n W)\to (\bar\xi_n W)\, U_{us}^\dagger$ and
$( W^\dagger D_{c}^{\perp\,\mu} W)\to U_{us} ( W^\dagger D_{c}^{\perp\,\mu} W)
U_{us}^\dagger$. We will elaborate on how RPI affects Wilson
coefficients in SCET in subsection~\ref{section_rpi} below. 

For Lagrangians and currents where the variable $v^\mu$ is not available we 
can not make use of the definitions in Eq.~(\ref{cdefn}). It is still
convenient to make use of a similar notation:
\begin{eqnarray} \label{cdefn2}
 ( \bar\xi_n W )_{z_1} 
  &=& \big[ \bar\xi_n W \delta(z_1\!-\!\bnP^\dagger) \big]\,,\nn\\
 ( W^\dagger D_{c}^{\perp\,\mu} W)_{z_2} 
 &=& \big[  W^\dagger D_{c}^{\perp\,\mu} W\: 
  \delta(z_2\!-\!\bnP^\dagger) \big]\,,
\end{eqnarray}
where we use the variables $z_i$ rather than $\omega_i$. Under a type-III
transformation the $z_i$ transform like $\bn$ so the delta function
is homogeneous (and compensated by an integration measure $dz_i$).

Using the scalings for fields and derivatives the power counting for an
arbitrary diagram, $\lambda^\delta$, can be determined entirely from its
operators using~\cite{bpspc}
\begin{eqnarray} \label{pc}
  \delta = 4 \!+\! \sum_k (k\!-\!4) [V_k^C \!+\! V_k^S \!+\! V_k^{SC}] 
     \!+\! (k\!-\!8) V_k^{U}\,. 
\end{eqnarray}
Here $V_k^{C,S,SC,U}$ count the number of order $\lambda^k$ operators which have
collinear fields, soft fields, both, or neither respectively. For any operator
the power of $k$ is derived by adding up the powers of $\lambda$ in its
components, so for instance $\bar\xi_n \bnslash in\mcdot D
\xi_n\sim \lambda^4$, counts as $V_4^C=1$. Since the operators are gauge 
invariant so is their value of $k$ and also the power counting of any diagram
using the result for $\delta$ in Eq.~(\ref{pc}). In this paper we focus on
operators with $V_k^S=V_k^{SC}=0$.

We have also found it convenient to define additional pure gluon operators.
In particular we will use the purely collinear field strength
\begin{eqnarray} \label{Bdefn}
 ig B_c^{\perp\mu} &=& [i\bn\mcdot D^c,i D_c^{\perp\mu}] \,.
\end{eqnarray}
We will also make use of the mixed tensors 
\begin{eqnarray}
  ig\: n\mcdot M = ign\mcdot B_c = [i\bn\mcdot D^c,in\mcdot D ]\,,\quad\qquad\, 
  ig \Mslash_\perp = [i\bn\mcdot D^c,i\Dslash_\perp^{\,us}]\,,\quad 
\end{eqnarray} 
In fact the operators $\Mslash_\perp$ and $n\cdot M$, together with
$ig\,\bn\mcdot M= [i\bn\mcdot D^c,i\bn\mcdot D^{us}]$ can be combined into a
single object closed under usoft Lorentz transformations, which transforms in
the desired way under the collinear and usoft gauge symmetries
\begin{eqnarray}
  ig \Mslash = [i\bn\mcdot D^c,i\Dslash^{\,us}
   +\frac{\bnslash}{2} gn\mcdot A_n]\,.
\end{eqnarray}
Finally the following results for manipulating covariant derivatives on Wilson
lines also prove to be useful
\begin{eqnarray} \label{toBD}
 (W^\dagger \DSppPl W) 
  &=& \big[W^\dagger \DSppPl W\big] - \cPslash_\perp^{\,\dagger} 
  = \Big[\frac{1}{\bnP} W^\dagger ig \Bslash_c^\perp W\Big] 
  -\cPslash_\perp^{\,\dagger} 
  \,,\nn\\
 (W^\dagger \DSppPr W) 
  &=& \big[W^\dagger \DSppPr W\big]  + \cPslash_\perp 
  = \Big[\frac{1}{\bnP} W^\dagger ig \Bslash_c^\perp W\Big] 
  +\cPslash_\perp  
  \,,
\end{eqnarray}

\subsection{Reduction in Spin Structures} \label{section_dirac}

Collinear quarks and heavy usoft quarks have spinors with only two non-zero 
components. In four component notation this is encoded in projection formulae
for the fields,
\begin{eqnarray}
 P_n \xi_n =\xi_n\,,\qquad P_v h_v = h_v \,,
\end{eqnarray}
where $P_n=(\nslash \bnslash)/4$ and $P_v=(1\!+\!\vslash)/2$. We also define the
orthogonal projector $P_\bn=(\bnslash \nslash)/4$ where $P_n+P_\bn=1$. A quark
bilinear with a heavy ultrasoft quark and light collinear quark therefore only
has four possible non-trivial Dirac structures. On the other hand if the heavy
ultrasoft quark is replaced by a massless ultrasoft quark which has a four
component spinor then their are eight possible Dirac structures. When generating
operators we should be careful not to include redundant Dirac
structures. Therefore, it is convenient to have a canonical basis which we can
project results onto to check their interdependence. For this purpose we choose
the basis
\begin{eqnarray} \label{basis}
 \bar\xi_n \Gamma_1\: h_v \,,\qquad 
 \Gamma_1 &=& \Big\{\frac{\bnslash}{2}, 
    \frac{\bnslash\gamma^5}{2}, \gamma_\perp^\mu \Big\}\,,\nn\\
 \bar\xi_n \Gamma_2\: q_{us} \,,\qquad 
 \Gamma_2 &=& \Big\{ 1, \frac{\bnslash}{2}, 
    \gamma^5, \frac{\bnslash\gamma^5}{2}, \gamma_\perp^\mu, 
     \frac{\bnslash\gamma_\perp^\mu}{2} \Big\}\,.
\end{eqnarray}
Any general Dirac structure can be projected onto a linear combination of terms
in these basis with the help of the following formulae
\begin{eqnarray} \label{proj1}
\bar\xi_n\Gamma\: h_v &=& \bar\xi_n \Gamma_1\: h_v\,,\nn \\[5pt]
 \Gamma_1 &=& 
  \frac{\bnslash}{2}\: {\rm tr}\Big[\frac{\nslash}{2} P_\bn\Gamma P_v\Big] 
  -\frac{\bnslash\gamma_5}{2}\: {\rm tr}\Big[\frac{\nslash}{2}\gamma_5 P_\bn 
   \Gamma P_v\Big]
  + \gamma_\perp^\mu\: {\rm tr}\Big[\gamma^\perp_\mu P_\bn \Gamma P_v \Big]\,.
\end{eqnarray}
and
\begin{eqnarray} \label{proj2}
\bar\xi_n\Gamma\: q_{us} &=& \bar\xi_n \Gamma_2\:q_{us}\,,\nn \\[5pt]
 \Gamma_2 &=& 
  {\bf 1}\: {\rm tr}\Big[\frac{\nslash\bnslash}{8}\Gamma\Big] 
  +\frac{\bnslash}{2}\: {\rm tr}\Big[\frac{\nslash}{4} \Gamma \Big] 
  +\gamma_5 \: {\rm tr}\Big[\frac{\nslash\bnslash\gamma_5 }{8}\Gamma\Big] 
  -\frac{\bnslash\gamma_5}{2}\: {\rm tr}\Big[\frac{\nslash\gamma_5}{4} 
   \Gamma\Big]\nn\\
  && + \gamma_\perp^\mu\: {\rm tr}\Big[ \frac{\gamma^\perp_\mu\nslash\bnslash}{8}
    \Gamma \Big]
  - \frac{\bnslash\gamma_\perp^\mu}{2}\: {\rm tr}\Big[\frac{\nslash
    \gamma^\perp_\mu}{4}     \Gamma \Big]
  \,.
\end{eqnarray}
The number of independent structures is quite logical, for $\bar\xi_n
\Gamma h_v$ each field is determined by two-component spinors and there are 
$2\times 2=4$ terms in the basis. For $\bar\xi_n \Gamma q_{us}$ only the
collinear spinor has two-components and there are $2\times 4=8$ terms in our
basis.  Our choice of basis in Eq.~(\ref{proj1}) differs from Ref.~\cite{bfps}
where the choice $\Gamma_1 = \{1,\gamma_5,\gamma_\perp^\mu\}$ was used, and
calculations were given in a frame where $v\mcdot n=1$ and $v_\perp=0$. When
$v^\mu$ is kept arbitrary we have found the basis in Eq.~(\ref{basis}) is more
convenient since it retains its orthonormality in an arbitrary frame.

The projections formulae in Eq.~(\ref{proj1}) can be used to reduce the possible
Dirac structures in constructing a complete basis of operators. It is convenient
to define $r_\perp^\mu \equiv i
\epsilon^{\mu\alpha\beta\gamma} n_\alpha \bn_\beta v_\gamma$ and $v_\perp^\mu =
v^\mu - n\mcdot v\, \bn^\mu/2 - \bn\mcdot v\, n^\mu/2$, since then
$v_\perp^\mu$, $r_\perp^\mu$, $n^\mu$, $\bn^\mu$ form a complete vector
basis. To reduce the Dirac structures we can use relations such as
\begin{eqnarray} \label{rel1}
 \vslash_\perp &\stackrel{\cdot}{=}& 1 - \frac{n\mcdot v}{2} \bnslash \,,\qquad
  \bnslash\vslash_\perp \stackrel{\cdot}{=}\bnslash - 2\bn\mcdot v\,,\qquad
  \gamma_\perp^\mu  \stackrel{\cdot}{=}\gamma_\mu 
   - \frac{n\mcdot v}{2} \bnslash \,,\nn\\
 r_\perp^\mu\gamma_5 & \stackrel{\cdot}{=}  & -2 v_\perp^\mu 
   +2 \gamma_\perp^\mu +\bnslash \gamma_\perp^\mu n\mcdot v\,,\qquad
 \bnslash r_\perp^\mu \gamma_5 \stackrel{\cdot}{=} 2\bnslash v_\perp^\mu 
   -4\bn\mcdot v\,\gamma_\perp^\mu -2\bnslash \gamma_\perp^\mu\,,\nn\\
 i\epsilon^{\mu\nu\alpha\beta}\!\!\!\!\! &&\!\!\!\!\! n_\alpha v_\beta 
  \,\frac{\bnslash\gamma_5}{2} 
   \stackrel{\cdot}{=} i\sigma_{\mu\nu} 
   - (v_\mu \gamma_\nu\!-\!v_\nu \gamma_{\mu}) -\frac{1}{2}(\bn_\mu n_\nu \!-\!
   n_\mu \bn_\nu) 
   - \frac{\bn\mcdot v}{2} (n_\mu \gamma_\nu\!-\!n_\nu \gamma_{\mu}) \nn\\
   &&\quad\qquad\quad 
   + \frac{n\mcdot v}{2} (\bn_\mu \gamma_\nu\!-\!\bn_\nu \gamma_{\mu}) \,,
  \qquad \ldots \,,
\end{eqnarray}
where the $\stackrel{\cdot}{=}$ indicates that these are only true between the
fields in Eq.~(\ref{proj1}). (The complete set of relations is rather lengthy
and is not shown.) The relations in Eq.~(\ref{rel1}) allow the structures on the
left to be traded for those on the right (with more than one iteration in some
cases). Using the projection formulae it is straightforward to show that the
most general Dirac structure possible for the LO scalar currents are
$\{1,\bnslash\}$, while the vector and axial-vector currents have the basis
shown in Eq.~(\ref{JVGamma}), and the tensor currents depend on the basis in
Eq.~(\ref{JTGamma})


\subsection{Reparameterization Invariance} \label{section_rpi}

The decomposition into collinear fields requires introducing two light-like
vectors $n$ and $\bn$, such that $n^2=\bn^2=0$ and $n\mcdot \bn=2$. These 
vectors break five of the six Lorentz generators. This part of the Lorentz
symmetry is restored order by order in the power counting by requiring invariance
under reparameterization transformations on $n$ and $\bn$~\cite{mmps}:  
\begin{eqnarray}\label{repinv}
\mbox{(I)} 
\left\{
\begin{tabular}{l}
$n_\mu \to n_\mu + \Delta_\mu^\perp$ \\
$\bn_\mu \to \bn_\mu$
\end{tabular}
\right.\qquad
\mbox{(II)} 
\left\{
\begin{tabular}{l}
$n_\mu \to n_\mu$ \\
$\bn_\mu \to \bn_\mu + \varepsilon_\mu^\perp$
\end{tabular}
\right.\qquad
\mbox{(III)} 
\left\{
\begin{tabular}{l}
$n_\mu \to (1+\alpha)\, n_\mu$ \\
$\bn_\mu \to (1-\alpha)\, \bn_\mu$
\end{tabular}
\right. \,,
\end{eqnarray}
where $\Delta^\perp\sim\lambda$, while $\varepsilon^\perp\sim \alpha \sim
\lambda^0$.  In general one has two options for constructing RPI invariants: i)
construct operators out of completely RPI invariant quantities and then expand
these in powers of $\lambda$, ii) construct operators order by order in
$\lambda$ and transform them to see what linear combinations are invariant, and
which operators are ruled out. In this paper we will adopt approach ii), since
starting with the most general gauge invariant sets and then reducing them
allows us to be confident that we do not miss operators that could arise at any
order in perturbation theory.

For our purposes it is convenient to divide the RPI transformations into two
subsets, those which include terms within the same order in $\lambda$ denoted by
$\delta_{\rm I}^{(\lambda^0)}$, $\delta_{\rm II}^{(\lambda^0)}$, and
$\delta_{\rm III}^{(\lambda^0)}$, and those which cause order $\lambda$
suppressed transformations denoted by $\delta_{\rm I}^{(\lambda)}$ and
$\delta_{\rm II}^{(\lambda)}$. All type-III transformations act within the same
order in $\lambda$ and it is easy to construct invariants under type-III. We
simply need to have the same number of $n$'s ($\bn$'s) in the numerator and
denominator, or have products of $n$ times $\bn$. The transformations of type-I
and type-II are more involved.  From Ref.~\cite{mmps} the transformations that
have terms of the same order in $\lambda$ are
\begin{eqnarray} \label{rpi0}
 && \delta_{\rm I}^{(\lambda^0)}: \qquad 
 n\mcdot {D}\to n\mcdot {D}+ \Delta^\perp\mcdot{D}_\perp\,,
   \quad
 {D}_\perp^\mu \to {D}_\perp^\mu -\frac{\Delta_\perp^\mu}{2}\ 
   \bn\mcdot {D} \,, \\
 && \delta_{\rm II}^{(\lambda^0)} : \qquad 
  \bn^\mu \to \bn^\mu +\varepsilon_\perp^\mu \,,\quad
  {D}_\perp^\mu \to {D}_\perp^\mu -\frac{n^\mu}{2}\,
    \varepsilon_\perp \mcdot {D}_\perp \nn\,, \quad
  \gamma_{\perp}^\mu \to \gamma_\perp^\mu-\frac{n^\mu}{2}\vepslash_\perp 
    -\frac{\varepsilon_\perp^\mu}{2} \nslash \,,
\end{eqnarray}
and the transformations that start one power down in $\lambda$ include
\begin{eqnarray} \label{rpi1}
 && \delta_{\rm I}^{(\lambda)}: \qquad 
  n^\mu \to n^\mu +\Delta_\perp^\mu \,,\quad
  {D}_\perp^\mu \to {D}_\perp^\mu -\frac{\bn^\mu}{2}\ 
   \Delta^\perp\mcdot {D}_\perp \,,\quad
  \bar\xi_n\to \bar\xi_n\Big(1+\frac{\bnslash\Slash{\Delslash}_\perp}{4}\Big)
  \,,\nn\\
 &&\qquad\qquad\ \gamma_{\perp}^\mu \to \gamma_\perp^\mu 
   -\frac{\Delta_\perp^\mu}{2} \bnslash
    -\frac{\bn^\mu}{2} \Delslash_\perp\,,\\
 && \delta_{\rm II}^{(\lambda)} : \qquad 
   \bn\mcdot{D}\to \bn\mcdot{D} + 
      \varepsilon_\perp\mcdot {D}_\perp \,,\quad
   {D}_\perp^\mu \to {D}_\perp^\mu -\frac{\varepsilon_\perp^\mu}{2}\,
    n \mcdot {D} \nn\,, \\
 &&\qquad\qquad\
  \bar\xi_n\to \bar\xi_n \Big(1 + \overleftarrow \Dslash_\perp
  \frac{1}{\bn\mcdot \overleftarrow {D}} \frac{\vepslash_\perp}{2} \Big)\,,\quad
  {\cal W} \to \bigg[\Big(1-\frac{1}{\bn\mcdot{D}}\: \varepsilon^\perp\mcdot 
  {D}^\perp \Big) {\cal W}\bigg] \,.
\end{eqnarray}
Here ${\cal W}$ is the RPI completed $W$, and is the Fourier transform with
respect to $y$ of a position space Wilson line involving $(\bn\mcdot
A_n\!+\!\bn\mcdot A_{us})(s\bn\!+\! x)$ taken from $s=-\infty$ to
$y$~\cite{bcdf}.  When expanded in $\lambda$, ${\cal W}= W + {\cal
  O}(\lambda^2)$, where $W$ involves only the $\bn\mcdot A_n$ field as in
Eq.~(\ref{W}).

If we start by considering LO operators then they must be invariant under the
$\delta^{(\lambda^0)}$ transformations in Eq.~(\ref{rpi0}) all by themselves.
The $\delta^{(\lambda)}$ transformations of the LO terms connect them to NLO
operator's $\delta^{(\lambda^0)}$ transformations. Since in the collinear sector
only $\delta^{(\lambda^{0},\lambda^{1})}$ terms exist this pattern repeats at
all higher orders in the power counting.  Note that here we will not need to
consider HQET RPI under the velocity $v^\mu$.  Since the transformation
$v^\mu\to v^\mu + \Delta_v^\mu$ where $\Delta_v^\mu\sim {\Lambda_{\rm
    QCD}/Q}\sim \lambda^2$, this type of RPI only needs to be taken into account
at one-higher order than the order we are working.  The combined SCET and HQET
RPI transformations were used in the ${\cal O}(\lambda^2)$ analysis of $J_{hl}$
in Ref.~\cite{bcdf}.

Finally we consider a new feature of RPI in SCET, namely how Wilson coefficients
are affected by reparameterization invariance. Our analysis is similar in spirit
to Ref.~\cite{rpiNeubert}, where heavy-to-heavy HQET currents with coefficients
depending on the change in velocity, $C(v\mcdot v')$, were analyzed.  If we
adopt the view of building invariants at all orders in $\lambda$ then the
coefficients in SCET must also be functions of invariants, such as operators
like
\begin{eqnarray}
  \bar\Psi_n  C(-i{\overleftarrow D}\mcdot {\cal V} )\: \Gamma\:  {\cal H}_v \,,
\end{eqnarray}
where $\Psi_n$, ${\cal H}_v$ are invariants including the quark fields $\xi_n$,
$h_{v}$, and ${\cal V}^\mu$ is the RPI version of the velocity
$v^\mu$~\cite{LM}.  When expanded in $\lambda$ the leading term involving the
covariant derivative in $C$ can be traded for $W$ and $\bnP$ using
Eq.~(\ref{DcW}), $C(-i n\mcdot v\: \bn\mcdot {\overleftarrow D}_c) = W C(n\mcdot
v \bnP^\dagger) W^\dagger$. Here we will use the opposite but equivalent
arrangement of starting with a current that is leading order in $\lambda$,
\begin{eqnarray}
  \bar\xi_n W\, C(n\mcdot v \bnP^\dagger)\, h_v \,,
\end{eqnarray}
and then determining how both the operators {\em and} coefficient transform
under RPI. We then determine which structures are required at one higher order
in $\lambda$ to cancel this change, and which allowed higher order
structures are left unconstrained.

\newpage
\section{Collinear-Ultrasoft Lagrangian}

In this section we discuss the mixed ultrasoft-collinear Lagrangians to ${\cal
  O}(\lambda^2)$. These actions are power suppressed~\cite{bps}, and start at
${\cal O}(\lambda)$~\cite{bcdf}.  In section~\ref{sect_uc_match} we consider the
derivation from integrating out components of the full theory field, which gives
a tree level derivation of the action (for further explanation of this approach
see Refs.~\cite{bfps,bpssoft,bcdf}). In Ref.~\cite{bcdf} this procedure was used
to derive a form for the mixed ultrasoft-collinear quark Lagrangian, but a
manifestly gauge invariant form was not determined.  In Ref.~\cite{Feldmann} the
analysis was extended to give manifestly gauge invariant operators in terms of
covariant derivatives. In section~\ref{sect_uc_match} we review the details of
how a derivation of a gauge invariant form of the action was carried out in
Ref.~\cite{bps4} where the result is purely in terms of field strengths.

However, since the analysis in section~\ref{sect_uc_match} is only valid at tree
level, it misses i) non-trivial Wilson coefficients in the tree level operators,
and ii) new operators whose coefficients can have zero tree-level matching.  In
Ref.~\cite{bcdf} point i) was addressed and it was shown diagrammatically that no
non-trivial Wilson coefficients are generated.  However, point ii) has not yet
been addressed, so additional operators could still be induced by matching at
some higher order in perturbation theory.  In section~\ref{sect_uc_rpi} we show
that both points i) and ii) can be simultaneously solved by using the full set
of symmetries of SCET when constructing operators. We also extend the derivation
to the mixed usoft-collinear pure gluon sector.

\subsection{Matching for ${\cal L}_{uc}$ at tree level, but all orders in
$\bn\mcdot A_n$ gluons} 
\label{sect_uc_match}

In this subsection we discuss in detail the matching calculation for the mixed
usoft-collinear quark Lagrangian~\cite{bps4}. The part of our discussion from
Eq.~(\ref{Luc0}) to (\ref{Lbcdf}) follows Ref.~\cite{bcdf}, but with our
momentum space notation. We start with the action ${\cal L}=\bar\psi i\Dslash\,
\psi$ and decompose it with SCET fields
\begin{eqnarray} \label{Luc0}
 {\cal L} 
  &=& \bar\xi_n \frac{\bnslash}{2} in\mcdot D\, \xi_n 
  + \bar\xi_n i\Dslash^\perp \xi_\bn + \bar\xi_n g\Aslash_c\: q_{us}
  + \bar q_{us}\: g\Aslash_c\: \xi_n + \bar q_{us} \: g\Aslash_c\: \xi_\bn 
  + \bar q_{us}\: i\Dslash_{us} \: q_{us} \nn \\
  && + \bigg[ \bar\xi_\bn i\Dslash^\perp \xi_n 
  + \bar\xi_\bn \frac{\nslash}{2} i\bn\mcdot D \xi_\bn 
  + \bar\xi_\bn g \Aslash_c q_{us} \bigg] \,,
\end{eqnarray}
where the $D$ is usoft plus collinear, $D_c$ is purely collinear, and collinear
momentum conservation has been enforced. Varying with respect to $\bar\xi_\bn$
gives an equation of motion to eliminate this field from the term in square
brackets
\begin{eqnarray}
  \xi_\bn = - \frac{1}{i\bn\mcdot D} \frac{\bnslash}{2} \Big[ 
    i\Dslash_\perp\xi_n + g \Aslash_n \: q_{us} \Big] \,,\qquad
  \bar\xi_\bn 
  = \Big[ \bar q_{us} g\Aslash_n - \bar\xi_n i\overleftarrow\Dslash_\perp 
    \Big] \frac{\bnslash}{2}  \frac{1}{i\bn\mcdot \overleftarrow D}  \,.
\end{eqnarray}
Plugging this into Eq.~(\ref{Luc0}) and expanding we find that the two collinear
quark terms exactly reproduce terms in the gauge invariant multipole expanded
action in Ref.~\cite{mmps}.\footnote{Note that in QED the ${\cal O}(\lambda)$
pure collinear quark Lagrangian can be written in terms of
$F_{\mu\nu}$~\cite{bcdf}. With the momentum space multipole expansion this
manipulation is not necessary to achieve a gauge invariant
result~\cite{chay2,rev1}.}  Using Eq.~(\ref{pc}) the terms with two ultrasoft
quarks and $\ge 2$ collinear gluons first show up at $\delta=3$, ie. ${\cal
O}(\lambda^3)$~\cite{bcdf}, and are therefore neglected.  The mixed
usoft-collinear quark terms are
\begin{eqnarray} \label{preExpn}
{\cal L}_{\xi q} = \Big[\bar\xi_n g \Aslash_n \,q_{us} 
  + \bar\xi_n \frac{\bnslash}{2}
  i\Dslash_\perp \frac{1}{i\bn\mcdot D} g \Aslash_n \, q_{us} \Big] +
  \Big[\bar q_{us} \, g \Aslash_n\, \xi_n 
  + \bar q_{us}\, g\Aslash_n 
  \frac{1}{i\bn\mcdot D} i\Dslash_\perp \frac{\bnslash}{2}\, \xi_n\Big] \,.
\end{eqnarray}
Taking Eq.~(\ref{preExpn}) and expanding to second order in $\lambda$ gives
\begin{eqnarray}
 {\cal L}_{\xi q}^{(1)} 
  &=& \bar\xi_n \Big( g\Aslash_\perp^c - i \Dslash_\perp^c 
   \frac{1}{i\bn\mcdot D_c}\: g\bn\mcdot A_c \Big) q_{us} \mbox{ + h.c.}
  \,, \\
 {\cal L}_{\xi q}^{(2)} &=& \bar\xi_n \frac{\bnslash}{2} \Big( gn\mcdot A^c + 
  i \Dslash_\perp^{\,c}  \frac{1}{i\bn\mcdot D_c}\: g\Aslash_\perp^{\, c} 
 \Big) q_{us} 
  - \bar\xi_n i\Dslash_\perp^{\,us} \frac{1}{i\bn\mcdot D_c}\: 
    g\bn\mcdot A^{c} \: q_{us}
  \mbox{ + h.c.}  \,, \nn
 \hspace{1.3cm}
\end{eqnarray}
where the superscripts denote the power suppression in $\lambda$ and
$\lambda^2$.  In each of the last terms in ${\cal L}_{\xi q}^{(1,2)}$ we use
$1-W= 1/(i\bn\mcdot D_c)\, g\bn\mcdot A_c$, and in the first term of ${\cal
L}_{\xi q}^{(2)}$ we write $g\Aslash_\perp^c = i\Dslash_\perp^c - 
\cPslash^\perp$ and then $q_{us}=[W+(1-W)] q_{us}$. Thus,
\begin{eqnarray} \label{Lbcdf}
 {\cal L}_{\xi q}^{(1)} &=& \bar\xi_n \big( i \Dslash_\perp^c W 
  - \cPslash_\perp \big) q_{us} \mbox{ + h.c.} 
  \,,\nn\\
 {\cal L}_{\xi q}^{(2)} &=& \bar\xi_n \frac{\bnslash}{2} \Big( gn\mcdot A^c 
  \!+\! i \Dslash_\perp^{\,c} \frac{1}{i\bn\mcdot D_c}\: (i\Dslash_\perp^{\, c}
  \!-\!\cPslash_\perp) \Big)[W\!+\!(1\!-\!W)] q_{us}
  + \bar\xi_n i\Dslash^{\,us}_\perp (W\!-\! 1)\: q_{us} \mbox{ + h.c.}\nn\\
%
 &=& \bar\xi_n \frac{\bnslash}{2} \Big( gn\mcdot A^c + 
   i \Dslash_\perp^{\,c} \frac{1}{i\bn\mcdot D_c}\: i\Dslash_\perp^{\, c} 
   \Big)W q_{us}  
 -\bar\xi_n \frac{\bnslash}{2} i \Dslash_\perp^{\,c} \frac{1}{i\bn\mcdot D_c}\:
   \cPslash_\perp \, q_{us} \nn \hspace{1.3cm} \\   
  && + \bar\xi_n \Big( \frac{\bnslash}{2} in\mcdot D^{us} + i\Dslash^{us}_\perp
   \Big) (W\!-\!1)\: q_{us}
  \mbox{ + h.c.} \,.
\end{eqnarray}
In manipulating ${\cal L}_{\xi q}^{(2)}$ we used the fact that integration by
parts is allowed on the $(1-W)q_{us}$ term and we can then use the equation of
motion for the collinear quark to give a term $(-in\mcdot D_{us})(1-W)$ which we
collected with the $i\Dslash^{\,us}_\perp(W-1)$ term.  The result in
Eq.~(\ref{Lbcdf}) agrees with Ref.~\cite{bcdf}, up to the fact that we performed
the multipole expansion in momentum space.

In Eq.~(\ref{Lbcdf}) we did not drop the $\cPslash_\perp\, q=0$ terms since we
want to make explicit the fact that it is the combination $(i \Dslash_\perp^c W
- \cPslash_\perp) = [i\Dslash_\perp^c W]$ which starts with at least
one-collinear gluon.  Written this way it appears that our ${\cal L}_{\xi
  q}^{(1)}$ is not collinear gauge invariant.  In the transformed result the
non-invariant term cancels if we use $\cPslash_\perp q_{us}=0$, but then it is
not explicit that the operator starts with one-collinear gluon, so ${\cal
  L}_{\xi q}^{(1)}$ has either one or the other explicit.  For ${\cal L}_{\xi
  q}^{(2)}$ Eq.~(\ref{Lbcdf}) still involves the gluon field $A_n^\mu$ so the
gauge invariance of this expression is not at all clear. However, the above
considerations indicates that it should be possible to write all the terms in
Eq.~(\ref{Lbcdf}) in terms of gluon field strengths, and thereby achieve a
manifestly gauge invariant action that starts with one-collinear gluon. This
derivation was carried out in Ref.~\cite{bps4}, but no details of the
calculation were described there. These details are described below in
Eqs.~(\ref{expnBM}) through (\ref{Luc2}).


To proceed we note that using Eq.~(\ref{Bdefn}), $ ig B^\nu_{c\perp} W =
[i\bn\mcdot D_c,iD_{c}^{\perp\nu}] W = i\bn\mcdot D_c\, iD_{c}^{\perp\nu} W 
-iD_{c}^{\perp\nu} W \bnP$. Making similar manipulations for $n\mcdot M$ and
$\Mslash_\perp$ we can write
\begin{eqnarray} \label{expnBM}
  ig \Bslash_c^{\perp} W &=&  
   i\bn\mcdot D_c\, (i\Dslash_c^{\perp} W-\cPslash_\perp) - 
   \{ i\Dslash_c^{\perp} W -\cPslash_c^{\perp} \}\bnP + 
   \{ g\bn\cdot A^c\} \cPslash_{\perp} \,. \nn\\
  ig n\mcdot M W  &=& i\bn\mcdot D_c\, in\mcdot D W-\bnP in\mcdot D_{us} -
 \{ in\mcdot D W-in\mcdot D_{us} \} \bnP \,,
 \nn\\
 i g \Mslash_\perp W &=& 
  i\bn\mcdot D_c i \Dslash_\perp^{us} W -\bnP i\Dslash_\perp^{us}
 + \{i\Dslash_\perp^{us} - i\Dslash_\perp^{us} W \}\bnP \,.
\end{eqnarray}
Now we take purely usoft fields on the right, and divide on the left by
$i\bn\mcdot D_c$.  In Eq.~(\ref{expnBM}) the terms in curly brackets start at
one-collinear gluon, so even in the presence of $1/(i\bn\mcdot D_c)$ these terms
are non-singular and can safely be dropped using that the label operators give
zero on the usoft field. This gives
\begin{eqnarray} \label{manipBM}
 \frac{1}{i\bn\mcdot D_c}\: ig \Bslash^{\perp} W q_{us} 
     &=& (i \Dslash^{\perp}_c W-\cPslash^{\perp} ) q_{us} \,,\nn\\
 \frac{1}{i\bn\mcdot D_c}\: ig n\mcdot M W q_{us} &=&
   (in\mcdot D W - in\mcdot D_{us}) q_{us} 
   - (W-1) in\mcdot D_{us} q_{us}   \,,\nn\\
 \frac{1}{i\bn\mcdot D_c}\: ig \Mslash_\perp W q_{us} 
     &=& i \Dslash_\perp^{us} (W-1)q_{us} \
   - (W-1)\: i \Dslash_\perp^{us} q_{us} \,.
\end{eqnarray}
These expressions allow us to write covariant derivatives acting on Wilson lines
in terms of field strengths.

Using Eq.~(\ref{manipBM}) for ${\cal L}_{\xi q}^{(1)}$ in Eq.~(\ref{Lbcdf}),
we arrive at the final result
\begin{eqnarray} \label{Luc1}
 {\cal L}_{\xi q}^{(1)} &=& 
    \bar\xi_n \: \frac{1}{i\bn\mcdot D_c}\: ig\Bslash_c^\perp
   W  q_{us} \mbox{ + h.c.} \,.
\end{eqnarray}
This form is particularly nice since it is explicitly collinear and usoft gauge
invariant and furthermore explicitly starts at one-collinear gluon due to the
$B_\perp$. To see the gauge invariance note that under a collinear gauge
transformation $U_c$ we have $\xi_n \to U_c\xi_n$, $W\to U_c W$, $\Bslash_\perp
\to U_c\, \Bslash_\perp\, U_c^\dagger$, and $(\bn\mcdot D_c)^{-1} \to U_c 
\: (\bn\mcdot D_c)^{-1}\: U_c^\dagger$ so all factors of $U_c$ cancel. Under an
ultrasoft gauge transformation $U_u$ we have $\xi_n \to U_u\xi_n$, $W\to U_u W
U_u^\dagger $, $\Bslash_\perp \to U_u\, \Bslash_\perp\, U_u^\dagger$,
$(\bn\mcdot D_c)^{-1} \to U_u \: (\bn\mcdot D_c)^{-1}\: U_u^\dagger$, and
$q_{us}\to U_u q_{us}$ so all factors of $U_{u}$ also cancel. In
Fig.~\ref{figLuc1} in section~\ref{sect_feyn} we show the one and two gluon
Feynman rules that follow from ${\cal L}_{uc}^{(1)}$ in Eq.~(\ref{Luc1}). A
non-trivial check on our manipulations is that the same Feynman rules can be
obtained from Eq.~(\ref{Lbcdf}) by using the free equations of motion.

We now proceed to further simplify ${\cal L}_{\xi q}^{(2)}$ in
Eq.~(\ref{Lbcdf}). Using Eq.~(\ref{manipBM}) leaves
\begin{eqnarray} \label{L2p2a}
  {\cal L}_{\xi q}^{(2)} &=&  \bar\xi_n  \frac{1}{i\bn\mcdot D_c}\: 
   ig\Mslash\: W \: q_{us} 
  + \bar\xi_n \frac{\bnslash}{2} i\Dslash_\perp^{\,c} 
  \frac{1}{(i\bn\mcdot D_c)^2}\:  ig \Bslash_\perp^{\, c} W \: q_{us}
  \hspace{0.cm}\nn\\
 && + \bar\xi_n  (W-1) i\Dslash_{us} \, q_{us} \mbox{ + h.c.} \,,
\end{eqnarray}
where in the first and last terms we used the fact that $\nslash \xi_n=0$ to
write a full $\gamma^\mu$ in $\Mslash\,$ and $\Dslash_{us}$. For the last term
in Eq.~(\ref{L2p2a}) we can now use the equation of motion for the usoft quark
field to give our final result
\begin{eqnarray} \label{Luc2}
 {\cal L}_{\xi q}^{(2)} &=&  \bar\xi_n  \frac{1}{i\bn\mcdot D_c}\: 
  ig\Mslash\: W \: q_{us} + 
 \bar\xi_n \frac{\bnslash}{2} i\Dslash_\perp^{\,c} 
 \frac{1}{(i\bn\mcdot D_c)^2}\:  ig \Bslash_\perp^{\, c} W \: q_{us} 
  \mbox{ + h.c.}
\end{eqnarray}

Again in this form the action is explicitly collinear and usoft gauge invariant
and furthermore explicitly starts at one-collinear gluon due to the field
strength $B$'s and $M$'s.  In the way we have written the result it is invariant
under usoft Lorentz transformations on $x^\mu$ which separately rotate
$\gamma_\mu$ and $D^\mu_{us}$.

Finally we note that the mixed usoft-collinear quark actions in
Eqs.~(\ref{Luc1}) and (\ref{Luc2}) proved to be important for the proof of a
factorization formula for heavy-to-light decays in Ref.~\cite{bps4}. In the next
section we analyze the most general possible basis for ${\cal L}_{\xi
  q}^{(1,2)}$ beyond tree level, which follow purely from symmetry
considerations and also discuss power suppressed terms in the collinear gluon
action.


\subsection{Most General Basis for ${\cal L}_{uc}$} \label{sect_uc_rpi}

The ultrasoft-collinear quark Lagrangian can be expanded in a power series in
the parameter $\lambda$. It is not possible to construct an invariant operator
that is dimension-4 and order $\lambda^0$. Therefore, we have the series
\begin{eqnarray}
  {\cal L}_{\xi q}= {\cal L}_{\xi q}^{(1)} + {\cal L}_{\xi q}^{(2)} + \ldots \,.
\end{eqnarray}
Since this is a Lagrangian, insertions of these operators do not inject
momentum, and we are free to integrate by parts as long as we are careful not to
generate singular terms.

To construct the most general quark action ${\cal L}_{\xi q}^{(1)}$ we can use a
single collinear quark field $\xi_n\sim \lambda$, an ultrasoft quark field
$q_{us}\sim\lambda^3$, and a $D_c^{\perp\mu}\sim\lambda$. These factors give a
dimension 4 operator, and from the power counting formula in Eq.~(\ref{pc}) they
give $\delta=1$ which is the correct order for ${\cal L}_{\xi q}^{(1)}$.  To
satisfy collinear gauge invariance without changing the order in the power
counting we make use of the Wilson line $W$ to write $(\bar\xi_n W)$ and
$(W^\dagger i D_c^{\perp\mu} W)$. Since the Lagrangian is a scalar we must dot
the index $\mu$ into another vector.  The possible Dirac structures are
restricted by the fact that $\nslash\xi_n=0$.  They are also restricted by
type-III RPI, for instance $\bnslash\gamma_\mu$ is not invariant and therefore
is ruled out (in the case of the heavy-to-light currents we can make use of the
product $n\mcdot v$, so $n\mcdot v\,\bnslash$ is allowed). Taking these
constraints into account leaves $\gamma_\mu$ as the only possibility. Thus, we
have reduced ${\cal L}_{uc}^{(1)}$ to the form
\begin{eqnarray}
 (\bar\xi_n W)\: \rho(\bnP,\mu)\: (W^\dagger \DSppP W)\: q_{us}\mbox{  + h.c.}\,,
\end{eqnarray}
where the coefficient $\rho$ is dimensionless and $\DSppP$ acts to the left or
right.  However, by type-III RPI invariance the coefficient $\rho(\bnP,\mu)$ can
not be a function of $\bnP$, leaving only $\rho(\mu)$.  Now $\rho$ is a
dimensionless function of the dimension-full parameter $\mu$ and can only be
equal to a constant (assuming no new dynamical scales like $\Lambda_{\rm QCD}$
are generated by renormalizing ${\cal L}_{uc}^{(1)}$). Now $\cPslash_\perp q =0$
since $q$ carries no perpendicular momenta of order $\lambda$, so using
Eq.~(\ref{toBD}) we see that $(W^\dagger\DSppPr W)$ can be traded for a
$\Bslash_c^\perp$ operator. For $\DSppPl$ we obtain the same $\Bslash_c^\perp$
operator, plus $(\bar\xi_n W)\Slash{\cP}_\perp^\dagger q=0$, which follows from
the fact that $q_{us}$, and by momentum conservation $(\bar\xi_n W)$, carries
zero collinear $\perp$ momentum. Fixing the constant $\rho=1$ by tree level
matching then leaves
\begin{eqnarray}
  {\cal L}^{(1)}_{\xi q} &=&  \bar\xi_n W 
    \frac{1}{\bnP} W^\dagger ig \Bslash_c^\perp W q_{us} \mbox{ + h.c.  }
  = \bar\xi_n \frac{1}{i\bn\mcdot D_c} ig \Bslash_c^\perp W q_{us}
    \mbox{ + h.c.  }\,.
\end{eqnarray}
In this form it is clear that the operator is collinear and usoft gauge
invariant and generates terms with $\ge 1$ collinear gluon as required by 
momentum conservation. Finally, it is easy to show that its order $\lambda^0$  
type-I and type-II variations vanish, since from Eq.~(\ref{rpi0}) only the 
transformation of $D^\perp_c$ must be considered and $\delta_I^{(\lambda^0)}
B_c^\perp \propto [i\bn\mcdot D_c, i\bn\mcdot D_c] =0$, while $\delta_{\rm
II}^{(\lambda^0)} \Bslash_c^\perp \propto \nslash$ which gives zero since
$\nslash\xi_n=0$.

The above line of reasoning can be repeated at ${\cal O}(\lambda^2)$. By power
counting and gauge invariance we can now have one $(W^\dagger i n\mcdot D W)$ or
two $(W^\dagger \DgppP W)$ factors with derivatives to the left or right. Again
we can use Eq.~(\ref{toBD}) to simplify the covariant derivative terms. We must
have an operator starting with one-collinear gluon, and again the $(W^\dagger D
W)$ factor next to $q_{us}$ must be in square brackets and can be turned into a
gluon field strength. Also by type-III RPI the Wilson coefficients must again be
numbers, except for operators with three or more invariant collinear products in
${\cal L}_{\xi q}^{(2)}$ where they can be functions of the ratio $z_i$ of
minus-momenta.  Taking into account these constraints leaves three possible
operators~\footnote{This assumes we have eliminated a possible four quark
  operator using the collinear gluon equations of motion~\cite{bps4}, $g^2
  (\bar\xi_nW T^A \bnslash W^\dagger \xi_n)1/\bnP^2 (\bar\xi_n W T^A\bnslash
  q_{us})= \bar\xi_n \bnslash/(2 i\bn\mcdot D_c)\:ig\, n\mcdot M W q_{us} + 2
  \bar\xi_n \bnslash/\{2(i\bn\mcdot D_c)^2\}\: [iD_{c\,\mu}^{\perp}, i g
  B_c^{\perp\mu}] W q_{us}$. This conclusion is not changed if we consider the
  most general possible four-quark operators allowed by all the symmetry
  constraints.}
\begin{eqnarray}
  {\cal L}_1^{(2)} &=& \rho_1\ (\bar\xi_n W) 
   \Big(W^\dagger \frac{1}{i\bn\mcdot D_c} ig\Mslash\ W\Big) q_{us} 
    \mbox{ + h.c.  } \,,\\
  {\cal L}_{2}^{(2)} &=& \int\!\!dz_1 dz_2\: \rho_2\Big(\frac{z_2}{z_1}\Big)\ 
    (\bar\xi_n W)_{z_1} \frac{\bnslash}{2}\:
      (W^\dagger \DSppPr W)_{z_2} \frac{1}{\bnP}
    \Big( W^\dagger \frac{1}{i\bn\mcdot D_c} ig \Bslash_c^{\perp} W 
     \Big) q_{us}
     \mbox{ + h.c.  }  \nn\\
  {\cal L}_{3}^{(2)} &=& \int\!\!dz_1 dz_2\: \rho_3\Big(\frac{z_2}{z_1}\Big)\ 
     (\bar\xi_n W)_{z_1} \frac{\bnslash}{2}\:
     (W^\dagger i\overrightarrow D^{\perp\mu}_c W)_{z_2} \frac{1}{\bnP}
    \Big( W^\dagger \frac{1}{i\bn\mcdot D_c} ig B_{c\mu}^{\perp}W 
    \Big)  \ q_{us} 
     \mbox{ + h.c.  }  \nn \,,
\end{eqnarray}
where $(\ldots)_{z_i} \equiv [\ldots \delta(z_i-\bnP^\dagger)]$, and we've used
our freedom to integrate by parts to make the perp covariant derivatives act to
the right. Note that the overall $\bn\mcdot p$ momentum is zero, so no $z_i$
label is used on the $\Bslash_c^\perp$ bracketed term (following the convention
in Ref.~\cite{cbis}). Again the presence or absence of factors of $\bnslash$ are
completely fixed by type-III RPI.  Now consider the type-I and type-II RPI
transformations. Computing the order $\lambda$ variations of ${\cal
L}^{(1)}_{uc}$ and simplifying the resulting expressions gives
\begin{eqnarray} \label{var1}
 \delta_{\rm I}^{(\lambda)} {\cal L}^{(1)}_{\xi q} &=&  - \bar\xi_n 
   \frac{\bnslash}{2}  \frac{1}{i\bn\mcdot D_c} ig\Bslash_\perp^c 
   \frac{\Delslash_\perp}{2} W\, q_{us} 
 -\bar\xi_n \frac{1}{i\bn\mcdot D_c} \big(\delta_{\rm I}^{(\lambda^0)} 
    ig \Mslash_\perp\big) W q_{us} \mbox{ + h.c.  }  \,,\\
 \delta_{\rm II}^{(\lambda)} {\cal L}^{(1)}_{\xi q}  &=& 
   - \bar\xi_n \Big\{ \DSppP \frac{1}{(i\bn\mcdot  D_c)^2}
     \: ig\Bslash_c^\perp \frac{\vepslash_\perp}{2} 
   \!+\!\frac{1}{i\bn\mcdot  D_c} \frac{\vepslash_\perp}{2} ig 
     n\mcdot M 
   \!+\! \frac{1}{i\bn\mcdot D_c} \big(\delta_{\rm II}^{(\lambda^0)} 
    ig \Mslash_\perp\big) \Big\} W q_{us} \mbox{ + h.c.}  \nn
\end{eqnarray}
The $\delta_{I,II}^{(\lambda^0)}\, ig \Mslash_\perp$ terms appear since
in the RPI transformations in Eq.~(\ref{rpi1}) it is the full $D_\perp^\mu$
which transforms.  For the order $\lambda^0$ variations of ${\cal L}^{(2)}_j$ we
find
\begin{eqnarray} \label{var2}
 \delta_{\rm I}^{(\lambda^0)} {\cal L}_1^{(2)} &=& \rho_1\ \bar\xi_n  \Big\{
   \frac{1}{i\bn\mcdot D_c} \frac{\bnslash}{2} \Delta_\perp\mcdot B_\perp 
  + \frac{1}{i\bn\mcdot D_c} \big(\delta_{\rm I}^{(\lambda^0)} ig\Mslash_\perp
    \big) \Big\} W q_{us} \mbox{ + h.c.} \,, \\
 \delta_{\rm II}^{(\lambda^0)} {\cal L}_1^{(2)} &=& \rho_1\ \bar\xi_n  \Big\{
  \frac{1}{i\bn\mcdot D_c} \frac{\vepslash_\perp}{2} ig\,n\mcdot M +
  \frac{1}{i\bn\mcdot D_c} \big(\delta_{\rm II}^{(\lambda^0)} ig\Mslash_\perp
    \big) \Big\} W q_{us} \mbox{ + h.c.} \,,\nn\\
 \delta_{\rm I}^{(\lambda^0)} {\cal L}_2^{(2)} &=& -\!\! \int\!\!dz_1 dz_2 
   \,\rho_2\Big(\frac{z_2}{z_1}\Big)\  (\bar\xi_n W)_{z_1} \delta(z_2) 
   \frac{\bnslash\Delslash_\perp}{4} \: \frac{1}{\bnP}
   \Big( W^\dagger \frac{1}{i\bn\mcdot D_c} ig \Bslash_c^{\perp} W 
    \Big) q_{us} 
     \mbox{ + h.c.}\,,  \nn\\
\delta_{\rm II}^{(\lambda^0)} {\cal L}_2^{(2)} &=& -\!\! \int\!\!dz_1 dz_2 
    \,\rho_2\Big(\frac{z_2}{z_1}\Big)\    (\bar\xi_n W)_{z_1} 
     (W^\dagger \DSppPr W)_{z_2} \frac{\vepslash_\perp}{2}\: \frac{1}{\bnP}
    \Big( W^\dagger \frac{1}{i\bn\mcdot D_c} ig \Bslash_c^{\perp} W 
    \Big) q_{us} \nn\\ 
  &&\hspace{-1.6cm} + \int\!\!dz_1 dz_2 \,\rho_2\Big(\frac{z_2}{z_1}\Big)\ 
   (\bar\xi_n W)_{z_1} 
      (W^\dagger \DSppPr W)_{z_2} \frac{1}{\bnP}
     \Big( W^\dagger \frac{1}{i\bn\mcdot D_c} 
     ig\, \varepsilon_\perp\!\mcdot B_c^{\perp} W \Big) q_{us} 
     \mbox{ + h.c.}\,,  \nn\\
 \delta_{\rm I}^{(\lambda^0)} {\cal L}_3^{(2)} &=& -\!\! \int\!\!dz_1 dz_2
     \,\rho_3\Big(\frac{z_2}{z_1}\Big)\ 
     (\bar\xi_n W)_{z_1} \delta(z_2) \frac{\bnslash}{4} \frac{1}{\bnP}
     \Big( W^\dagger \frac{1}{i\bn\mcdot D_c} ig \Delta^\perp\!\mcdot 
      B_c^{\perp} W \Big) q_{us} 
     \mbox{ + h.c.}\,,  \nn\\
\delta_{\rm II}^{(\lambda^0)} {\cal L}_3^{(2)} &=& \int\!\!dz_1 dz_2 
    \,\rho_3\Big(\frac{z_2}{z_1}\Big)\    (\bar\xi_n W)_{z_1} 
     \frac{\vepslash_\perp}{2}\: 
     (W^\dagger i\overrightarrow D_c^{\perp\mu} W)_{z_2} \frac{1}{\bnP}
    \Big( W^\dagger \frac{1}{i\bn\mcdot D_c} ig B_{c\mu}^{\perp} W \Big) 
    q_{us} \,. \nn
\end{eqnarray}
Comparing Eqs.~(\ref{var1}) and (\ref{var2}) we see that it is not possible to
form an invariant involving ${\cal L}_3^{(2)}$ so $\rho_3=0$, while an invariant
can be formed from ${\cal L}_{uc}^{(1)}+{\cal L}_1^{(2)}+{\cal L}_2^{(2)}$ by
taking $\rho_1=1$ and $\rho_2(z_2/z_1)=1$.  Since $\rho_2$ is independent of
$z_1/z_2$ the integrals over $z_{1,2}$ can be performed. Therefore, we can write
our final result for the first two orders in the usoft-collinear quark
Lagrangian as
\begin{eqnarray} \label{Lxiq_final}
 {\cal L}^{(1)}_{\xi q}  &=&  
    \bar\xi_n \frac{1}{i\bn\mcdot D_c} ig \Bslash_c^\perp W q_{us}
    \mbox{ + h.c.  }\,, \nn\\
 {\cal L}^{(2a )}_{\xi q} &=& \bar\xi_n  \frac{1}{i\bn\mcdot D_c}\: 
     ig \Mslash\ \: W \, q_{us} \mbox{ + h.c.} \,, \nn\\
 {\cal L}^{(2b)}_{\xi q} &=&  \bar\xi_n \frac{\bnslash}{2} 
    i\Dslash_\perp^{\,c} \frac{1}{(i\bn\mcdot D_c)^2}\:  
    ig \Bslash_\perp^{\, c} W \: q_{us} \mbox{ + h.c.\hspace{1.5cm}}
\end{eqnarray}
These terms agree {\em exactly} with the result from tree level matching in
Eqs.~(\ref{Luc1}) and (\ref{Luc2}).  The analysis here shows that no other terms
are induced by matching at any order in $\alpha_s$. 

Next we proceed to analyze power suppressed terms in the collinear gluon
action. Starting with the LO collinear gluon action~\cite{bpssoft}, ${\cal
L}_{cg}^{(0)}$ and making it RPI
invariant with $iD^\mu = iD_c^\mu + iD_{us}^\mu$ gives
\begin{eqnarray} \label{Lcg}
 {\cal L}_{cg} &=&  \frac{1}{2 g^2}\, {\rm tr}\ \Big\{ 
    \big[i D^\mu \,, i D^\nu  \big]^2 \Big\}\,.
\end{eqnarray}
It is straightforward to see that no other gauge invariant pure glue 
dimension-4 operators are possible. We could build a more general gauge
invariant operator out of a string of $m$ terms $[{\cal W} iD^{\mu_i} {\cal
  W}^\dagger]_{z_i}$, with $m-4$ factors of $1/\bnP$ to make up the mass
dimensions.  However, type-III RPI then demands $m-4$ factors of $\bn_{\mu_i}$
in the numerators which using $\bn\mcdot D {\cal W}= 0$ collapses the operator
to the case $m=4$. Finally, since ${\cal W}$ transforms under type-II RPI as in
Eq.~(\ref{rpi1}), but $iD^\mu$ does not we find that these operators must have
Wilson coefficients $C(z_i)$ that are independent of the $z_i$ parameters. In
this case all factors of ${\cal W}$ cancel out and we are left with
Eq.~(\ref{Lcg}) (after performing the $z_i$ integrals and fixing the coefficient
at tree level).  Expanding Eq.~(\ref{Lcg}) we see that the order $\lambda$ and
$\lambda^2$ suppressed terms are
\begin{eqnarray}
 {\cal L}_{cg}^{(1)} &=& \frac{2}{g^2}\: {\rm tr} 
  \Big\{ \big[i {\cal D}^\mu , iD_c^{\perp\nu} \big] 
         \big[i {\cal D}_\mu , iD_{us\,\nu}^\perp \big] \Big\}\,,\\
 {\cal L}_{cg}^{(2)} &=&  \frac{1}{g^2}\: {\rm tr} 
  \Big\{ \big[i {\cal D}^\mu , iD_{us}^{\perp\nu} \big] 
         \big[i {\cal D}_\mu , iD_{us\,\nu}^\perp \big] \Big\}
  + \frac{1}{g^2}\: {\rm tr} 
  \Big\{ \big[i D_{us}^{\perp\mu} , iD_{us}^{\perp\nu} \big] 
         \big[i {D}_{c\mu}^\perp , i{D}_{c\nu}^\perp \big] \Big\}\nn\\
 &+&\frac{1}{g^2}\: {\rm tr} 
  \Big\{ \big[i {\cal D}^\mu , i n\mcdot D \big] 
         \big[i {\cal D}_\mu , i \bn\mcdot D_{us} \big] \Big\}
  + \frac{1}{g^2}\: {\rm tr} 
  \Big\{ \big[i D_{us}^{\perp\mu} , iD_{c}^{\perp\nu} \big] 
         \big[i {D}_{c\mu}^\perp , i{D}_{us\nu}^\perp \big] \Big\}\nn
  \,,
\end{eqnarray}
where ${\cal D}^\mu = D_c^\mu + \bn^\mu n\mcdot D_{us} /2$.

In Ref.~\cite{bpssoft} the gauge fixing terms in the LO gluon action were given
in a general covariant gauge. We do not bother to consider the possibility of
other leading order gauge fixing terms since we have some residual freedom to
choose these terms however we like.  In an RPI invariant form the terms from
Ref.~\cite{bpssoft} are
\begin{eqnarray}
 {\cal L}_{cg} &=&   
  2\, {\rm tr} \Big\{ \, \bar c_{n}\,  \Big[ iD_\mu , \Big[ iD^\mu 
      \,, c_{n}\,\Big]\Big] \Big\}  
   + \frac{1}{\alpha}\, {\rm tr}\ \Big\{ [i D_\mu\,, A_{n}^\mu]
   [i D_\nu\,, A_{n}^\nu]\Big\}\,, 
\end{eqnarray}
and the subleading terms in their expansion are
\begin{eqnarray}
 {\cal L}_{cg}^{(1)} &=& 
   2\, {\rm tr} \Big\{ \, \bar c_{n}\,  \Big[ iD^{us}_{\perp\mu} , 
    \Big[ iD_c^{\perp\mu}  \,, c_{n}\,\Big]\Big] \Big\}  
   +  2\, {\rm tr} \Big\{ \, \bar c_{n}\,  \Big[ iD^{c}_{\perp\mu} , 
    \Big[ iD_{us}^{\perp\mu}  \,, c_{n}\,\Big]\Big] \Big\}  \nn\\
  && + \frac{2}{\alpha}\, {\rm tr}\ \Big\{ 
  [i D_{\perp\mu}^{us}\,, A^{\perp\mu}_{n}] 
  [i D_\nu\,, A_{n}^\nu] \Big\}\,, \\
{\cal L}_{cg}^{(2)} &=& 
   2\, {\rm tr} \Big\{ \, \bar c_{n}\,  \Big[ iD^{us}_{\perp\mu} , 
    \Big[ iD_{us}^{\perp\mu}  \,, c_{n}\,\Big]\Big] \Big\}  
   +  \, {\rm tr} \Big\{ \, \bar c_{n}\,  \Big[ i \bn\mcdot D^{us} , 
    \Big[ i n\mcdot D  \,, c_{n}\,\Big]\Big] \Big\}  
   \nn\\
&&   +  \, {\rm tr} \Big\{ \, \bar c_{n}\,  \Big[ in\mcdot D , 
    \Big[ i \bn\mcdot D_{us} \,, c_{n}\,\Big]\Big] \Big\}  \nn\\
  && + \frac{1}{\alpha}\, {\rm tr}\ \Big\{ 
  [i D_{\perp\mu}^{us}\,, A^{\perp\mu}_{n}] 
  [i D_{\perp\nu}^{us}\,, A^{\perp\nu}_{n}] \Big\}
  + \frac{1}{\alpha}\, {\rm tr}\ \Big\{ 
  [i \bn\mcdot D^{us}\,, n\mcdot A_{n}] 
  [i D_\nu\,, A_{n}^\nu] \Big\} \,.
\end{eqnarray}


\section{Most General Basis for Heavy-to-Light currents}

In this section we give our derivation of the most general basis of
heavy-to-light currents at ${\cal O}(\lambda)$. The scalar current is given in
great detail, and forms the basis of the analysis for the other Dirac
structures.  Expanding the heavy-to-light currents in powers of $\lambda$ we
write
\begin{eqnarray}
  J_{hl} &=& {J}^{(d)} + {K}^{(d)} + \ldots \,,
\end{eqnarray}
for the LO currents ($J^{(d)}$), and NLO currents ($K^{(d)}$).  The superscript
denotes whether the current is scalar ($d=s$), pseudo-scalar ($d=p$), vector
($d=v$), axial-vector ($d=a$), or tensor ($d=t$). For the preliminary basis
where only constraints from gauge invariance, power counting, and type-III RPI
invariance are imposed we use a calligraphic notation ${\cal J}^{(d)} + {\cal
K}^{(d)} +\ldots$, and then switch to Roman for the final basis that is
invariant under all the type-I and type-II constraints.  We will also make use
of a convolution notation
\begin{eqnarray} \label{Jconv}
  J^{(d)} = \int\!\! d\omega \: C^{(d)}\Big(\frac{\omega}{m_b},
  \frac{\mu}{m_b}\Big)\: J^{(d)}(\omega)  \,,
\end{eqnarray}
where $J^{(d)}(\omega)$ contains fields and operators with the notation in
Eq.~(\ref{cdefn}) and the Wilson coefficients $C^{(d)}(\hat\omega,\mu/m_b)$ are
numerical functions of the convolution parameter (where
$\hat\omega=\omega/m_b$).


\subsection{Scalar Currents} \label{sectJs}

From gauge invariance and power counting the most general leading order
heavy-to-light current has the form $\bar\xi_n W \Gamma h_v$~\cite{bfps}. For
$v_\perp\ne 0$ the most general allowed scalar spin structures from
section~\ref{section_dirac} are then $\Gamma=\{1,\bnslash\}$. Type-III RPI
demands that the $\bnslash$ is accompanied by either a $n\mcdot v$ or a
$1/\bn\mcdot v$. Thus after imposing constraints i)-iv) of
section~\ref{sect_const} we are left with the possible leading order currents
\begin{eqnarray} \label{Js_ginv}
  {\cal J}_{1}^{(s)}=c_1^{(s)}\ \bar\xi_n W h_v
    \,,\qquad
  {\cal J}_{2}^{(s)}=c_2^{(s)}\ \bar\xi_n W n\mcdot v \frac{\bnslash}{2} h_v
   \,,\qquad
  {\cal J}_{3}^{(s)}=c_3^{(s)}\ \bar\xi_n W \frac{\bnslash}{2\,\bn\mcdot v}h_v 
   \,,
\end{eqnarray}
where $c_i$ are dimensionless Wilson coefficients.  With type-III RPI invariance
the $c_i$ can only depend on the combination $(n\mcdot v\: \bnP)$, the b-quark
mass $m_b$, and the renormalization scale $\mu$.  Now consider the order
$\lambda^0$ type-I and II RPI transformations in Eq.~(\ref{rpi0}). Since none of
the operators in Eq.~(\ref{Js_ginv}) involve quantities that have $\delta_{\rm
I}^{(\lambda^0)}$ transformations they are invariant under type-I at this
order. However, under the analogous type-II transformations
\begin{eqnarray}
  \delta_{\rm II}^{(\lambda^0)} {\cal J}_1^{(s)} &=& 0 \,,\qquad
  \delta_{\rm II}^{(\lambda^0)} {\cal J}_2^{(s)} = c_2^{(s)} \ \bar\xi_n W 
     \Big(n\mcdot v \frac{\vepslash_\perp}{2}\Big) h_v \,,\nn\\
  \delta_{\rm II}^{(\lambda^0)} {\cal J}_3^{(s)} &=& c_3^{(s)} \ \bar\xi_n W 
     \Big( \frac{\vepslash_\perp}{2\bn\mcdot v} - \frac{\bnslash}
     {2(\bn\mcdot v)^2} \varepsilon_\perp\mcdot v \Big) h_v \,.
\end{eqnarray}
Thus, it is not possible to form an invariant involving the currents ${\cal
J}_{2,3}^{(s)}$, and only the current ${\cal J}_1^{(s)}$ is allowed.  Therefore,
we can rewrite our final result for the most general leading order scalar
current as
\begin{eqnarray} \label{Js0}
  J_0^{(s)} = C_0^{(s)}\Big( \frac{-n\mcdot v\,\bnP}{m_b},\frac{\mu}{m_b}\Big)\  
  \bar\xi_n W h_v \,.
\end{eqnarray}
Since the Wilson coefficient is dimensionless it can only be a function of the
dimensionless ratios of parameters as shown. The minus sign in the first
variable is included so that $\bnP$ gives the total outgoing momentum of
$\bar\xi_nW$.  Switching to the convolution notation in Eq.~(\ref{Jconv}) and
defining $\hat\omega=\omega/m_b$ we can write Eq.~(\ref{Js0}) as
\begin{eqnarray}\label{Js_final}
  J_0^{(s)} &=& \int\!\! d\omega \: C_0^{(s)}(\hat\omega,\mu/m_b)
     \: J_0^{(s)}(\omega)\,,\nn\\
  J_0^{(s)}(\omega) &=& (\bar\xi_n W)_{\omega}\:  h_v \,.
\end{eqnarray}
Thus our notation is that $J_0^{(s)}$ contains the Wilson coefficient, while
$J_0^{(s)}(\omega)$ is purely the field operator.  With the convolution notation
in Eq.~(\ref{Js_final}) the Wilson coefficients are just numerical functions
which do not transform under RPI. We will often suppress the dependence of
Wilson coefficients on $\mu/m_b$ in what follows.

Next consider currents that are suppressed by a power of $\lambda$. At this
order the only additional structure we can use is a $D_c^{\perp\,\mu} \sim
\lambda$, where the derivative acts to the left or to the right. To form the
most general collinear gauge invariant we take $(W^\dagger D_c^{\perp\,\mu} W)$,
which we then insert between the $(\bar\xi_n W)$ and $h_v$ to satisfy the usoft
gauge invariance.  Since the two collinear factors are invariant by themselves
they can have arbitrary labels $\omega_{1,2}$. Thus we have operators with the
structure
\begin{eqnarray} \label{preK}
  (\bar\xi_n W)_{\omega_1} \Gamma \,
    (W^\dagger\! \DgppP W)_{\omega_2}\, \frac{1}{\bnP^\dagger}\,  h_v \,.
\end{eqnarray}
The factor of $1/\bnP^\dagger$ is included to make the Wilson coefficients
dimensionless.  To make a scalar current the $\mu$ superscript in
Eq.~(\ref{preK}) can be dotted into a $\gamma_\mu$ or $v_\mu$ in $\Gamma$. In
either case the most general remaining Dirac structure involves either $1$ or
$\bnslash$ as follows from section~\ref{section_dirac}. Thus, combining the
constraints from gauge invariance, spin reduction, and type-III RPI leaves eight
${\cal O}(\lambda)$ suppressed currents
\begin{eqnarray}
  {\cal K}_{j}^{(s)} = \int\!\!d\omega_1 d\omega_2\  
    b_j^{(s)}(\hat\omega_1,\hat\omega_2) \ 
    {\cal K}_j^{(s)}(\omega_1,\omega_2) \,,
\end{eqnarray}
where $j=1,\ldots,8$, and the $b_i$ coefficients are dimensionless functions of
$\hat\omega_{1,2}=\omega_{1,2}/m_b$ and $\mu/m_b$. The eight operators are
\begin{eqnarray} \label{Ks1_ginv}
  {\cal K}_{\{1,2\}}^{(s)}(\omega_1,\omega_2) &=&
    (\bar\xi_n W)_{\omega_1}
    (W^\dagger\! \DSppPl W)_{\omega_2} \frac{1}{\bnP^\dagger}
    \Big\{\frac{\bnslash}{2} \,, \frac{1}{n\mcdot v} \Big\} h_v
    \,,\nn \\
  {\cal K}_{\{3,4\}}^{(s)}(\omega_1,\omega_2) &=&  (\bar\xi_n W)_{\omega_1}   
    (W^\dagger\! \DvppPl W)_{\omega_2} \frac{1}{\bnP^\dagger} 
     \Big\{\frac{\bnslash}{2} \,, \frac{1}{n\mcdot v} \Big\} h_v
    \,,\nn \\
  {\cal K}_{\{5,6\}}^{(s)}(\omega_1,\omega_2) &=&  (\bar\xi_n W)_{\omega_1} 
    (W^\dagger\! \DSppPr W)_{\omega_2} \frac{1}{\bnP^\dagger}
    \Big\{\frac{\bnslash}{2} \,, \frac{1}{n\mcdot v} \Big\} h_v
    \,,\nn \\
  {\cal K}_{\{7,8\}}^{(s)}(\omega_1,\omega_2) &=& (\bar\xi_n W)_{\omega_1}     
    (W^\dagger\! \DvppPr W)_{\omega_2} \frac{1}{\bnP^\dagger} 
    \Big\{\frac{\bnslash}{2} \,, \frac{1}{n\mcdot v} \Big\} h_v
    \,.
\end{eqnarray}
Note that the dependence of the Wilson coefficients on the labels $\omega_i$
account for insertions of $1/\bnP^\dagger$ in all possible locations.  Just as
for the leading currents we cannot use $\bn\mcdot v$ to form a type-III
invariant in Eq.~(\ref{Ks1_ginv}) as it leads to currents which can not be made
invariant under type-II transformations (the transformed currents would depend
on $\epsilon_\perp\mcdot v/(\bn\mcdot v)^2$ in a way that could not be
canceled).

%
%
Next consider the type-I transformations for the currents in
Eq.~(\ref{Ks1_ginv}). For these subleading currents only the $\lambda^0$
transformations are necessary since we are only working to order $\lambda$.
Under type-I only $D_c^\perp$ transforms and we have
\begin{eqnarray} \label{tIWDW}
  \delta_{\rm I}^{(\lambda^0)} (W^\dagger i\!\DgppPl W)_{\omega_2}
    &=& -\frac{\Delta_\perp^\mu}{2}\, 
    (W^\dagger i\bn\,\!\cdot\!\! \lD W)_{\omega_2} 
     = +\frac{\Delta_\perp^\mu}{2}\, \bnP^\dagger\,
     \delta(\omega_2\!-\!n\mcdot v \bnP^\dagger)
     \,,\nn\\
  \delta_{\rm I}^{(\lambda^0)} (W^\dagger i\!\DgppPr W)_{\omega_2}
    &=& -\frac{\Delta_\perp^\mu}{2}\, 
     (W^\dagger i\bn\,\cdot\!\! \rD W)_{\omega_2}
     = -\frac{\Delta_\perp^\mu}{2}\, \bnP\, 
     \delta(\omega_2\!-\!n\mcdot v\bnP^\dagger)\,.
\end{eqnarray}
Since $\bnP h_v=0$ it is easy to see that each of ${\cal K}_{5,6,7,8}^{(s)}$ are
type-I invariant all by themselves. The other currents do transform, and using
Eq.~(\ref{tIWDW}) gives 
\begin{eqnarray} \label{KItrnsfm}
  \delta_{\rm I}^{(\lambda^0)}{\cal K}_{1}^{(s)}(\omega_1,\omega_2)  &=&
     \delta(\omega_2)\ 
     (\bar\xi_n W)_{\omega_1} \Big(\frac{-\bnslash\Delslash_\perp}{4}\Big) h_v
     \,,\nn\\ \qquad
  \delta_{\rm I}^{(\lambda^0)}{\cal K}_{2}^{(s)}(\omega_1,\omega_2)   &=&
     \delta(\omega_2)\ 
    (\bar\xi_n W)_{\omega_2} \Big(\frac{\Delslash_\perp}{2n\mcdot v}\Big) 
     h_v\,, \nn \\
  \delta_{\rm I}^{(\lambda^0)}{\cal K}_{3}^{(s)}(\omega_1,\omega_2)   &=&
     \delta(\omega_2)\ 
     (\bar\xi_n W)_{\omega_1}  \Big(\frac{\bnslash\Delta_\perp\mcdot v}{4}\Big) 
     h_v \,,\nn\\
  \delta_{\rm I}^{(\lambda^0)}{\cal K}_{4}^{(s)}(\omega_1,\omega_2) &=&
     \delta(\omega_2)\ 
     (\bar\xi_n W)_{\omega_1} \Big(\frac{\Delta_\perp\mcdot v}{2n\mcdot v}\Big)
     h_v \,,\nn \\
  \delta_{\rm I}^{(\lambda^0)}{\cal K}_{5,6,7,8}^{(s)}(\omega_1,\omega_2)  
     &=& 0\,.
\end{eqnarray}
The delta functions $\delta(\omega_2)$ cause only the coefficients
$b_{1,2,3,4}^{(s)}(\hat\omega_1,0)$ to appear in the transformation of ${\cal
K}_{\{1,2,3,4\}}^{(s)}$. We also need the order $\lambda$ variation of the LO
current in Eq.~(\ref{Js_final}). In this computation we must be careful to note
that the $\delta(\omega\!-\!n\mcdot v \bnP^\dagger)$ in $(\bar\xi_n W)_{\omega}$
depends on $n\mcdot v$, and therefore also transforms
\begin{eqnarray} \label{tIdelta}
 \delta_I^{(\lambda^0)}\, \delta(\omega\!-\!n\mcdot v \bnP^\dagger) = - v\mcdot
 \Delta_\perp \bnP^\dagger\: \delta'(\omega\!-\!n\mcdot v \bnP^\dagger) =
 -\frac{v\mcdot \Delta_\perp}{n\mcdot v} \, \frac{d}{d\omega}\:\omega \:
 \delta(\omega\!-\!n\mcdot v \bnP^\dagger) \,.
\end{eqnarray} 
Using Eq.~(\ref{rpi1}) we find a term from transforming the delta function and a
term from transforming the collinear quark field
\begin{eqnarray} \label{JItrnsfm2}
 \delta_{\rm I}^{(\lambda)} J_0^{(s)}(\omega) =  
     (\bar\xi_n W)_{\omega}\,   \Big(\frac{\bnslash 
     \Delslash_\perp}{4}\Big) h_v
    - \frac{v\mcdot \Delta_\perp}{n\mcdot v}\frac{d}{d\omega}\:\omega\, 
    (\bar\xi_n W)_{\omega}\,  h_v 
    \,.
\end{eqnarray}
Demanding invariance under the transformations in Eqs.~(\ref{KItrnsfm}) and
(\ref{JItrnsfm2}) gives non-trivial constraints on the Wilson coefficients in
$J_0^{(s)}$ and ${\cal K}_i^{(s)}$.  From Eq.~(\ref{KItrnsfm}) the currents
${\cal K}_{\{1,2,3,4\}}^{(s)}$ are invariant by themselves provided that
$b_{\{1,2,3,4\}}^{(s)}(\hat\omega_1,0)=0$. However, in Eq.~(\ref{JItrnsfm2})
the first term can only be canceled by a ${\cal K}_1^{(s)}$ with
$b_{1}^{(s)}(\hat\omega,0)= C_0^{(s)}(\hat\omega)$. To cancel the second term we
integrate by parts to give a $\omega d/d\omega=\hat\omega d/d\hat\omega$ acting
on $C_0^{(s)}$. This term can then be canceled by
$b_4^{(s)}(\hat\omega_1,0)=-2\hat\omega\: d/d\hat\omega\,
C_0^{(s)}(\hat\omega)$. Thus, the summary of type-I invariants is 
$ J_0 + {\cal K}_1^{(s)} + {\cal K}_4^{(s)} \,,
$
$   {\cal K}_{\{5,6,7,8\}}^{(s)} \,,
$
$   {\cal K}_{\{2,3\}}^{(s)} \,,  
$
with any $b_{5,6,7,8}^{(s)}(\hat\omega_1,\hat\omega_2)$ and coefficients
\begin{eqnarray} \label{SIb}
 b_1^{(s)}(\hat\omega,0) &=& C_0^{(s)}(\hat\omega) 
   \,,\qquad
 b_{2,3}^{(s)}(\hat\omega,0) = 0
   \,,\qquad 
 b_4^{(s)}(\hat\omega,0) = -2\hat\omega\: d/d\hat\omega\, C_0^{(s)}(\hat\omega)
   \,.
\end{eqnarray}

%
%
Now consider the type-II transformations. The analog of Eq.~(\ref{tIdelta}) is
\begin{eqnarray}
\delta_{\rm II}^{(\lambda)}\, \delta(\omega\!-\!n\mcdot v \bnP^\dagger) 
 = - n\mcdot v\, \varepsilon_\perp\!\mcdot\cP^\dagger_\perp\:
\delta'(\omega\!-\!n\mcdot v \bnP^\dagger) 
 = - n\mcdot v\, \varepsilon_\perp\!\mcdot\cP^\dagger_\perp
 \frac{d}{d\omega} \:
 \delta(\omega\!-\!n\mcdot v \bnP^\dagger) \,.
\end{eqnarray}
For the order $\lambda$ variation of the LO current $J_0^{(s)}(\omega)$ we have
terms from the transformation of the delta function, the collinear quark field,
and the Wilson line $W$
\begin{eqnarray} \label{J0_II}
  \delta_{\rm II}^{(\lambda)} J_0(\omega) &=& 
     n\mcdot v\:
    \varepsilon_\perp\mcdot \cP_\perp\: \frac{d}{d\omega}\:
   (\bar\xi_n W)_{\omega}   \: h_v +
  \Big(\bar\xi_n \DSppPl
    \frac{1}{i\bn\mcdot \lD} 
    \frac{\vepslash_\perp}{2}  W\Big)_{\omega}\, h_v \nn\\
  && - \Big( \bar\xi_n  \frac{1}{i\bn\mcdot D_c}i\,\varepsilon_\perp\mcdot 
    \rD^{\perp} W\Big)_{\omega}\, h_v    \,.
\end{eqnarray}
In the subleading currents ${\cal K}_i^{(s)}$ both $\bn^\mu$ and $D^\perp_c$
have $\lambda^0$ transformations.\footnote{Since $\delta_{\rm II}\bnP={\cal
O}(\lambda)$ the transformation of the delta functions in ${\cal K}_i^{(s)}$
only appear at one higher order.} A $\DSppP$ transforms to give a $\nslash$, so
since $\nslash \xi_n=0$ it is easy to see that ${\cal K}_{\{2,6\}}^{(s)}$ are
invariant under type-II transformations at this order. The transformations for
the remaining currents are more involved
\begin{eqnarray} \label{J1_II}
 \delta_{\rm II}^{(\lambda^0)}  {\cal K}_{\{2,6\}}^{(s)}(\omega_1,\omega_2) 
  &=& 0 \,,\nn\\
 \delta_{\rm II}^{(\lambda^0)}  {\cal K}_{1}^{(s)}(\omega_1,\omega_2) 
  &=&    
    (\bar\xi_n W)_{\omega_1}  (W^\dagger\! \DSppPl W)_{\omega_2} 
    \frac{\vepslash_\perp}{2} \frac{1}{\bnP^\dagger} h_v
    \,, \\
 \delta_{\rm II}^{(\lambda^0)} {\cal K}_{3}^{(s)}(\omega_1,\omega_2)  
   &=& 
   (\bar\xi_n W)_{\omega_1} \Big\{
    \frac{\vepslash_\perp}{2} (W^\dagger  \DvppPl W)_{\omega_2}
    - \frac{n\mcdot v\,\bnslash}{4} (W^\dagger i\varepsilon_\perp \mcdot\!
    \lD^\perp W)_{\omega_2} \Big\} \frac{1}{\bnP^\dagger}  h_v
    \,,\nn \\
 \delta_{\rm II}^{(\lambda^0)}  {\cal K}_{4}^{(s)}(\omega_1,\omega_2)  
   &=& -\frac{1}{2}\: 
    (\bar\xi_n W)_{\omega_1} 
    (W^\dagger i\varepsilon_\perp \mcdot\! \lD^\perp W)_{\omega_2} 
    \frac{1}{\bnP^\dagger}\, h_v  \,,\nn \\
 \delta_{\rm II}^{(\lambda^0)}  {\cal K}_{5}^{(s)}(\omega_1,\omega_2)  
   &=& 
    (\bar\xi_n W)_{\omega_1} (W^\dagger\! \DSppPr W)_{\omega_2}
    \frac{\vepslash_\perp}{2} \frac{1}{\bnP^\dagger}  h_v
    \,,\nn \\
 \delta_{\rm II}^{(\lambda^0)}  {\cal K}_{7}^{(s)}(\omega_1,\omega_2)
    &=& 
    (\bar\xi_n W)_{\omega_1} \Big\{
    \frac{\vepslash_\perp }{2} (W^\dagger  \DvppPr W)_{\omega_2}
    - \frac{n\mcdot v\,\bnslash}{4} (W^\dagger i\varepsilon_\perp\! \mcdot\!
    \rD^\perp W)_{\omega_2} \Big\} \frac{1}{\bnP^\dagger} h_v
    \,,\nn \\
 \delta_{\rm II}^{(\lambda^0)}  {\cal K}_{8}^{(s)}(\omega_1,\omega_2)
    &=&  -\frac{1}{2}\: 
    (\bar\xi_n W)_{\omega_1} 
    (W^\dagger i\varepsilon_\perp \mcdot\! \rD^\perp W)_{\omega_2}
    \frac{1}{\bnP^\dagger} h_v \,,\nn
\end{eqnarray}
It is straightforward to see that it is not possible to form a type-II invariant
using only the currents ${\cal K}_{\{1,3,4,5,7,8\}}^{(s)}$.  However, it is
possible to form an invariant taking a combination of ${\cal
K}_{\{1,4,8\}}^{(s)}$ with $J_0^{(s)}$. To facilitate this we rewrite 
Eq.~(\ref{J0_II}) as
\begin{eqnarray} \label{J0_IIa}
  \delta_{\rm II}^{(\lambda^1)} J_0^{(s)}(\omega) &=& \int\!\! d\omega_1 \Big[
    n\mcdot v\: \frac{d}{d\omega}\:(\bar\xi_n W)_{\omega_1}  \Big\{ 
   (W^\dagger i\varepsilon_\perp\!\mcdot\! \rD^\perp W)_{\omega-\omega_1} +
   (W^\dagger i\varepsilon_\perp\!\mcdot\! \lD^\perp W)_{\omega-\omega_1}
  \Big\}  \: h_v \nn\\
  &&\hspace{-2cm}
   - (\bar\xi_n W)_{\omega_1} (W^\dagger \DSppPl W)_{\omega-\omega_1} \,
    \frac{1}{\bnP^\dagger}\, \frac{\vepslash_\perp}{2} \, h_v 
   - (\bar\xi_n W)_{\omega_1} \frac{1}{\omega_1\!-\!\omega}  
   (W^\dagger i\varepsilon_\perp\!\mcdot  \rD^{\perp} W)_{\omega-\omega_1}
   \, h_v \Big]
\,.
\end{eqnarray}
To derive Eq.~(\ref{J0_IIa}) we used $\varepsilon_\perp\mcdot \cP_\perp\:
\bar\xi_n W h_v = \bar\xi_n i\varepsilon_\perp\mcdot \rD^\perp W h_v + \bar\xi_n
i\varepsilon_\perp \mcdot \lD^\perp W h_v$, and the fact that the $\omega_1$
integration can be carried out with the delta function in $(\bar\xi_n
W)_{\omega_1}$ to get back a product of operators with momentum label $\omega$
where the intermediate $W W^\dagger$ cancel out. Now for the $d/d\omega$ terms in
Eq.~(\ref{J0_IIa}) we can integrate by parts in
$[C_0(\hat\omega)\,\delta_{\rm II}^{(\lambda^1)} J_0^{(s)}(\omega)]$ so that the
derivative acts on the Wilson coefficient $C_0^{(s)}(\hat\omega)$. It is then
evident that the third term in Eq.~(\ref{J0_IIa}) can be canceled by
$\delta_{\rm II}^{(\lambda^0)} {\cal K}_1^{(s)}$ with
$b_1^{(s)}(\hat\omega_1,\hat\omega\! - \!\hat\omega_1)= C_0^{(s)}(\hat\omega)$,
the second term is canceled by $\delta_{\rm II}^{(\lambda^0)} {\cal K}_4^{(s)}$
with $b_4^{(s)}(\hat\omega_1,\hat\omega\! - \!\hat\omega_1)= -2\hat\omega
d/d\hat\omega\, C_0^{(s)}(\hat\omega)$, and the first and fourth terms are
canceled by $\delta_{\rm II}^{(\lambda^0)} {\cal K}_8^{(s)}$ with
$b_8^{(s)}(\hat\omega_1,\hat\omega\! - \!\hat\omega_1)= -2\hat\omega\,
d/d\hat\omega\, C_0^{(s)}(\hat\omega) - 2 \hat \omega/(\hat\omega_1\! -
\!\hat\omega)\, C_0^{(s)}(\hat\omega) $. Therefore, type-II RPI rules out the
operators ${\cal K}_{3,5,7}^{(s)}$ and leaves only the invariants
\begin{eqnarray} \label{SII}
  && 
  J_0^{(s)} + {\cal K}_1^{(s)} + {\cal K}_4^{(s)}+{\cal K}_8^{(s)}\,,
   \qquad
  {\cal K}_2^{(s)} \,,
\qquad 
  {\cal K}_6^{(s)} \,.
\end{eqnarray}
For type-II invariance their coefficients can have any
$b_{\{2,4\}}^{(s)}(\hat\omega_1,\hat\omega_2)$, but require
\begin{eqnarray} \label{SIIb}
 && b_1^{(s)}(\hat\omega_1,\hat\omega\!-\!\omega_1)
   = C_0^{(s)}(\hat\omega)
   \,,\qquad 
 b_4^{(s)}(\hat\omega,\hat\omega\!-\!\omega_1) 
   = -2\hat\omega\: \frac{d}{d\hat\omega}\, C_0^{(s)}(\hat\omega)
   \,, \\
 && b_8^{(s)}(\hat\omega_1,\hat\omega\!-\!\omega_1) 
   = -2\hat\omega\: \frac{d}{d\hat\omega}\, C_0^{(s)}(\hat\omega)
   -\frac{2\hat\omega}{(\hat\omega_1\!-\!\hat\omega)}\: C_0^{(s)}(\hat\omega)
   \,. \nn 
\end{eqnarray}

The restrictions on the Wilson coefficients are summarized in
Table~\ref{table_Ks}.  Comparing the invariants in Eqs.~(\ref{SIb}) and
(\ref{SIIb}), we see that the combinations in Eq.~(\ref{SII}) are the most
general combinations invariant under type-I and type-II with the restrictions in
Eq.~(\ref{SIIb}) plus $b_2^{(s)}(\hat\omega,0)=0$.
\begin{table}[t!]
\begin{center}
\begin{tabular}{|c|c|c||l|}
\hline\hline
  & \hspace{0.1cm} {\parbox{2cm}{\vspace{0.2cm} RPI-I\\[-25pt] 
  \ \ \ $b_i^{(s)}(\hat\omega_1,0)=$\vspace{0.1cm}}}
  \hspace{0.05cm} 
  & {\parbox{2cm}{\vspace{0.2cm} RPI-II\\[-25pt] 
  \ \ \ $b_i^{(s)}(\hat\omega_1,\hat\omega_2)=$\vspace{0.1cm}}}
  & {\hspace{1.3cm}\mbox{Combined Constraints}\hspace{-1.3cm}}
  \\
\hline
\hspace{0.1cm}$b_1^{(s)}$\hspace{0.1cm} 
  & $C_0^{(s)}(\hat\omega_1)$ 
  & $C_0^{(s)}(\hat\omega_1\!+\!\hat\omega_2)$ 
  & \hspace{0.2cm} $b_1^{(s)}(\hat\omega_1,\hat\omega_2)=
  C_0^{(s)}(\hat\omega_1\!+\!\hat\omega_2)$ \\
$b_2^{(s)}$ & $0$ & $b_2^{(s)}(\hat\omega_1,\hat\omega_2)$ 
  & \hspace{0.2cm} $b_2^{(s)}(\hat\omega_1,\: 0_{\phantom{2}})=0$ \\
$b_3^{(s)}$ & $0$ & $0$ 
  & \hspace{0.2cm} $b_3^{(s)}(\hat\omega_1,\hat\omega_2)=0$ \\
$b_4^{(s)}$ & $-2\hat\omega_1 C^{(s)\prime}_0(\hat\omega_1)$ 
 & $-2(\hat\omega_1\!+\!\hat\omega_2)\: 
  C^{(s)'}_0(\hat\omega_1\!+\!\hat\omega_2)$
 & \hspace{0.1cm} $b_4^{(s)}(\hat\omega_1,\hat\omega_2)
  =-2(\hat\omega_1\!+\!\hat\omega_2)\:
  C^{(s)'}_0(\hat\omega_1\!+\!\hat\omega_2)$\hspace{0.1cm} 
 \\ \hline
$b_5^{(s)}$ & $b_5^{(s)}(\hat\omega_1,0)$ & $0$ 
  & \hspace{0.2cm} $b_5^{(s)}(\hat\omega_1,\hat\omega_2)=0$ \\
$b_6^{(s)}$ & $b_6^{(s)}(\hat\omega_1,0)$  
  & $b_6^{(s)}(\hat\omega_1,\hat\omega_2)$  
  & \hspace{0.2cm} 
  $b_6^{(s)}(\hat\omega_1,\hat\omega_2)$ \ \mbox{unconstrained} \\
$b_7^{(s)}$ & $b_7^{(s)}(\hat\omega_1,0)$ & $0$ 
  & \hspace{0.2cm} $b_7^{(s)}(\hat\omega_1,\hat\omega_2)=0$ \\
$b_8^{(s)}$ & $b_8^{(s)}(\hat\omega_1,0)$ 
  & $-2(\hat\omega_1\!+\!\hat\omega_2)\: 
   C^{(s)'}_0(\hat\omega_1\!+\!\hat\omega_2)$  
  & \hspace{0.2cm} $b_8^{(s)}(\hat\omega_1,\hat\omega_2)
   =-2(\hat\omega_1\!+\!\hat\omega_2)\: 
   C^{(s)'}_0(\hat\omega_1\!+\!\hat\omega_2)$ \\
 & & 
 \hspace{0.2cm} $+2 \big(1 + \frac{\hat\omega_1}{\hat\omega_2}\big)\: 
  C_0^{(s)}(\hat\omega_1\!+\!\hat\omega_2)$ \hspace{0.1cm}
 & 
 \hspace{2.2cm} 
  $+2 \big(1 + \frac{\hat\omega_1}{\hat\omega_2}\big)\:
   C_0^{(s)}(\hat\omega_1\!+\!\hat\omega_2)$ \hspace{0.1cm}
 \\
\hline\hline
\end{tabular}
\end{center} 
\caption{\setlength\baselineskip{12pt} 
Summary of RPI constraints on the coefficients of the scalar currents in
Eq.~(\ref{Ks1_ginv}).  The first column shows the constraints from type-I RPI on
$b_i^{(s)}(\hat\omega_1,0)$, the second column shows the constraint on
$b_i^{(s)}(\hat\omega_1,\hat\omega_2)$ from type-II RPI and the third column
gives the combined constraint.  A generic entry, like $b_2^{(s)}(\hat\omega_1,
\hat\omega_2)$ in the second row of the RPI-II column, indicates no constraint. 
The final currents are displayed in Eq.~(\ref{Ks_final}), and are defined so
that they automatically satisfy these constraints.
\label{table_Ks}
}
\end{table}
However, with this constraint on $b_2^{(s)}$ the operator ${\cal K}_2^{(s)}$ is
actually identical to ${\cal K}_6^{(s)}$ with an unconstrained coefficient
$b_6^{(s)}$. To see this note that within square brackets $[W^\dagger \DSppPr
W]_{\omega_2} = [W^\dagger \DSppPl W]_{\omega_2}= -[W^\dagger ig \Bslash_c^\perp
W]_{\omega_2}/\omega_2 $, so the difference comes from $\DSppPl$ acting also on
$(\bar\xi_n W)$ in ${\cal K}_2^{(s)}$. However, since the factor $(\bar\xi_n W)$
on the left is a collinear color singlet we can write
\begin{eqnarray} \label{toB}
 (W^\dagger \DSppPl W)_{\omega_2} 
  = \big[W^\dagger \DSppPl W \big]_{\omega_2} 
  - \Slash{\cP}_\perp^\dagger (W^\dagger  W)_{\omega_2}
  = [W^\dagger \DSppPl W]_{\omega_2} 
  - \Slash{\cP}_\perp^\dagger\: \delta(\omega_2) 
  \,,
\end{eqnarray}
and the last term vanishes since $b_2(\omega_1,0)=0$. Given this result and the
constraints in Eq.~(\ref{SIIb}) it is convenient to define 
\begin{eqnarray}
 K_1^{(s)}(\omega) &=& \int\!\! d\omega_1\
   {\cal K}_1^{(s)}(\omega_1,\omega\!-\!\omega_1) \,, \nn\\
 K_2^{(s)}(\omega) &=& \hat\omega \int\!\! d\omega_1\ 
   \big\{{\cal K}_4^{(s)}(\omega_1,\omega\!-\!\omega_1) 
    +{\cal K}_8^{(s)}(\omega_1,\omega\!-\!\omega_1)\big\} \,,\nn\\
 K_3^{(s)}(\omega) &=& \omega \int\!\! d\omega_1\ {(\omega_1\!-\!\omega)^{-1}}\, 
   {\cal K}_8^{(s)}(\omega_1,\omega\!-\!\omega_1) \,,\nn\\
 K_4^{(s)}(\omega_1,\omega_2) &=& {(\hat\omega_1\!+\!\hat\omega_2)}\:
   {\cal K}_6^{(s)}(\omega_1,\omega_2). 
\end{eqnarray}
From RPI it is only these operators that can ever appear.  RPI has ruled out
some currents and restricted $K_{1,2,3}^{(s)}$ to only depend on one parameter.
Once we know this, we can simply {\em forget} about the ${\cal K}_i^{(s)}$ and
work directly with the $K_i^{(s)}$.  Using capital $B$'s for their Wilson
coefficients, our final basis of subleading scalar operators $K_{1-4}^{(s)}$ is
\begin{eqnarray} 
  \nn\\[-8pt]
  {K}_{1-3}^{(s)} &=& \int\!\!d\omega \ \ 
    B_{1-3}^{(s)}(\hat\omega) \ \ 
    {K}_{1-3}^{(s)}(\omega)\,,\nn\\
   {K}_{4}^{(s)} &=& \int\!\!d\omega_1 d\omega_2\ \
    B_{4}^{(s)}(\hat\omega_1,\hat\omega_2) \ \
    {K}_{4}^{(s)}(\omega_1,\omega_2)\,, \\
 && \psframe[linewidth=2pt,linecolor=red,framearc=.3](-3.,-0.2)(8,2.7)
 \nn
\end{eqnarray}
where
\begin{eqnarray} \label{Ks_final}
  K_{1}^{(s)}(\omega) &=& -
    \Big(\bar\xi_n \frac{\bnslash}{2} \DSppPl W\Big)_{\!\omega} \:
    \frac{1}{\bnP^\dagger}\: h_v
    \,, \\[4pt]
  K_{2}^{(s)}(\omega) &=&  
    \frac{v\mcdot \cP_\perp}{m_b}\ \big(\bar\xi_n W \big)_{\omega}\:
    h_v\,,\nn\\[4pt]
  K_{3}^{(s)}(\omega) &=& 
    \Big(\bar\xi_n\, \frac{1}{n\mcdot v\,i\bn\mcdot D_c}\DvppPr\, W 
    \Big)_{\!\omega} \: h_v\,,\nn \\[4pt]
  K_4^{(s)}(\omega_1,\omega_2) &=& \frac{1}{m_b}\, 
    \big(\bar\xi_n W\big)_{\omega_1} 
    \Big(\frac{1}{\bnP} W^\dagger ig \Bslash_c^\perp W\Big)_{\omega_2}\,  h_v 
    \,,\nn \\
 && \psframe[linewidth=2pt,linecolor=red,framearc=.3](-3.5,-0.2)(7.5,5.2)
   \nn
\end{eqnarray}
and the RPI type-I and type-II constraints on the Wilson coefficients become
\begin{eqnarray}\label{rpiBs}
  B_1^{(s)}(\hat\omega) = C_0^{(s)}(\hat\omega)   \,, \qquad 
  B_2^{(s)}(\hat\omega)  = -2 C_0^{(s)\,\prime}(\hat\omega) \,, \qquad 
  B_3^{(s)}(\hat\omega)  = -2 C_0^{(s)}(\hat\omega)  \,.
\end{eqnarray}
The prime here denotes a derivative with respect to $\hat\omega$.  Thus, we
conclude that there are $4$ subleading ${\cal O}(\lambda)$ scalar heavy-to-light
currents.  The coefficient $B_4^{(s)}(\hat\omega_1,\hat\omega_2)$ is completely
unconstrained, \OMIT{ $B_4^{(s)}(\hat\omega_1,\hat\omega_2) = 
b_4(\hat\omega_1,\hat\omega_2)/(\hat\omega_1\!+\!\hat\omega_2)$,} while the 
other coefficients are fixed from RPI invariance.

In Ref.~\cite{bcdf} it was noted that $J_0^{(s)}$, $K_1^{(s)}$, and $K_3^{(s)}$
are connected by RPI, and for these operators our results agree with taking
their matching calculation in Eq.~(120) and multiplying by a common Wilson
coefficient. The operator $K_2^{(s)}$ does not appear in Ref.~\cite{bcdf},
because the derivative on its coefficient causes it to vanish at tree
level.  Our three-body operator $K_4^{(s)}$ is also new. In the limit that
$B_4(\hat\omega_1,\hat\omega_2)$ depends {\em only} on the sum $\hat\omega_+ =
\hat\omega_1 + \hat\omega_2$ we can switch variables to $\hat\omega_+,
\hat\omega_1$ and reduce $K_4^{(s)}$ to a two-body operator. At tree level this
is always possible since the Wilson coefficient is independent of the
$\omega_i$. To see how the reduction works we write
\begin{eqnarray} \label{Kstwo}
  \int\!\! d\omega_1 \: K_4^{(s)}(\omega_1,\omega_+-\omega_1) 
  &=& \frac{1}{m_b}\, 
    \big(\bar\xi_n W \frac{1}{\bnP} W^\dagger ig \Bslash_c^\perp 
    W\Big)_{\omega_+}\,  h_v  \nn \\
  &=& \frac{1}{m_b}\, 
    \Big(\bar\xi_n  [\DSppPr  W ]\Big)_{\omega_+}\,  h_v  \,,
\end{eqnarray}
where in the last line we used Eq.~(\ref{toBD}). The derivative structure of
this two-body operator is similar to that of operators in
Ref.~\cite{bpspc,bcdf}, however the specific spin structure appearing in
Eq.~(\ref{Kstwo}) does not appear from matching the QCD scalar current at tree
level. Beyond tree level $B_4$ can depend separately on $\omega_1$ and
$\omega_2$ and the reduction in Eq.~(\ref{Kstwo}) is not valid.


\subsection{Vector Currents}

The steps for deriving the general set of vector currents are very similar to
the steps for the scalar currents in the previous section, so our presentation 
will be more concise.

At LO gauge invariance plus power counting allows currents of the form
$\bar\xi_n W Z_\mu\, h_v$. Imposing type-III RPI invariance allows the
Dirac structures
\begin{eqnarray} \label{JVGamma}
  Z_\mu&=&\Big\{ \gamma_\mu\,, {\gamma_\mu \bnslash n\mcdot v}\,,
   \frac{\gamma_\mu\bnslash}{\bn\mcdot v}\,, {n_\mu\bnslash }\,, 
   \frac{n_\mu}{n\mcdot v}\,, {n_\mu\bn\mcdot v}\,, v_\mu, 
   {v_\mu\bnslash n\mcdot v  }\,, \frac{v_\mu \bnslash }{\bn\mcdot v}\,,\nn\\
   &&\ \ \bn_\mu n\mcdot v, \frac{\bn_\mu}{\bn\mcdot v}\,, 
   \bn_\mu \bnslash (n\mcdot v)^2\,, 
   \frac{\bn_\mu\bnslash n\mcdot v}{\bn\mcdot v}\,,
   \frac{\bn_\mu \bnslash}{(\bn\mcdot v)^2}
     \Big\} \,.
\end{eqnarray}
However it is easy to show that all structures involving $\bn$ are ruled out by
the $\delta_{\rm II}^{(\lambda^0)}$ transformations. Thus, at leading order
there are only three allowed vector currents
\begin{eqnarray}
  {J}_{1-3}^{(v)} &=& \int\!\!d\omega \ \ 
    C_{1-3}^{(v)}(\hat\omega) \ \ 
    {J}_{1-3}^{(v)}(\omega)\,,
\end{eqnarray}
where the coefficients are functions of $\hat\omega=\omega/m_b$ and $\mu/m_b$,
and
\begin{eqnarray} \label{Jvlo}
 J_{1-3}^{(v)}(\omega) &=& ( \bar\xi_n W)_{\omega} 
  \Big\{ \gamma_\mu\,, v_\mu \,, \frac{n_\mu}{n\mcdot v} \Big\}  h_v
  \,.
\end{eqnarray}
The choices $\{1,2,3\}$ correspond to the three different Dirac structures, 
and our basis in Eq.~(\ref{Jvlo}) agrees with Ref.~\cite{chay}.
 
\begin{table}[t!]
\begin{center}
\begin{tabular}{|c|c|c||l|}
\hline\hline
 & \hspace{0.1cm} {\parbox{2cm}{\vspace{0.2cm} RPI-I\\[-25pt] 
  \ \ \ $b_i^{(v)}(\hat\omega_1,0)=$\vspace{0.1cm}}}
  \hspace{0.05cm} 
  & {\parbox{2cm}{\vspace{0.2cm} RPI-II\\[-25pt] 
  \ \ \ $b_i^{(v)}(\hat\omega_1,\hat\omega_2)=$\vspace{0.1cm}}}
  & {\hspace{1.3cm}\mbox{Combined Constraints}\hspace{-1.3cm}}
  \\
\hline
\hspace{0.1cm}$b_{1-3}^{(v)}$\hspace{0.1cm} 
  & $C_{1-3}^{(v)}(\hat\omega_1)$ 
  & $C_{1-3}^{(v)}(\hat\omega_1\!+\!\hat\omega_2)$ 
  & \hspace{0.2cm} $b_{1-3}^{(v)}(\hat\omega_1,\hat\omega_2)=
  C_{1-3}^{(v)}(\hat\omega_1\!+\!\hat\omega_2)$ \\
$b_{4-6}^{(v)}$ & $0$ & $b_{4-6}^{(v)}(\hat\omega_1,\hat\omega_2)$ 
  & \hspace{0.2cm} $b_{4-6}^{(v)}(\hat\omega_1,\: 0_{\phantom{2}})=0$ \\
$b_7^{(v)}$ & $0$ & $0$ 
  & \hspace{0.2cm} $b_7^{(v)}(\hat\omega_1,\hat\omega_2)=0$ \\
$b_8^{(v)}$ & \hspace{0.1cm}$-2\, C_3^{(v)}(\hat\omega_1)$ 
  \hspace{0.1cm}
  & $b_8^{(v)}(\hat\omega_1,\hat\omega_2)$ 
  & \hspace{0.2cm} $b_8^{(v)}(\hat\omega_1,0)=-2\, C_3^{(v)}(\hat\omega_1)$ \\
$b_{9-11}^{(v)}$ & $0$ & $0$ 
  & \hspace{0.2cm} $b_{9-11}^{(v)}(\hat\omega_1,\hat\omega_2)=0$ \\
$b_{12,13}^{(v)}$ & $-2\,\hat\omega_1\, C^{(v)\prime}_{1,2}(\hat\omega_1)$ 
 & $-2(\hat\omega_1\!+\!\hat\omega_2)\: 
  C^{(v)'}_{1,2}(\hat\omega_1\!+\!\hat\omega_2)$
 & \hspace{0.1cm} $b_{12,13}^{(v)}(\hat\omega_1,\hat\omega_2)
  =-2(\hat\omega_1\!+\!\hat\omega_2)\:
  C^{(v)'}_{1,2}(\hat\omega_1\!+\!\hat\omega_2)$\hspace{0.1cm} 
 \\
$b_{14}^{(v)}$ & $-2\,\hat\omega_1\, C^{(v)\prime}_{3}(\hat\omega_1)$ 
 & $-2(\hat\omega_1\!+\!\hat\omega_2)\: 
  C^{(v)'}_{3}(\hat\omega_1\!+\!\hat\omega_2)$
 & \hspace{0.1cm} $b_{14}^{(v)}(\hat\omega_1,\hat\omega_2)
  =-2(\hat\omega_1\!+\!\hat\omega_2)\:
  C^{(v)'}_{3}(\hat\omega_1\!+\!\hat\omega_2)$\hspace{0.1cm} 
 \\
&  $+2\, C_3^{(v)}(\hat\omega_1)$ 
  &  $-b_8^{(v)}(\hat\omega_1,\hat\omega_2)$ \hspace{1.8cm}
  & \hspace{2.3cm} $-b_8^{(v)}(\hat\omega_1,\hat\omega_2)$ 
  \\ \hline
$b_{15-17}^{(v)}$ & $b_{15-17}^{(v)}(\hat\omega_1,0)$ & $0$ 
  & \hspace{0.2cm} $b_{15-17}^{(v)}(\hat\omega_1,\hat\omega_2)=0$ \\
$b_{18-20}^{(v)}$ & $b_{18-20}^{(v)}(\hat\omega_1,0)$  
  & $b_{18-20}^{(v)}(\hat\omega_1,\hat\omega_2)$  
  & \hspace{0.2cm} 
  $b_{18-20}^{(v)}(\hat\omega_1,\hat\omega_2)$ \ \mbox{unconstrained} \\
$b_{21}^{(v)}$ & $b_{21}^{(v)}(\hat\omega_1,0)$ & $0$ 
  & \hspace{0.2cm} $b_{21}^{(v)}(\hat\omega_1,\hat\omega_2)=0$ \\
$b_{22}^{(v)}$ & $b_{22}^{(v)}(\hat\omega_1,0)$  
  & $b_{22}^{(v)}(\hat\omega_1,\hat\omega_2)$  
  & \hspace{0.2cm} 
  $b_{22}^{(v)}(\hat\omega_1,\hat\omega_2)$ \ \mbox{unconstrained} \\
$b_{23-25}^{(v)}$ & $b_{23-25}^{(v)}(\hat\omega_1,0)$  
  & $0$  
  & \hspace{0.2cm} 
  $b_{23-25}^{(v)}(\hat\omega_1,\hat\omega_2)=0$ \\
$b_{26,27}^{(v)}$ & $b_{26,27}^{(v)}(\hat\omega_1,0)$ 
  & $-2(\hat\omega_1\!+\!\hat\omega_2)\: 
   C^{(v)'}_{1,2}(\hat\omega_1\!+\!\hat\omega_2)$  
  & \hspace{0.2cm} $b_{26,27}^{(v)}(\hat\omega_1,\hat\omega_2)
   =-2(\hat\omega_1\!+\!\hat\omega_2)\: 
   C^{(v)'}_{1,2}(\hat\omega_1\!+\!\hat\omega_2)$ \\
 & & 
 \hspace{0.2cm} $-2\, C_{1,2}^{(v)}(\hat\omega_1\!+\!\hat\omega_2)$ 
 \hspace{1.4cm}
 & \hspace{2.6cm} 
  $-2 \, C_{1,2}^{(v)}(\hat\omega_1\!+\!\hat\omega_2)$ \hspace{0.1cm}
 \\
\OMIT{
$b_{27}^{(v)}$ & $b_{27}^{(v)}(\hat\omega_1,0)$ 
  & $-2(\hat\omega_1\!+\!\hat\omega_2)\: 
   C^{(v)'}_2(\hat\omega_1\!+\!\hat\omega_2)$  
  & \hspace{0.2cm} $b_{27}^{(v)}(\hat\omega_1,\hat\omega_2)
   =-2(\hat\omega_1\!+\!\hat\omega_2)\: 
   C^{(v)'}_2(\hat\omega_1\!+\!\hat\omega_2)$ \\
 & & 
 \hspace{0.2cm} $-2 \big(1 + \frac{\hat\omega_1}{\hat\omega_2}\big)\: 
  C_2^{(v)}(\hat\omega_1\!+\!\hat\omega_2)$ \hspace{0.1cm}
 & 
 \hspace{2.5cm} 
  $-2 \big(1 + \frac{\hat\omega_1}{\hat\omega_2}\big)\:
   C_2^{(v)}(\hat\omega_1\!+\!\hat\omega_2)$ \hspace{0.1cm}
 \\
}
$b_{28}^{(v)}$ & $b_{28}^{(v)}(\hat\omega_1,0)$ 
  & $-2(\hat\omega_1\!+\!\hat\omega_2)\: 
   C^{(v)'}_3(\hat\omega_1\!+\!\hat\omega_2)$  
  & \hspace{0.2cm} $b_{28}^{(v)}(\hat\omega_1,\hat\omega_2)
   =-2(\hat\omega_1\!+\!\hat\omega_2)\: 
   C^{(v)'}_3(\hat\omega_1\!+\!\hat\omega_2)$ \\
 & & 
 \hspace{0.2cm} $+2 \big(1 + \frac{\hat\omega_1}{\hat\omega_2}\big)\: 
  C_3^{(v)}(\hat\omega_1\!+\!\hat\omega_2)$ \hspace{0.1cm}
 & 
 \hspace{2.5cm} 
  $+2 \big(1 + \frac{\hat\omega_1}{\hat\omega_2}\big)\:
   C_3^{(v)}(\hat\omega_1\!+\!\hat\omega_2)$ \hspace{0.1cm}
 \\
 & & \hspace{0.2cm} $-b_{22}(\hat\omega_1,\hat\omega_2)$ \hspace{2.2cm}
  &  \hspace{2.5cm} $-b_{22}(\hat\omega_1,\hat\omega_2)$ \\
\hline\hline
\end{tabular}
\end{center} 
\caption{\setlength\baselineskip{12pt} 
Summary of RPI constraints on the coefficients of the vector currents in 
Eq.~(\ref{Kv1_ginv}). The first column shows the constraints from type-I RPI
on $b_i^{(s)}(\omega,0)$, the second column shows the constraint on 
$b_i^{(s)}(\omega_1,\omega_2)$ from type-II RPI and the third column gives the 
combined constraint.
The final currents are displayed in Eq.~(\ref{Vfinal}),
and are defined so that they automatically satisfy these constraints.
\label{table_Kv}
}
\end{table}

At NLO the power counting only allows a single $D_\perp$ to appear. For the
possible spin structures it is easy to see that type-II RPI invariance does not
allow the vector index to be in an $\bn^\mu$ or any factors of $\bn\mcdot v$ to
appear just as for the leading currents.  Imposing the constraints from gauge
invariance, spin reduction, and type-III RPI then leaves 28 ${\cal
O}(\lambda)$ suppressed currents [$j=1,\ldots,28$]
\begin{eqnarray}
  {\cal K}_{j}^{(v)} &=& \int\!\!d\omega_1 d\omega_2 \ \ 
    b^{(v)}_{j}(\hat\omega_1,\hat\omega_2) \ \ 
    {\cal K}_{j}^{(v)}(\omega_1,\omega_2)\,,
\end{eqnarray}
with operators
\begin{eqnarray} \label{Kv1_ginv}
 {\cal K}_{1-6}^{(v)}(\omega_1,\omega_2) &=& 
    (\bar\xi_n W)_{\omega_1}\, (W^\dagger\! \DSppPl W)_{\omega_2}\,
    \,\frac{1}{\bnP^\dagger}\,
    \Gamma_{1-6}^\mu\,   h_v \,,\nn \\
 {\cal K}_{7,8}^{(v)}(\omega_1,\omega_2)  &=& 
     (\bar\xi_n W)_{\omega_1}\,  (W^\dagger i\!\!\DgppPl W)_{\omega_2}
     \,\frac{1}{\bnP^\dagger}\,
     \Big\{ \frac{\bnslash}{2} \,, \frac{1}{n\mcdot v} \Big\} h_v 
     \,, \\
  {\cal K}_{9-14}^{(v)}(\omega_1,\omega_2) &=& 
     (\bar\xi_n W)_{\omega_1} (W^\dagger\! \DvppPl W)_{\omega_2}
    \,\frac{1}{\bnP^\dagger}\,
    \Gamma_{1-6}^\mu\, h_v \,,\nn \\[4pt]
  {\cal K}_{15-20}^{(v)}(\omega_1,\omega_2)  &=& 
    (\bar\xi_n W)_{\omega_1}\, (W^\dagger\! \DSppPr W)_{\omega_2} \,
    \,\frac{1}{\bnP^\dagger}\,
    \Gamma_{1-6}^\mu\, h_v   \,,\nn \\
  {\cal K}_{21,22}^{(v)}(\omega_1,\omega_2)  &=& 
    (\bar\xi_n W)_{\omega_1}  (W^\dagger i\!\!\DgppPr W)_{\omega_2}
    \,\frac{1}{\bnP^\dagger}\,
    \Big\{ \frac{\bnslash}{2} \,, \frac{1}{n\mcdot v} \Big\} h_v 
    \,,\nn \\[1pt]
  {\cal K}_{23-28}^{(v)}(\omega_1,\omega_2)  &=& 
     (\bar\xi_n W)_{\omega_1} (W^\dagger\! \DvppPr W)_{\omega_2} 
     \,\frac{1}{\bnP^\dagger}\,
     \Gamma_{1-6}^\mu\, h_v \,,\nn
\end{eqnarray}
where the six $\Gamma_i^\mu$ matrices are
\OMIT{$\Gamma_{9-14}^\mu = \Gamma_{15-20}^\mu = \Gamma_{23-28}^\mu = 
\Gamma_{1-6}^\mu$ and}
\begin{eqnarray} \label{Gmu}
 \Gamma_{1-6}^\mu &=& \Big\{ \frac{\bnslash\gamma^\mu}{2}\,,
 \frac{\bnslash v^\mu}{2}\,, \frac{\bnslash n^\mu}{2 n\mcdot v}\,,
 \frac{\gamma^\mu}{n\mcdot v}\,, \frac{v^\mu}{n\mcdot v}\,,
 \frac{n^\mu}{(n\mcdot v)^2} \Big\} \,.
\end{eqnarray}

%
%
Working out the transformations of the leading and subleading currents in
a similar way as was done for the scalar currents we find
that the type-I invariants are 
$   J_1^{(v)} + {\cal K}_1^{(v)} + {\cal K}_{12}^{(v)} \,,
$
$   J_2^{(v)} + {\cal K}_2^{(v)} + {\cal K}_{13}^{(v)} \,,
$
$   J_3^{(v)} + {\cal K}_3^{(v)} + {\cal K}_{8}^{(v)} + {\cal K}_{14}^{(v)}\,,
$
$   {\cal K}_j^{(v)}\,,
$
$   {\cal K}_k^{(v)}\,,  
$
where $j=\{4,5,6,7,9,10,11\}$ and $k=\{15,\ldots,28\}$. Type-I invariance
allows any coefficients $b_k^{(v)}(\hat\omega_1,\hat\omega_2)$, but restricts
$b_{1-14}^{(v)}(\hat\omega_1,\hat\omega_2)$ as shown in the second column of 
Table~\ref{table_Kv}.

\OMIT{
\begin{eqnarray} \label{VIb}
  b_{\{1,2,3\}}^{(v)}(\hat\omega,0) &=& C_{\{1,2,3\}}^{(v)}(\hat\omega)\,,
   \qquad\quad\ 
  b_j^{(v)}(\hat\omega_1,0) = 0 \,,\qquad\ \ 
  b_{8}^{(v)}(\hat\omega,0) = -2 C_3^{(v)}(\hat\omega)\nn\\[5pt]
  b_{\{12,13\}}^{(v)}(\hat\omega,0) &=& -2\hat\omega\,
     C_{\{1,2\}}^{(v)\,\prime}(\hat\omega) \: \,, \qquad
  b_{14}^{(v)}(\hat\omega,0) = -2\hat\omega\,
     C_{3}^{(v)\,\prime}(\hat\omega) +2 C_3^{(v)}(\hat\omega)
  \,.
\end{eqnarray}
}
%
%

Looking at the transformations under type-II we find that the invariants are
independent of the $\{\gamma_\mu,v_\mu,n_\mu\}$ choice. Our results for the
type-II invariants are
$   J_1^{(v)} + {\cal K}_1^{(v)} + {\cal K}_{12}^{(v)}+ {\cal K}_{26}^{(v)}\,,
$
$   J_2^{(v)} + {\cal K}_2^{(v)} + {\cal K}_{13}^{(v)}+ {\cal K}_{27}^{(v)}\,,
$
$   J_3^{(v)} + {\cal K}_3^{(v)} + {\cal K}_{8}^{(v)} + {\cal K}_{14}^{(v)}
    + ({\cal K}_{22}^{(v)} + {\cal K}_{28}^{(v)})\,,
$
$  ({\cal K}_{22}^{(v)}-{\cal K}_{28}^{(v)})
$,
while ${\cal K}_\ell^{(v)}$ for $\ell=\{4,5,6,18,19,20\}$ are invariant by
themselves.   For
these combinations type-II invariance allows any
$b_\ell^{(v)}(\hat\omega_1,\hat\omega_2)$, but restricts
$b^{(v)}_{1-3,7,9-17,21-28}$ as shown in the third column of
Table~\ref{table_Kv}. Furthermore, currents ${\cal K}_m^{(v)}$ with
$m={\{7,9,10,11,15,16,17,21,23,24,25\}}$ are ruled out (ie. $b_m^{(v)}=0$).
\OMIT{
requires
\begin{eqnarray} \label{VIIb}
  b_{\{1,2,3\}}^{(v)}(\hat\omega_1,\hat\omega\!-\!\hat\omega_1) 
   &=& C_{\{1,2,3\}}^{(v)}(\hat\omega)\,,
   \qquad\ 
  b_{8}^{(v)}(\hat\omega_1,\hat\omega\!-\!\hat\omega_1)
   + b_{14}^{(v)}(\hat\omega_1,\hat\omega\!-\!\hat\omega_1) 
   = -2 \hat\omega\, C_3^{(v)\,\prime}(\hat\omega) \,, \nn\\[5pt]
  b_{\{12,13\}}^{(v)}(\hat\omega_1,\hat\omega\!-\!\hat\omega_1) 
   &=& -2\hat\omega\, C_{\{1,2\}}^{(v)\,\prime}(\hat\omega) \: \,, \nn\\
  && \hspace{-3.9cm}
   b_{22}^{(v)}(\hat\omega_1,\hat\omega\!-\!\hat\omega_1)
   + b_{28}^{(v)}(\hat\omega_1,,\hat\omega\!-\!\hat\omega_1) = 
   -2\hat\omega\, C_{3}^{(v)\,\prime}(\hat\omega) 
    -\frac{2\hat\omega}{(\hat\omega_1\!-\!\hat\omega)} \:
    C_{3}^{(v)}(\hat\omega) \,, \nn \\
  && \hspace{-3.9cm} 
   b_{\{26,27\}}^{(v)}(\hat\omega_1,\hat\omega\!-\!\hat\omega_1) 
    = -2\hat\omega\, C_{\{1,2\}}^{(v)\,\prime}(\hat\omega) 
    -\frac{2\hat\omega}{(\hat\omega_1\!-\!\hat\omega)} \:
    C_{\{1,2\}}^{(v)}(\hat\omega)
    \,.
\end{eqnarray}
}

It is easy to see that the type-I and type-II conditions in Table~\ref{table_Kv}
are compatible. The combined set of constraints are given by those in the fourth
column.  \OMIT{ together with any $b_k^{(v)}(\hat\omega_1,\hat\omega_2)$ with
  $k=\{18,19,20,28\}$, and any coefficients satisfying $b_8(\hat\omega,0) =
  -2C_3^{(v)}(\hat\omega)$ and $b_{\{4,5,6\}}^{(v)}(\hat\omega,0)=0$.}  Using
Eq.~(\ref{toB}) we can show that the constrained ${\cal K}_{\{4,5,6\}}^{(v)}$
are redundant with ${\cal K}_{\{18,19,20\}}^{(v)}$ respectively, just as was
done for the scalar current with ${\cal K}_2^{(s)}$ and ${\cal K}_6^{(s)}$. We
can also use Eq.~(\ref{toB}) to convert $K_8^{(v)}(\hat\omega_1,\hat\omega_2)
-K_{14}^{(v)}(\hat\omega_1,\hat\omega_2)$ into a term proportional to
$\delta(\omega_2)$ and a term that is redundant with
$K_{22}^{(v)}(\hat\omega_1,\hat\omega_2)-K_{28}^{(v)}(\hat\omega_1,
\hat\omega_2)$.

From the combined constraints we can then define a new complete set
of allowed vector operators, $K_{1-14}^{(v)}$.
\OMIT{
\begin{eqnarray}
  K_{\{1,2,3\}}^{(v)}(\omega) &=& \int\!\! d\omega_1\
   {\cal K}_{\{1,2,3\}}^{(v)}(\omega_1,\omega\!-\!\omega_1) \,, \nn\\
 K_4^{(v)}(\omega) &=&  \int\!\! d\omega_1\ 
   \big\{{\cal K}_8^{(v)}(\omega_1,\omega\!-\!\omega_1) 
    -{\cal K}_{14}^{(v)}(\omega_1,\omega\!-\!\omega_1)\big\} \,,\nn\\
 K_{\{5,6,7\}}^{(v)}(\omega) &=& \hat\omega \int\!\! d\omega_1\ 
   \big\{{\cal K}_{\{12,13,14\}}^{(v)}(\omega_1,\omega\!-\!\omega_1) 
    +{\cal K}_{\{26,27,28\}}^{(v)}(\omega_1,\omega\!-\!\omega_1)\big\} \,,\nn\\
 K_{\{8,9,10\}}^{(v)}(\omega) &=& \omega \int\!\! d\omega_1\ 
   {(\omega_1\!-\!\omega)^{-1}}\, 
   {\cal K}_{\{26,27,28\}}^{(v)}(\omega_1,\omega\!-\!\omega_1) \,,\nn\\
 K_{\{11,12,13\}}^{(v)}(\omega_1,\omega_2) &=& 
   {(\hat\omega_1\!+\!\hat\omega_2)}\:
   {\cal K}_{\{18,19,20\}}^{(v)}(\omega_1,\omega_2)\,,\nn\\
 K_{14}^{(v)}(\omega_1,\omega_2) &=& 
   {(\hat\omega_1\!+\!\hat\omega_2)}\:
   \big\{ {\cal K}_{22}^{(v)}(\omega_1,\omega_2) -
     {\cal K}_{28}^{(v)}(\omega_1,\omega_2) \big\} \,. 
\end{eqnarray}
}
Therefore after imposing type-I and type-II RPI plus all other constraints
we are left with our final set of allowed vector current operators
\begin{eqnarray} \label{Kvint}
   \nn\\[-5pt]
  {K}_{1-10}^{(v)} &=& \int\!\!d\omega \ \ 
    B_{1-10}^{(v)}(\hat\omega) \ \ 
    {K}_{1-10}^{(v)}(\omega)\,,\nn\\
   {K}_{11-14}^{(v)} &=& \int\!\!d\omega_1 d\omega_2\ \
    B_{11-14}^{(v)}(\hat\omega_1,\hat\omega_2) \ \
    {K}_{11-14}^{(v)}(\omega_1,\omega_2)\,,\\
 && \psframe[linewidth=2pt,linecolor=red,framearc=.3](-3.2,-0.2)(10,2.6)
 \nn
\end{eqnarray}
where
\begin{eqnarray} \label{Vfinal}
  K_{1-3}^{(v)}(\omega) &=&  
     -\Big(\bar\xi_n \frac{\bnslash}{2} \DSppPl W\Big)_{\!\omega} \,
     \frac{1}{\bnP^\dagger}\: 
     \Big\{ \gamma_\mu\,, v_\mu\,, \frac{n_\mu}{n\mcdot v} \Big\} h_v
    \,, \\[4pt]
   K_{4}^{(v)}(\omega) &=& 
    \Big(g^{\mu\alpha}\!-\!\frac{n^\mu v^\alpha}{n\mcdot v}\Big) 
     \big(\bar\xi_n  \:i\!\!\DgppPld\, W\big)_{\!\omega}\, 
     \frac{1}{\bnP^\dagger}\,\frac{1}{n\mcdot v}\:  h_v \,,\nn \\[4pt]
  K_{5-7}^{(v)}(\omega) &=&  
    \frac{v\mcdot \cP_\perp}{m_b}\: 
    \big( \bar\xi_n W \big)_{\omega} 
    \Big\{ \gamma_\mu\,, v_\mu\,, \frac{n_\mu}{n\mcdot v} \Big\}h_v\,,\nn\\[4pt] 
  K_{8-10}^{(v)}(\omega) &=& 
    \Big( \bar\xi_n\, 
    \frac{1}{n\mcdot v\: i\bn\mcdot D_c}  \DvppPr\: W\Big)_{\!\omega} 
   \Big\{ {\gamma_\mu}\,, {v_\mu}\,, 
    \frac{n_\mu}{n\mcdot v} \Big\} h_v
    \,,\nn \\[4pt]
  K_{11-13}^{(v)}(\omega_1,\omega_2) &=&  
   \frac{1}{m_b}\,\big( \bar\xi_n W \big)_{\omega_1}
    \Big\{ \gamma_\mu\,, v_\mu\,, \frac{n_\mu}{n\mcdot v} \Big\}
    \Big( \frac{1}{\bnP} W^\dagger ig \Bslash_c^\perp W\Big)_{\!\omega_2}
     h_v 
    \,,\nn\\[4pt]
  K_{14}^{(v)}(\omega_1,\omega_2) &=&  \frac{1}{m_b}\,
    \big(\bar\xi_n W\big)_{\!\omega_1} 
    \Big(g^{\mu\alpha}\!-\!\frac{n^\mu v^\alpha}{n\mcdot v}\Big) 
    \Big(\frac{1}{\bnP}
    W^\dagger ig B_{c\,\alpha}^\perp W\Big)_{\!\omega_2}\, h_v 
    \,. \nn\\
 && \psframe[linewidth=2pt,linecolor=red,framearc=.3](-4.3,-0.2)(9.5,7.6)
   \nn
\end{eqnarray}
The coefficients $B_{\{11,12,13,14\}}^{(v)}$ in Eq.~(\ref{Kvint}) depend on two
parameters $\omega_{1,2}$ and are unconstrained. The remaining coefficients
depend on only one parameter and are fixed by reparameterization invariance
\begin{eqnarray} \label{rpiBv}
  B_{1-3}^{(v)}(\hat\omega) &=& C_{1-3}^{(v)}(\hat\omega) \,, 
    \qquad
  B_{5-7}^{(v)}(\hat\omega)  = -2 C_{1-3}^{(v)\,\prime}(\hat\omega) \,, 
    \qquad
  B_{8-10}^{(v)}(\hat\omega)  = -2 C_{1-3}^{(v)}(\hat\omega)  \,,\nn \\
   B_4^{(v)}(\hat\omega)  &=& -2 C_3^{(v)}(\hat\omega)  \,.
\end{eqnarray} 

The form of the currents $K_{1,2}^{(v)}$ agree with Ref.~\cite{chay}, and if we
take a frame where $n\mcdot v=1$ and $v_\perp=0$ then $K_{3,4}^{(v)}$ also
agree.  Ref.~\cite{chay} looked at type-I RPI of the vector currents and our
constraints on $B_{1-4}^{(v)}$ agree with the ones found there. (We note that
the authors of Ref.~\cite{chay} also checked these results with explicit
one-loop computations.) At tree-level one matches onto the currents
$K_{1,8}^{(v)}$ and the two-body limit of $K_{13}^{(v)}$ (using the analog of
Eq.~(\ref{Kstwo})), and we agree with Ref.~\cite{bcdf} on the form of these
currents and the RPI constraint between $K_{1}^{(v)}$ and $K_8^{(v)}$.  The
structures in Eq.~(\ref{Vfinal}) which are new and which only appear beyond tree
level are $K_{5-7,9-12,14}^{(v)}$ and the three-body form of $K_{13}^{(v)}$.


\subsection{Pseudoscalar and Axial-Vector Currents}

The results for the pseudoscalar and axial-vector heavy-to-light currents can be
directly obtained from the analysis for the scalar and vector currents
respectively. The analysis is identical except for the extra $\gamma_5$ in the
Dirac structure.  For the pseudo-scalar currents we have the basis
$\{\bnslash/2\,, 1/n\mcdot v\}\gamma_5$, while for the axial-vector currents we
have $\Gamma^\mu_{\!\{1,\ldots,6\}}\:\gamma_5$ where $\Gamma_j^\mu$ is defined
in Eq.~(\ref{Gmu}). At LO the most general allowed pseudoscalar current are thus
\begin{eqnarray}\label{Jps_final}
  J_0^{(p)} &=& \int\!\! d\omega \: C_0^{(p)}(\hat\omega,\mu/m_b)
     \: J_0^{(p)}(\omega)\,,\nn\\
  J_0^{(p)}(\omega) &=& (\bar\xi_n W)_{\omega_1} \gamma_5 h_v \,,
\end{eqnarray}
while the axial-vector currents are
\begin{eqnarray}
  {J}_{1-3}^{(a)} &=& \int\!\!d\omega \ \ 
    C_{1-3}^{(a)}(\hat\omega) \ \ 
    {J}_{1-3}^{(a)}(\omega)\,,\nn\\
 J_{1-3}^{(a)}(\omega) &=& ( \bar\xi_n W)_{\omega} 
  \Big\{ \gamma_\mu\,, -v_\mu \,, -\frac{n_\mu}{n\mcdot v} \Big\}\gamma_5 h_v
  \,.
\end{eqnarray}

At NLO we again have eight possible pseudo-scalar currents ${\cal K}_j^{(p)}$
and 28 possible axial-vector currents ${\cal K}_j^{(a)}$ before imposing all
type-I and type-II constraints. After imposing the RPI constraints we find
results very similar to those in Eq.~(\ref{Ks_final}) and (\ref{Vfinal}). Thus
for the final NLO pseudoscalar currents we have
\begin{eqnarray}
 \nn \\[-5pt]
  {K}_{1-3}^{(p)} &=& \int\!\!d\omega \ \ 
    B_{1-3}^{(p)}(\hat\omega) \ \ 
    {K}_{1-3}^{(p)}(\omega)\,,\nn\\
   {K}_{4}^{(p)} &=& \int\!\!d\omega_1 d\omega_2\ \
    B_{4}^{(p)}(\hat\omega_1,\hat\omega_2) \ \
    {K}_{4}^{(p)}(\omega_1,\omega_2)\,, \\
 && \psframe[linewidth=2pt,linecolor=red,framearc=.3](-3.,-0.2)(8,2.7)
 \nn
\end{eqnarray}
where\\[-5pt]
\begin{eqnarray} \label{Pfinal}
  K_{1}^{(p)}(\omega) &=& 
    -\Big(\bar\xi_n \frac{\bnslash}{2} \DSppPl W\Big)_{\!\omega} \:
    \frac{1}{\bnP^\dagger}\:\gamma_5 h_v
    \,, \\[4pt]
  K_{2}^{(p)}(\omega) &=&  
    \frac{v\mcdot \cP_\perp}{m_b}\ \big(\bar\xi_n W \big)_{\omega}\:
    \gamma_5 h_v\,,\nn\\[4pt]
  K_{3}^{(p)}(\omega) &=& 
    \Big(\bar\xi_n\, \frac{1}{n\mcdot v\,i\bn\mcdot D_c}\DvppPr\, W 
    \Big)_{\!\omega} \: \gamma_5 h_v\,,\nn \\[4pt]
  K_4^{(p)}(\omega_1,\omega_2) &=& \frac{1}{m_b}\, 
    \big(\bar\xi_n W\big)_{\omega_1} \gamma_5
    \Big(\frac{1}{\bnP} W^\dagger ig \Bslash_c^\perp W\Big)_{\omega_2}\,  
     h_v 
    \,,\nn  \\
 && \psframe[linewidth=2pt,linecolor=red,framearc=.3](-3.5,-0.2)(7.5,5.2)
   \nn
\end{eqnarray}
and the RPI type-I and type-II constraints on the Wilson coefficients are
\begin{eqnarray} \label{rpiBp}
  B_1^{(p)}(\hat\omega) = C_0^{(p)}(\hat\omega)   \,, \qquad 
  B_2^{(p)}(\hat\omega)  = -2 C_0^{(p)\,\prime}(\hat\omega) \,, \qquad 
  B_3^{(p)}(\hat\omega)  = -2 C_0^{(p)}(\hat\omega)  \,.
\end{eqnarray}
For the final axial-vector NLO currents we find
\begin{eqnarray} \label{Kaint}
  \nn \\[-5pt]
  {K}_{1-10}^{(a)} &=& \int\!\!d\omega \ \ 
    B_{1-10}^{(a)}(\hat\omega) \ \ 
    {K}_{1-10}^{(a)}(\omega)\,,\nn\\
   {K}_{11-14}^{(a)} &=& \int\!\!d\omega_1 d\omega_2\ \
    B_{11-14}^{(a)}(\hat\omega_1,\hat\omega_2) \ \
    {K}_{11-14}^{(a)}(\omega_1,\omega_2)\,,\\
 && \psframe[linewidth=2pt,linecolor=red,framearc=.3](-3.5,-0.2)(10,2.6)
 \nn
\end{eqnarray}
where\\[-5pt]
\begin{eqnarray} \label{Afinal}
  K_{1-3}^{(a)}(\omega) &=&  
     - \Big(\bar\xi_n \frac{\bnslash}{2} \DSppPl W\Big)_{\!\omega} \,
     \frac{1}{\bnP^\dagger}\: 
     \Big\{ \gamma_\mu\,, -v_\mu\,, -\frac{n_\mu}{n\mcdot v} \Big\} \gamma_5 h_v
    \,, \\[4pt]
   K_{4}^{(a)}(\omega) &=& 
     -\Big(g^{\mu\alpha}\!-\!\frac{n^\mu v^\alpha}{n\mcdot v}\Big) 
     \big(\bar\xi_n  \:i\!\!\DgppPld\, W\big)_{\!\omega}\, 
     \frac{1}{\bnP^\dagger}\,\frac{1}{n\mcdot v}\:  \gamma_5 h_v \,,\nn \\[4pt]
  K_{5-7}^{(a)}(\omega) &=&  
    \frac{v\mcdot \cP_\perp}{m_b}\: 
    \big( \bar\xi_n W \big)_{\omega} 
    \Big\{ \gamma_\mu\,, -v_\mu\,, -\frac{n_\mu}{n\mcdot v} \Big\}\gamma_5 
    h_v\,,\nn\\[4pt] 
  K_{8-10}^{(a)}(\omega) &=& \Big( \bar\xi_n\, 
    \frac{1}{n\mcdot v\: i\bn\mcdot D_c}  \DvppPr\: W\Big)_{\!\omega} 
   \Big\{ {\gamma_\mu}\,, -{v_\mu}\,, 
    -\frac{n_\mu}{n\mcdot v} \Big\} \gamma_5  h_v
    \,,\nn \\[4pt]
  K_{11-13}^{(a)}(\omega_1,\omega_2) &=&  
    \frac{1}{m_b}\,\big( \bar\xi_n W \big)_{\omega_1}
    \Big\{ \gamma_\mu\,, -v_\mu\,, -\frac{n_\mu}{n\mcdot v} \Big\}\gamma_5
    \Big( \frac{1}{\bnP} W^\dagger ig \Bslash_c^\perp W\Big)_{\!\omega_2}
      h_v 
    \,,\nn\\[4pt]
  K_{14}^{(a)} &=&  -\frac{1}{m_b}\, \big(\bar\xi_n W\big)_{\!\omega_1}  
    \Big(g^{\mu\alpha}\!-\!\frac{n^\mu v^\alpha}{n\mcdot v}\Big) \gamma_5
    \Big(\frac{1}{\bnP}
    W^\dagger ig B_{c\,\alpha}^\perp W\Big)_{\!\omega_2}\, h_v 
    \,. \nn\\
 && \psframe[linewidth=2pt,linecolor=red,framearc=.3](-4.1,-0.2)(10.,7.6)
   \nn
\end{eqnarray}
The coefficients $B_{\{11,12,13,14\}}^{(a)}$ in Eq.~(\ref{Kvint}) depend on two
parameters $\omega_{1,2}$ and are unconstrained. The remaining coefficients
depend on only one parameter and are fixed by reparameterization invariance
\begin{eqnarray} \label{rpiBa}
  B_{1-3}^{(a)}(\hat\omega) &=& C_{1-3}^{(a)}(\hat\omega)\,, \qquad 
  B_{5-7}^{(a)}(\hat\omega) = -2 C_{1-3}^{(a)\,\prime}(\hat\omega)\,, \qquad 
  B_{8-10}^{(a)}(\hat\omega) = -2 C_{1-3}^{(a)}(\hat\omega) \,, \nn \\
  B_4^{(a)}(\hat\omega) &=& -2 C_2^{(a)}(\hat\omega) \,.
\end{eqnarray} 

The form of the pseudo-scalar and axial-vector currents are very analogous to the
scalar and vector currents, so rather than comparing with the literature we
simply refer to the comparisons in the proceeding sections for which part of 
our results were previously known.


\subsection{Tensor Currents} \label{sectJt}


At leading order in $\lambda$, there are four tensor currents, defined as
\begin{eqnarray}
J_{1-4}^{\mu\nu}(\omega) =\big(\bar\xi_n W\big)_{\!\omega}\
 \Gamma_{1-4}^{\mu\nu} h_v \,,
\end{eqnarray}
where the most general allowed Dirac structures are
\begin{eqnarray} \label{JTGamma}
\Gamma_{1-4}^{\mu\nu} = \bigg\{i\sigma^{\mu\nu}\,,\,
\gamma^{\mbox{\tiny $[$}\mu,} v^{\nu\mbox{\tiny$]$}}\,, \,
\frac{1}{n\mcdot v}\gamma^{\mbox{\tiny$[$}\mu,} n^{\nu\mbox{\tiny$]$}}\,,\,
\frac{1}{n\mcdot v} n^{\mbox{\tiny$[$}\mu,} v^{\nu\mbox{\tiny$]$}}\bigg\}\,,
\end{eqnarray}
where $\gamma_{\mbox{\tiny $[$}\mu,} v_{\nu\mbox{\tiny$]$}}=\gamma_\mu
v_\nu-\gamma_\nu v_\mu$ etc.  As before, no $\bn_\mu$ can appear at leading
order from type-II RPI.

\begin{table}[t!]
\begin{center}
\begin{tabular}{|c|c|c||l|}
\hline\hline
 & \hspace{0.1cm} {\parbox{2cm}{\vspace{0.2cm} RPI-I\\[-25pt] 
  \ \ \ $b_i^{(t)}(\hat\omega_1,0)=$\vspace{0.1cm}}}
  \hspace{0.05cm} 
  & {\parbox{2cm}{\vspace{0.2cm} RPI-II\\[-25pt] 
  \ \ \ $b_i^{(t)}(\hat\omega_1,\hat\omega_2)=$\vspace{0.1cm}}}
  & {\hspace{1.3cm}\mbox{Combined Constraints}\hspace{-1.3cm}}
  \\
\hline
\hspace{0.1cm}$b_{1-4}^{(t)}$\hspace{0.1cm} 
  & $C_{1-4}^{(t)}(\hat\omega_1)$ 
  & $C_{1-4}^{(t)}(\hat\omega_1\!+\!\hat\omega_2)$ 
  & \hspace{0.2cm} $b_{1-4}^{(t)}(\hat\omega_1,\hat\omega_2)=
  C_{1-4}^{(t)}(\hat\omega_1\!+\!\hat\omega_2)$ \\
$b_{5-8}^{(t)}$ & $0$ & $b_{5-8}^{(t)}(\hat\omega_1,\hat\omega_2)$ 
  & \hspace{0.2cm} $b_{5-8}^{(t)}(\hat\omega_1,\: 0_{\phantom{2}})=0$ \\
$b_{9-11}^{(t)}$ & $0$ & $0$ 
  & \hspace{0.2cm} $b_{9-11}^{(t)}(\hat\omega_1,\hat\omega_2)=0$ \\
$b_{12}^{(t)}$ & \hspace{0.1cm}$2\, C_3^{(t)}(\hat\omega_1)$ 
  & $b_{12}^{(t)}(\hat\omega_1,\hat\omega_2)$ 
  & \hspace{0.2cm} $b_{12}^{(t)}(\hat\omega_1,0)=2\, C_3^{(t)}(\hat\omega_1)$\\
$b_{13}^{(t)}$ & \hspace{0.1cm}$-2\, C_4^{(t)}(\hat\omega_1)$ 
  \hspace{0.1cm}
  & $b_{13}^{(t)}(\hat\omega_1,\hat\omega_2)$ 
  & \hspace{0.2cm} $b_{13}^{(t)}(\hat\omega_1,0)=-2\, C_4^{(t)}(\hat\omega_1)$\\
$b_{14}^{(t)}$ & $0$  & $b_{14}^{(t)}(\hat\omega_1,\hat\omega_2)$ 
  & \hspace{0.2cm} $b_{14}^{(t)}(\hat\omega_1,0)=0$ \\
$b_{15-18}^{(t)}$ & $0$ & $0$ 
  & \hspace{0.2cm} $b_{15-18}^{(t)}(\hat\omega_1,\hat\omega_2)=0$ \\
$b_{19,20}^{(t)}$ & $-2\,\hat\omega_1\, C^{(t)\prime}_{1,2}(\hat\omega_1)$ 
 & $-2(\hat\omega_1\!+\!\hat\omega_2)\: 
  C^{(t)'}_{1,2}(\hat\omega_1\!+\!\hat\omega_2)$
 & \hspace{0.1cm} $b_{19,20}^{(t)}(\hat\omega_1,\hat\omega_2)
  =-2(\hat\omega_1\!+\!\hat\omega_2)\:
  C^{(t)'}_{1,2}(\hat\omega_1\!+\!\hat\omega_2)$\hspace{0.1cm} 
 \\
$b_{21,22}^{(t)}$ & $-2\,\hat\omega_1\, C^{(t)\prime}_{3,4}(\hat\omega_1)$ 
 & $-2(\hat\omega_1\!+\!\hat\omega_2)\: 
  C^{(t)'}_{3,4}(\hat\omega_1\!+\!\hat\omega_2)$
 & \hspace{0.1cm} $b_{21,22}^{(t)}(\hat\omega_1,\hat\omega_2)
  =-2(\hat\omega_1\!+\!\hat\omega_2)\:
  C^{(t)'}_{3,4}(\hat\omega_1\!+\!\hat\omega_2)$\hspace{0.1cm} 
 \\
&  $+2\, C_{3,4}^{(t)}(\hat\omega_1)$ 
  &  $\pm b_{12,13}^{(t)}(\hat\omega_1,\hat\omega_2)$ \hspace{1.2cm}
  & \hspace{2.6cm} $\pm b_{12,13}^{(t)}(\hat\omega_1,\hat\omega_2)$ 
  \\ \hline
%
%
$b_{23-26}^{(t)}$ & $b_{23-26}^{(t)}(\hat\omega_1,0)$ & $0$ 
  & \hspace{0.2cm} $b_{23-26}^{(t)}(\hat\omega_1,\hat\omega_2)=0$ \\
$b_{27-30}^{(t)}$ & $b_{27-30}^{(t)}(\hat\omega_1,0)$  
  & $b_{27-30}^{(t)}(\hat\omega_1,\hat\omega_2)$  
  & \hspace{0.2cm} 
  $b_{27-30}^{(t)}(\hat\omega_1,\hat\omega_2)$ \ \mbox{unconstrained} \\
$b_{31-33}^{(t)}$ & $b_{31-33}^{(t)}(\hat\omega_1,0)$ & $0$ 
  & \hspace{0.2cm} $b_{31-33}^{(t)}(\hat\omega_1,\hat\omega_2)=0$ \\
$b_{34-36}^{(t)}$ & $b_{34-36}^{(t)}(\hat\omega_1,0)$  
  & $b_{34-36}^{(t)}(\hat\omega_1,\hat\omega_2)$  
  & \hspace{0.2cm} 
  $b_{34-36}^{(t)}(\hat\omega_1,\hat\omega_2)$ \ \mbox{unconstrained} \\
$b_{37-40}^{(t)}$ & $b_{37-40}^{(t)}(\hat\omega_1,0)$  
  & $0$  
  & \hspace{0.2cm} 
  $b_{37-40}^{(t)}(\hat\omega_1,\hat\omega_2)=0$ \\
$b_{41,42}^{(t)}$ & $b_{41,42}^{(t)}(\hat\omega_1,0)$ 
  & $-2(\hat\omega_1\!+\!\hat\omega_2)\: 
   C^{(t)'}_{1,2}(\hat\omega_1\!+\!\hat\omega_2)$  
  & \hspace{0.2cm} $b_{41,42}^{(t)}(\hat\omega_1,\hat\omega_2)
   = -2(\hat\omega_1\!+\!\hat\omega_2)\: 
   C^{(t)'}_{1,2}(\hat\omega_1\!+\!\hat\omega_2)$ \\
 & & 
 \hspace{0.05cm} $+2\,  \big(1 \!+\! \frac{\hat\omega_1}{\hat\omega_2}\big)\:
  C_{1,2}^{(t)}(\hat\omega_1\!+\!\hat\omega_2)$ 
 & \hspace{2.8cm} 
  $+2 \,\big(1 \!+\! \frac{\hat\omega_1}{\hat\omega_2}\big)\:
  C_{1,2}^{(t)}(\hat\omega_1\!+\!\hat\omega_2)$ \hspace{0.1cm}
 \\
$b_{43,44}^{(t)}$ & $b_{43,44}^{(t)}(\hat\omega_1,0)$ 
  & $-2(\hat\omega_1\!+\!\hat\omega_2)\: 
   C^{(t)'}_{3,4}(\hat\omega_1\!+\!\hat\omega_2)$  
  & \hspace{0.2cm} $b_{43,44}^{(t)}(\hat\omega_1,\hat\omega_2)
   =-2(\hat\omega_1\!+\!\hat\omega_2)\: 
   C^{(t)'}_{3,4}(\hat\omega_1\!+\!\hat\omega_2)$ \\
 & & 
 \hspace{0.2cm} $+2 \big(1 \!+\! \frac{\hat\omega_1}{\hat\omega_2}\big)\: 
  C_{3,4}^{(t)}(\hat\omega_1\!+\!\hat\omega_2)$ \hspace{0.1cm}
 & 
 \hspace{2.5cm} 
  $+2 \big(1 + \frac{\hat\omega_1}{\hat\omega_2}\big)\:
   C_{3,4}^{(t)}(\hat\omega_1\!+\!\hat\omega_2)$ \hspace{0.1cm}
 \\
 & & \hspace{0.2cm} $\pm b_{34,35}(\hat\omega_1,\hat\omega_2)$ \hspace{1.5cm}
  &  \hspace{2.5cm} $\pm b_{34,35}(\hat\omega_1,\hat\omega_2)$ \\
\hline\hline
\end{tabular}
\end{center} 
\caption{\setlength\baselineskip{12pt} 
Summary of RPI constraints on the coefficients of the tensor currents in 
Eq.~(\ref{Kt1_ginv}). The first column shows the constraints from type-I RPI
on $b_i(\omega,0)$, the second column shows the constraint on 
$b_i(\omega_1,\omega_2)$ from type-II RPI and the third column gives the 
combined constraint. Each $\pm$ refers to the first and second terms in their
row respectively. The final currents are displayed in 
Eqs.~(\ref{Tfinal1},\ref{Tfinal2}), and are defined so that they automatically 
satisfy these constraints. 
\label{table_Kt}
}
\end{table}
At $O(\lambda)$, 44 currents can be written down before imposing the 
RPI constraints. They can be chosen as
\begin{eqnarray} \label{Kt1_ginv}
{\cal K}_{1-8}^{(t)} &=& (\bar\xi_n W)_{\omega_1}
  (W^\dagger \DSppPl W)_{\omega_2} 
 \left\{\frac{\bnslash}{2}\Gamma_{1-4}^{\mu\nu}\,,
 \frac{1}{n\mcdot v}\Gamma_{1-4}^{\mu\nu}\right\}
 \frac{1}{\overline{\cal P}^\dagger} \, h_v \,, \\
{\cal K}_{9-14}^{(t)} &=& (\bar\xi_n W)_{\omega_1} (W^\dagger
 i\overleftarrow D{}_{c\perp}^{[\mu\,,} W)_{\omega_2} \Gamma_{1-6}^{\nu ]}
 \frac{1}{\overline{\cal P}^\dagger}\, h_v\,,\nn \\
{\cal K}_{15-22}^{(t)} &=& (\bar\xi_n W)_{\omega_1} (W^\dagger
  \DvppPl  W)_{\omega_2} 
  \left\{\frac{\bnslash}{2}\Gamma_{1-4}^{\mu\nu}\,,
  \frac{1}{n\mcdot v}\Gamma_{1-4}^{\mu\nu}\right\}
  \frac{1}{\overline{\cal P}^\dagger} \, h_v \,,\nn
\end{eqnarray}
and
\begin{eqnarray}
{\cal K}_{23-30}^{(t)} &=& (\bar\xi_n W)_{\omega_1} 
 (W^\dagger \DSppPr  W)_{\omega_2} \left\{\frac{\bnslash}{2}
 \Gamma_{1-4}^{\mu\nu}\,,
 \frac{1}{n\mcdot v}\Gamma_{1-4}^{\mu\nu}\right\}
 \frac{1}{\overline{\cal P}^\dagger} \, h_v \,, \\
{\cal K}_{31-36}^{(t)} &=&  (\bar\xi_n W)_{\omega_1}(W^\dagger
 i{\overrightarrow D}_{\! c\perp}^{[\mu\,,} W)_{\omega_2} \Gamma_{1-6}^{\nu ]}
 \frac{1}{\overline{\cal P}^\dagger}\, h_v \nn \\
{\cal K}_{37-44}^{(t)} &=& (\bar\xi_n W)_{\omega_1} 
  (W^\dagger \DvppPr  W)_{\omega_2} 
  \left\{\frac{\bnslash}{2}\Gamma_{1-4}^{\mu\nu}\,,
  \frac{1}{n\mcdot v}\Gamma_{1-4}^{\mu\nu}\right\}
  \frac{1}{\overline{\cal P}^\dagger} \, h_v\,. \nn
\end{eqnarray}
The Dirac matrix with one index $\Gamma_{1-6}^\mu$ is defined as in (\ref{Gmu})
\begin{eqnarray}
\Gamma_{1-6}^\mu = \left\{
\frac{\bnslash}{2}\left(\gamma_\mu, v_\mu, \frac{n_\mu}{v\mcdot n}\right)\,,
\frac{1}{n\mcdot v}\left(\gamma_\mu, v_\mu, \frac{n_\mu}{v\mcdot n}\right)
\right\}
\end{eqnarray}
The constraints from type-I and type-II RPI are derived as before. The final
constraints on the Wilson coefficients $b_{1-44}$ are shown in
Table~\ref{table_Kt}.

After imposing the constraints from the table one finds the final minimal set of
tensor heavy-light currents in the effective theory at $O(\lambda)$
\begin{eqnarray}
 \nn\\[-5pt]
K^{(t)}_{1-14} &=& 
\int \mbox{d}\omega B_{1-14}^{(t)}(\omega) K_{1-14}^{\mu\nu}(\omega)\,,\\
K^{(t)}_{15-21} &=& \int \mbox{d}\omega_1 \mbox{d}\omega_2
B_{15-21}^{(t)}(\omega_1,\omega_2) K_{15-21}^{\mu\nu}(\omega_1,\omega_2)\,.
 \nn \\
 && \psframe[linewidth=2pt,linecolor=red,framearc=.3](-3.,-0.2)(8,2.7)
 \nn
\end{eqnarray}

There are 14 independent $O(\lambda)$ two-body operators given explicitly by
\begin{eqnarray} \label{Tfinal1}
\nn\\[-5pt]
K_{1-4}^{(t)}(\omega) &=& -\Big(\bar\xi_n \frac{\bnslash}{2}  \DSppPl W
 \Big)_\omega \:\frac{1}{\bnP^\dagger}
 \Gamma_{1-4}^{\mu\nu}  h_v \,, \\
K_{5,6}^{(t)}(\omega) &=& \Big(\bar\xi_n \,i\overleftarrow D{}_{c\perp}^{\alpha}
 W\Big)_\omega  \: \frac{1}{\bnP^\dagger}\left\{
 g_{\alpha [\mu\,,} \gamma_{\nu ]} \!+\! v_\alpha \Gamma_3^{\mu\nu}\,,\,
 g_{\alpha [\mu\,,} v_{\nu ]} \!-\! v_\alpha \Gamma_4^{\mu\nu}\right\}
 h_v \,, \hspace{1.5cm}\nn \\
K_{7-10}^{(t)}(\omega) &=& \frac{1}{m_b}(v\mcdot {\cal P}_\perp )
(\bar\xi_n W)_\omega \Gamma_{1-4}^{\mu\nu}  h_v \,,\nn \\
K_{11-14}^{(t)}(\omega) &=& 
 \Big(\bar\xi_n \frac{1}{\bn\mcdot iD_c} \DvppPr W\Big)_\omega 
 \frac{1}{n\mcdot v} \Gamma_{1-4}^{\mu\nu}  h_v\,. \nn \\
 && \psframe[linewidth=2pt,linecolor=red,framearc=.3](-3,-0.2)(11.2,4.6)
   \nn
\end{eqnarray}
Their coefficients are fixed by reparameterization invariance in terms of the
${\cal O}(\lambda^0)$ Wilson coefficients $C_{1-4}^{(t)}(\omega)$ as
\begin{eqnarray} \label{rpiBt}
B_{1-4}^{(t)}(\omega) &=&  C_{1-4}^{(t)}(\omega)\,,\qquad\quad
B_5^{(t)}(\omega) = 2C_3^{(t)}(\omega)\,,\qquad 
B_6^{(t)}(\omega) = -2C_4^{(t)}(\omega) \,,\nn \\
B_{7-10}^{(t)}(\omega) &=& -2 C^{(t)\prime}_{1-4}(\omega)\,,\quad
B_{11-14}^{(t)}(\omega) = -2 C_{1-4}^{(t)}(\omega)\,.
\end{eqnarray}

In addition, there are 7 three-body collinear operators given by
\begin{eqnarray} \label{Tfinal2}
 \nn\\[-5pt]
 K_{15-18}^{(t)}(\omega_1,\omega_2) &=& \frac{1}{m_b}
   \big(\bar\xi_n W\big)_{\!\omega_1}\:
   \Gamma_{1-4}^{\mu\nu}
  \Big( \frac{1}{\bnP} W^\dagger ig \Bslash_{c,\perp} W\Big)_{\omega_2} 
   h_v \,,\nn \\
 K_{19}^{(t)}(\omega_1,\omega_2) &=& \frac{1}{m_b}
  \big(\bar\xi_n W\big)_{\!\omega_1}\:
  (g_{\alpha [\mu\,,} \gamma_{\nu ]} + v_\alpha \Gamma_3^{\mu\nu})
  \Big(\frac{1}{\bnP} W^\dagger ig B_{c,\perp}^\alpha W\Big)_{\omega_2} 
   h_v \,,\nn \\
 K_{20}^{(t)}(\omega_1,\omega_2) &=& \frac{1}{m_b}
  \big(\bar\xi_n W\big)_{\!\omega_1}\:
  (g_{\alpha [\mu\,,} v_{\nu ]} - v_\alpha \Gamma_4^{\mu\nu})
  \Big(\frac{1}{\bnP} W^\dagger ig B_{c,\perp}^\alpha W \Big)_{\omega_2} 
   h_v\,, \\
 K_{21}^{(t)}(\omega_1,\omega_2) &=& \frac{1}{m_b}
  \big(\bar\xi_n W\big)_{\!\omega_1}\
  g_{\alpha [\mu\,,} {n_{\nu ]}} \frac{1}{n\mcdot v}
  \Big( \frac{1}{\bnP} W^\dagger ig B_{c,\perp}^\alpha W \Big)_{\omega_2} 
   h_v\,. \nn \\
 && \psframe[linewidth=2pt,linecolor=red,framearc=.3](-3.9,-0.2)(9.8,4.6)
   \nn
\end{eqnarray}
Their coefficients $B_{15-21}^{(t)}(\hat\omega_1,\hat\omega_2)$ are not
constrained by any symmetry of the effective theory and have to be determined by
an explicit matching calculation.

At tree-level one matches onto the currents $K_{1,11}^{(t)}$ and the two-body
limit of $K_{17}^{(t)}$ (using the analog of Eq.~(\ref{Kstwo})), and we agree
with Ref.~\cite{bcdf} on the form of these currents and the RPI constraint
between $K_{1}^{(t)}$ and $K_{11}^{(t)}$.  The remaining operators in
Eqs.~(\ref{Tfinal1}) and (\ref{Tfinal2}) are new and only appear beyond tree
level (including the three body structure of $K_{11}^{(t)}$).


\section{Summary for Coefficients, Operators, and Feynman Rules
}

In this section we summarize results that should be useful for future
phenomelogical applications. In Section~\ref{sect_match} we summarize the full
set of known matching results and compare with the literature, in
section~\ref{sect_Jhlnoperp} we give simplified expressions for our basis of
currents in the frame $v_\perp=0$, $n\mcdot v=1$, and in section~\ref{sect_feyn}
we give Feynman rules for the subleading currents and ${\cal L}_{\xi
  q}^{(1,2)}$.


\subsection{Matching results for the currents} \label{sect_match}

In this subsection we summarize the one-loop matching results for the LO and NLO
$J_{hl}$ Wilson coefficients. With $\hat\omega_i=\omega/m_b$ these coefficients
are defined in previous sections as $C_i^{(d)}(\hat\omega,\mu/m_b)$ and
$B_i^{(d)}(\hat\omega_i,\mu/m_b)$ respectively, where $(d)$ denotes whether
the current is a scalar, pseudoscalar, vector, axial-vector, or a tensor.

For the LO currents the basis we use is different (though equivalent) to the
basis used in Ref.~\cite{bfps}. Since the one-loop matching for the LO
coefficients can be found in Ref.~\cite{bfps} it is useful to have the explicit
relation between our basis of coefficients $C_i^{(d)}$ and the coefficients
$C_j$ ($j=1$-$12$) that can be found there. We find
\begin{eqnarray} \label{switch}
&&\phantom{x}\hspace{-0.8cm}
   C_0^{(s)} = C_1 \,,\qquad 
   C_1^{(v)} = C_3 \,, \qquad\qquad
   C_1^{(a)} = C_6 \,, \qquad\qquad
   C_1^{(t)} = C_{10} \,,
   \\
&&\phantom{x}\hspace{-0.8cm}
   C_0^{(p)} = C_2 \,,\qquad
   C_2^{(v)} = C_5 \,,\qquad\qquad
   C_2^{(a)} = C_8 \,,\qquad\qquad
   C_2^{(t)} = -C_{12} \,,\qquad 
   \nn\\
&&\phantom{x}\hspace{-0.8cm}
 \phantom{C_0^{(p)} = C_2 \,,\,}\qquad
  C_3^{(v)} = C_4\!-\!C_3 \,,\qquad
  C_3^{(a)} = C_7\!-\!C_6 \,,\qquad
  C_3^{(t)} = C_{10}\!-\!C_9 \,,\quad
    \nn\\
 &&\phantom{x}\hspace{-0.8cm}
  \phantom{C_0^{(p)} = C_2 \,,\,\qquad
  C_3^{(v)} = C_4\!-\!C_3 \,,\qquad
  C_3^{(a)} = C_6\!-\!C_7 \,,\qquad}
  C_4^{(t)} = C_{12}\!+\!C_{10}\!-\!C_{11} \,.\nn
\end{eqnarray}
At tree level the matching between QCD and SCET is scheme independent. Matching
with the full QCD currents $\bar u \{1,\gamma_5,\gamma_\mu,\gamma_\mu\gamma_5,
i\sigma_{\mu\nu} \} b$ we find
\begin{eqnarray}
  && C_0^{(s)} = C_0^{(p)} = C_1^{(v)} = C_1^{(a)} = C_1^{(t)} = 1\,, \nn\\ &&
  C_2^{(v)} = C_3^{(v)} = C_2^{(a)} = C_3^{(a)} = C_2^{(t)} = C_3^{(t)} =
  C_4^{(t)} = 0 \,.
\end{eqnarray}
At one-loop we use the $\overline{\rm MS}$ scheme with naive dimensional
regularization (NDR) and match at $\mu=m_b$ to determine the
$C_i^{(d)}(\hat\omega,\mu/m_b)$. Using Eq.~(\ref{switch}) and results in
Ref.~\cite{bfps} one finds\OMIT{\footnote{Only $C_0^{(p)}$, $C_1^{(a)}$ depend 
on the choice of the anticommuting $\gamma_5$ in NDR.}}
\begin{eqnarray} \label{matchC}
C_{0}^{(s,p)}(\hat\omega,1) 
 &=& 1 - \frac{\alpha_s(m_b)C_F}{4\pi} \bigg\{ 2 \ln^2 (\hat\omega) 
  + 2 {\rm Li}_2(1\!-\!\hat\omega) - \frac{2\ln(\hat\omega)}{1-\hat\omega} +\frac{\pi^2}{12} 
  \bigg\} \,, \nn \\*
C_{1}^{(v,a)}(\hat\omega,1) 
 &=& 1 - \frac{\alpha_s(m_b)C_F}{4\pi} \bigg\{ 2\!\ln^2(\hat\omega) 
  + 2 {\rm Li}_2(1\!-\!\hat\omega) 
  + \ln(\hat\omega) \Big( \frac{3\hat\omega-2}{1-\hat\omega}\Big)
  + \frac{\pi^2}{12} + 6\bigg\} , \nn \\*
C_{1}^{(t)}(\hat\omega,1) 
 &=& 1 - \frac{\alpha_s(m_b)C_F}{4\pi} \bigg\{ 2\!\ln^2(\hat\omega) 
  + 2 {\rm Li}_2(1\!-\!\hat\omega) 
  +  \ln(\hat\omega) \Big( \frac{4\hat\omega-2}{1-\hat\omega} \Big) 
  + \frac{\pi^2}{12} + 6 \bigg\} , \nn \\
C_{2}^{(v,a)} (\hat\omega,1) 
 &=& \frac{\alpha_s(m_b)C_F}{4\pi}\:  \bigg\{ \frac{2}{(1-\hat\omega)} 
  + \frac{2\hat\omega\ln(\hat\omega)} {(1-\hat\omega)^2} \bigg\} \,, \nn \\
C_{2}^{(t)}(\hat\omega,1)  
 &=& 0  \,, \nn \\
C_{3}^{(v,a)}(\hat\omega,1) 
 &=& \frac{\alpha_s(m_b)C_F}{4\pi} \bigg\{ 
  \frac{(1-2\hat\omega)\hat\omega \ln(\hat\omega) }{(1-\hat\omega)^2}
  - \frac{\hat\omega}{1-\hat\omega} \bigg\} 
 \,, \nn \\*
C_{3}^{(t)}(\hat\omega,1) 
 &=& \frac{\alpha_s(m_b)C_F}{4\pi} \bigg\{ 
    \frac{-2\hat\omega \ln(\hat\omega)}{1-\hat\omega} \bigg\} 
 \,, \nn \\*
C_{4}^{(t)}(\hat\omega,1)  
 &=& 0 \,,
\end{eqnarray}
where $C_F=4/3$ for color $SU(3)$. To determine the coefficients for scales
$m_b\Lambda_{\rm QCD} < \mu^2 < m_b^2$ we require their anomalous
dimensions.\footnote{The full NLO result requires a two-loop anomalous dimension
  which uses information from Ref.~\cite{Korchemsky})} The LO and NLO anomalous
dimensions are universal and the running of these coefficients is given in
Ref.~\cite{bfps} (or for the case $\hat\omega=1$ in Ref.~\cite{bfl}).

At NLO in $\lambda$ tree level matching of the QCD current $\bar u \Gamma b$
onto SCET gives~\cite{chay,bpspc,bcdf}
\begin{eqnarray} \label{J1abtree}
 J^{(1a)}_{tree} &=& -\, \bar\xi_{n,p'} \frac{\bnslash}{2} \DSppPl W 
  \frac{1}{\bnP^\dagger}\, \Gamma\: h_v, \nn\\
 J^{(1b)}_{tree} &=& -\, \bar\xi_{n,p'} \Gamma\, \frac{\nslash}{2}\: \DSppPr W 
  \frac{1}{n\mcdot v\:m_b}\,  h_v\,, \nn\\
 J^{(1c)}_{tree} &=& -\, \frac{2}{n\mcdot v}\: \bar\xi_{n,p'} \Gamma\: 
  \frac{1}{i\bn\mcdot \overrightarrow D_c} \DvppPr W   h_v\,.
\end{eqnarray}
The normalization in Eq.~(\ref{J1abtree}) was first derived for $J^{(1a)}$ in
Ref.~\cite{chay}, and for $J^{(1b,1c)}$ in Ref.~\cite{bcdf}.  Comparing
Eq.~(\ref{J1abtree}) with our basis of currents we see that for any choice of
$\Gamma$ the $J^{(1a,1c)}_{tree}$ match onto a subset of our two-body currents.
On the other hand $J^{(1b)}_{tree}$ does not appear in the basis of two-body
currents. Instead it is obtained from the projection of a subset of the
three-body currents for cases where the corresponding coefficients
$B_i(\hat\omega_2,\hat\omega_2)$ depend only on the sum
$\hat\omega_1+\hat\omega_2$. This is certainly the case at tree level since the
coefficients are $\hat\omega_i$ independent. The three-body structure of the
currents can only show up at the level of one-loop matching. 

Using Eq.~(\ref{J1abtree}) to determine the tree-level value of the NLO Wilson
coefficients of the operators in Eqs.~(\ref{Ks_final},\ref{Vfinal},
\ref{Pfinal}, \ref{Afinal}, \ref{Tfinal1}, \ref{Tfinal2}) we find
\begin{eqnarray} \label{Bitree}
&&\phantom{x}\hspace{-0.2cm}
  B_1^{(s,p)} =1\,, \qquad\quad
  B_1^{(v,a)} = 1 \,, \qquad\quad
  B_{9,10}^{(v,a)} = 0 \,,\qquad\quad
  B_1^{(t)} = 1 \,,\qquad\quad
  B_{12-14}^{(t)} = 0 \,,
  \hspace{1.4cm} \nn \\
&&\phantom{x}\hspace{-0.2cm}
  B_2^{(s,p)} =0\,,\qquad\quad
  B_{2-4}^{(v,a)} = 0 \,,\qquad\quad
  B_{11,12}^{(v,a)} = 0 \,,\qquad\quad\!\!
  B_{2-6}^{(t)} = 0 \,,\qquad\quad\,
  B_{15,16}^{(t)} = 0 \,,
  \nn\\ 
&&\phantom{x}\hspace{-0.2cm}
  B_3^{(s,p)} = -2\,, \quad\quad\:
  B_{5-7}^{(v,a)} = 0 \,,\qquad\quad
  B_{13}^{(v,a)} = -1 \,,\qquad\!\!
  B_{7-10}^{(t)} = 0 \,,\qquad\qquad\!
  B_{17}^{(t)} = 1 \,,
  \nn\\
&&\phantom{x}\hspace{-0.2cm}
  B_4^{(s,p)} = 0\,,\qquad\quad
  B_8^{(v,a)} = -2\,,\qquad\:
  B_{14}^{(v,a)} = 0 \,,\qquad\quad\,
  B_{11}^{(t)} = -2 \,,\qquad
  B_{18-21}^{(t)} = 0 \,.
\end{eqnarray}
These results are in agreement with the RPI constraints in Eqs.~(\ref{rpiBs},
\ref{rpiBv}, \ref{rpiBp}, \ref{rpiBa}, \ref{rpiBt}). Coefficients in
Eq.~(\ref{Bitree}) that are zero indicate that the corresponding currents can
not be inferred at tree level since they are first matched onto at one-loop (or
beyond).  The full one-loop matching for all the ${\cal O}(\lambda)$ currents is
not currently known from direct computations. However, many of the NLO
coefficients are fixed in terms of the LO coefficients by RPI, namely
$B_{0-3}^{(s,p)}(\hat\omega)$, $B_{1-10}^{(v,a)}(\hat\omega)$, and
$B_{1-14}^{(t)}(\hat\omega)$. Summarizing
Eqs.~(\ref{rpiBs},\ref{rpiBv},\ref{rpiBp},\ref{rpiBa},\ref{rpiBt}) we have
\begin{eqnarray} \label{rpitot}
  B_1^{(s,p)} &=& C_0^{(s,p)}   \,, \qquad 
  B_2^{(s,p)}  = -2 C_0^{(s,p)\,\prime} \,, \qquad 
  B_3^{(s,p)}  = -2 C_0^{(s,p)}  \,,\nn\\
  B_{1-3}^{(v,a)} &=& C_{1-3}^{(v,a)}\,, \qquad 
  B_4^{(v,a)} = -2 C_3^{(v,a)} \,, \qquad
  B_{5-7}^{(v,a)} = -2 C_{1-3}^{(v,a)\,\prime}\,, \qquad 
  B_{8-10}^{(v,a)} = -2 C_{1-3}^{(v,a)} \,,\nn \\  
  B_{1-4}^{(t)} &=&  C_{1-4}^{(t)}\,,\qquad\quad
  B_5^{(t)} = 2C_3^{(t)}\,,\qquad\qquad\,
  B_6^{(t)} = -2C_4^{(t)} \,,\qquad\quad
  B_{7-10}^{(t)} = -2 C^{(t)\prime}_{1-4}\,,\nn\\
&&\hspace{-1.8cm}
  B_{11-14}^{(t)} = -2 C_{1-4}^{(t)}\,,
\end{eqnarray}
where to save space the $\hat\omega$ and $\mu$ dependence of the expressions on
both sides of these equalities is suppressed.  These results can be used to
determine the matching for these coefficients at $\mu=m_b$ using
Eq.~(\ref{matchC}). They also imply that the anomalous dimensions of these
coefficients are determined by the anomalous dimension of the leading order
coefficients~\cite{bfps}, so their values for scales $m_b\Lambda_{\rm QCD} <
\mu^2 < m_b^2$ is known.

For the coefficients of the 3-body operators,
$B_{4}^{(s,p)}(\hat\omega_1,\hat\omega_2)$,
$B_{11-14}^{(v,a)}(\hat\omega_1,\hat\omega_2)$, and
$B_{15-21}^{(t)}(\hat\omega_1,\hat\omega_2)$, neither the one-loop matching
results, nor even the LO anomalous dimensions are currently known.

Finally, we note that it is possible to relate the pseudoscalar and axial-vector
coefficients from the scalar and vector coefficients. For massless quarks the
QCD diagrams and SCET diagrams change in a trivial way under the chiral
transformation, $q\to \gamma_5 q$, and $\xi_n\to \gamma_5\xi_n$, provided we
work in a scheme such as NDR.  Therefore in this scheme the Wilson coefficients
of operators with and without $\gamma_5$ are related (see for example
Ref.~\cite{bfps} for the relations between LO coefficients).
\OMIT{
~\cite{bfps}, and we have
\begin{eqnarray}
  C_{0}^{(p)} = C_0^{(s)}\,,\quad
  C_{1-3}^{(a)} = C_{1-3}^{(v)} \,,\quad 
  B_4^{(p)}= B_4^{(s)} \,,\quad
  B_{11-14}^{(a)} =  B_{11-14}^{(v)}\,.
\end{eqnarray}
}
In other renormalization schemes these coefficients may differ.


\subsection{Summary of ${\cal O}(\lambda)$ currents in the frame $v_\perp=0$, 
$n\mcdot v=1$}
\label{sect_Jhlnoperp}

In sections~\ref{sectJs} through \ref{sectJt} we have derived the most general
basis of heavy-to-light current to ${\cal O}(\lambda)$ in an arbitrary
frame. However, for applications it is often most convenient to pick a frame
where $v_\perp=0$ and $v\mcdot n=1$.  In this frame the currents
$K_{2,3}^{(s,p)}$, $K_{5-10}^{(v,a)}$, and $K_{7-14}^{(t)}$ drop out. Thus there
are only (2,2,8,8,13) order ${\cal O}(\lambda)$ heavy-to-light currents which
are (scalar, pseudo-scalar, vector, axial-vector, tensor).  In this section we
summarize our results with this choice of basis vectors.

In this frame our leading order results for the $J_{hl}$ currents with a
complete set of Dirac structures can be summarized as
\begin{eqnarray} 
  J^{(0)} &=&  \int\!\! d\omega\ C_i^{(d)}(\hat\omega)\: J^{(0)}_i(\omega) 
    \,,\nn\\
  J^{(0)}_i(\omega) &=& (\bar\xi_n W)_{\omega}\, \Gamma_i^{(d)}\, h_v \,,
\end{eqnarray}
where in $\Gamma_i^{(d)}$ the $(d)$ specifies the type of current
(scalar,vector,$\ldots$), the $i$ specifies the member of the complete set of
possible structures of that type, and the Wilson coefficients are
$C_i^{(d)}(\hat\omega)$. For the minimum basis of Dirac structures we found
\begin{eqnarray}
  \Gamma_0^{(s)} &=& 1 \,,\qquad 
  \Gamma_0^{(p)} = \gamma_5 \,, \nn\\
  \Gamma_{\{1,2,3\}}^{(v)} &=& 
    \big\{ \gamma_\mu\: ,\: v_\mu\: ,\: {n_\mu} \big\}\nn\\
  \Gamma_{\{1,2,3\}}^{(a)} &=& 
    \big\{ \gamma_\mu\gamma_5 \: ,\: v_\mu \gamma_5 \: ,\:
    {n_\mu} \gamma_5 \big\}\nn\\
  \Gamma_{\{1,2,3,4\}}^{(t)} &=& 
    \big\{ i\sigma_{\mu\nu} \: ,\: \gamma_{\mbox{\tiny$[$}\mu,} 
    v_{\nu\mbox{\tiny$]$}} \: , \: 
    \gamma_{\mbox{\tiny$[$}\mu,} n_{\nu\mbox{\tiny$]$}}\: ,\:  
    n_{\mbox{\tiny$[$}\mu, } v_{\nu\mbox{\tiny$]$}} \big\} \,,
\end{eqnarray}
which is simply a linear combination of the basis in Ref.~\cite{bfps}.  At order
$\lambda$ our corresponding results for $J_{hl}$ in this frame can be summarized
as
\begin{eqnarray} \label{J1aJ1b}
  J^{(1a)} &=&  \int\!\! d\omega\ \ B_i^{(d)}(\hat\omega)\: \ J^{(1a)}_i(\omega)
    \,,\\
  J^{(1b)} &=&  \int\!\! d\omega_1\, d\omega_2\ \
     B_i^{(d)}(\hat\omega_1,\hat\omega_2)\: \ 
     J^{(1b)}_i(\omega_1,\omega_2) \,,\nn\\
  J^{(1a)}_i(\omega) &=& 
     \big(\bar\xi_n  \:i\!\!\DgppPld\, W\big)_{\!\omega}\, 
     \frac{1}{\bnP^\dagger}\, \Upsilon^{(d)\alpha}_{i}\:  h_v   \,,\nn \\
  J^{(1b)}_i(\omega_1,\omega_2) &=&   \frac{1}{m_b}\,
    \big(\bar\xi_n W\big)_{\!\omega_1} \Theta_{i}^{(d)\alpha} 
   \Big(\frac{1}{\bnP}
    W^\dagger ig B_{c\,\alpha}^\perp W\Big)_{\!\omega_2}\,  h_v 
    \,, \nn 
\end{eqnarray}
where
\begin{eqnarray}
  \Upsilon_{1}^{(s)\alpha} \!&=&\! \gamma^\alpha_\perp \frac{\bnslash}{2}  \,,
    \qquad 
   \Theta_{4}^{(s)\alpha} =\gamma^\alpha_\perp \,, \qquad 
   \Upsilon_{1}^{(p)\alpha} =\gamma^\alpha_\perp \frac{\bnslash}{2}
    \gamma_5\,, \qquad 
   \Theta_{4}^{(p)\alpha} = \gamma_5 \gamma^\alpha_\perp\,,
    \\
  \Upsilon_{1-4}^{(v)\alpha} \!&=&\! \Big\{ 
    \gamma^\alpha_\perp \frac{\bnslash}{2} \gamma^\mu 
     , \: \gamma^\alpha_\perp \frac{\bnslash}{2}  v^\mu
     , \: \gamma^\alpha_\perp \frac{\bnslash}{2}  n^\mu 
     , \: g^{\alpha\mu}_\perp \Big\}  \,,\qquad 
  \Theta_{11-14}^{(v)\alpha} = \Big\{  \gamma^\mu \gamma^\alpha_\perp
    , \:  v^\mu \gamma^\alpha_\perp\: , \:  n^\mu \gamma^\alpha_\perp
    , \: g_\perp^{\mu\alpha} \Big\}  \,,\nn\\
 \Upsilon_{1-4}^{(a)\alpha} \!\!&=&\!\! \Big\{ 
    \gamma^\alpha_\perp \frac{\bnslash}{2}  \gamma^\mu 
     , \, \gamma^\alpha_\perp  \frac{\bnslash}{2} v^\mu
     , \, \gamma^\alpha_\perp \frac{\bnslash}{2}  n^\mu 
     , \, g_\perp^{\alpha\mu} \Big\}\gamma_5  \,,\quad 
  \Theta_{11-14}^{(a)\alpha} \!=\! \Big\{  \gamma^\mu\gamma_5 \gamma^\alpha_\perp
     , \,  v^\mu\gamma_5\gamma^\alpha_\perp 
     , \,  n^\mu\gamma_5 \gamma^\alpha_\perp
   , , \, g^{\mu\alpha}_\perp \gamma_5 \Big\}  ,\nn\\
 \Upsilon_{1-6}^{(t)\alpha} \!&=&\! \bigg\{i\gamma^\alpha_\perp
    \frac{\bnslash}{2} \sigma^{\mu\nu}\,,\, 
    \gamma^\alpha_\perp \frac{\bnslash}{2}
     \gamma^{\mbox{\tiny $[$}\mu,} v^{\nu\mbox{\tiny$]$}}\,,\, 
    \gamma^\alpha_\perp \frac{\bnslash}{2}
     \gamma^{\mbox{\tiny$[$}\mu,} n^{\nu\mbox{\tiny$]$}}\,,\,
    \gamma^\alpha_\perp n^{\mbox{\tiny$[$}\mu,} v^{\nu\mbox{\tiny$]$}}\,,\,
    g_\perp^{\alpha \mbox{\tiny$[$}\mu\,,} \gamma^{\nu\mbox{\tiny$]$}} 
      \,,\,
    g_\perp^{\alpha\mbox{\tiny$[$}\mu\,,} v^{\nu\mbox{\tiny$]$}} 
   \bigg\}\,, \nn\\
  \Theta_{15-21}^{(t)\alpha} &=&  \bigg\{i \sigma^{\mu\nu}\gamma^\alpha_\perp
    \,,\,
    \gamma^{\mbox{\tiny $[$}\mu,} v^{\nu\mbox{\tiny$]$}}\gamma^\alpha_\perp\,,\, 
    \gamma^{\mbox{\tiny$[$}\mu,} n^{\nu\mbox{\tiny$]$}}\gamma^\alpha_\perp\,,\,
    n^{\mbox{\tiny$[$}\mu,} v^{\nu\mbox{\tiny$]$}} \gamma^\alpha_\perp\,,\,
    g_\perp^{\alpha \mbox{\tiny$[$}\mu\,,} \gamma^{\nu\mbox{\tiny$]$}} 
     \,,\,
    g_\perp^{\alpha\mbox{\tiny$[$}\mu\,,} v^{\nu\mbox{\tiny$]$}} 
     \,, 
    g_\perp^{\alpha\mbox{\tiny$[$}\mu\,,} {n^{\nu\mbox{\tiny$]$}}}
   \bigg\}\,. \nn
\end{eqnarray}
Note that due to Eq.~(\ref{toBD}) the form of $J^{(1b)}$ in Eq.~(\ref{J1aJ1b})
is identical to the form of the currents that was used in Ref.~\cite{bps4} 
since $[W^\dagger i \DgppPrd W]=[1/\bnP\: W^\dagger ig B_{c\,\alpha}^\perp W]$.


\subsection{Feynman rules for  $J_{hl}$ and ${\cal L}_{\xi q}$
} 
\label{sect_feyn}

\begin{figure}[t!]
\begin{eqnarray}
  && \hspace{-1.5cm}
  \begin{picture}(20,60)(20,-20) \mbox{\epsfxsize=3.6truecm
  \hbox{\epsfbox{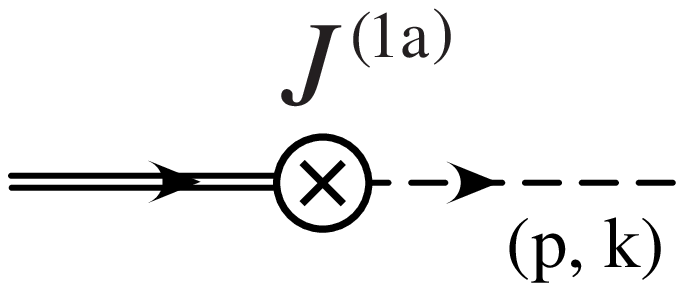}} } \end{picture} \hspace{2.8cm} \mbox{\large
  \raisebox{1.1cm}{=} \hspace{0.3cm} 
  \raise30pt 
  \hbox{$ \displaystyle  - i\, B_i^{(d)}(\bn\mcdot \hat p)\: 
  \frac{p^\perp_\alpha\:\Upsilon_i^{(d)\alpha}}{\bn\mcdot p}
  $ } } \nn\\[-20pt] 
 && \hspace{-1.5cm}
  \begin{picture}(20,83)(20,3)\mbox{\epsfxsize=3.6truecm
  \hbox{\epsfbox{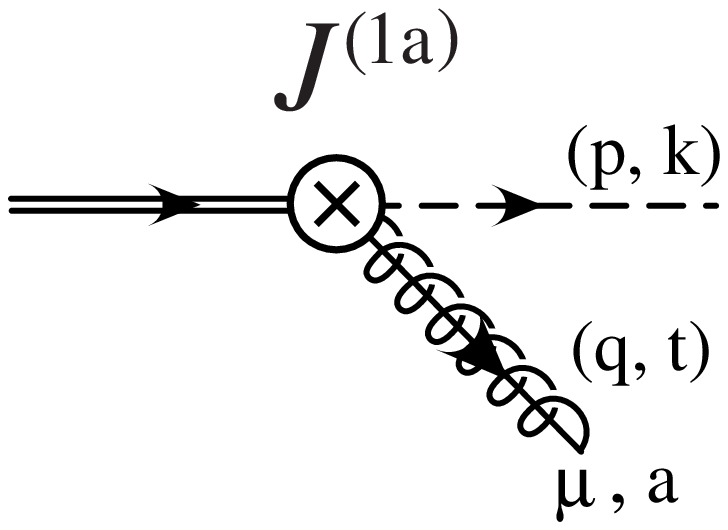}} } 
   \end{picture} \hspace{2.8cm}
  \parbox{9cm}{\mbox{\large \raisebox{3.4cm}{=} \hspace{0.1cm} 
  \raise95pt
  \hbox{ \large $\displaystyle 
  -i\,  B_i^{(d)}\big(\bn\mcdot (\hat p\!+\!\hat q)\big)\:
  \frac{g\, T^a}{\bn\mcdot (p\!+\!q)}  \:
  \bigg[ \Upsilon_i^{(d)\mu}
   +  \frac{\bn^\mu\: p^\perp_\alpha \Upsilon_i^{(d)\alpha} }{\bn\mcdot q} 
  \bigg] $ }}} \nn 
\end{eqnarray}
\vspace{-2.4cm} {\caption[1]{\setlength\baselineskip{10pt} Feynman rules for the
    $O(\lambda)$ currents $J^{(1a)}$ in Eq.~(\ref{J1aJ1b}) with zero and one
    gluon (the fermion spinors are suppressed).  For the collinear particles we
    show their (label,residual) momenta, where label momenta are $p,q\sim
    \lambda^{0,1}$ and residual momenta are $k,t\sim \lambda^2$. Momenta with a
    hat are normalized to $m_b$, $\hat p = p/m_b$ etc.}
\label{figJ1a} }
\end{figure}   
\begin{figure}[t!]
\begin{eqnarray}
  && \hspace{-1.5cm}
  \begin{picture}(20,60)(20,-20) \mbox{\epsfxsize=3.6truecm
  \hbox{\epsfbox{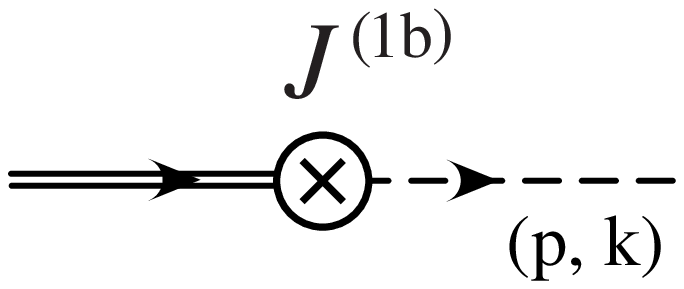}} } \end{picture} \hspace{2.8cm} \mbox{\large
  \raisebox{1.1cm}{=} \hspace{0.3cm} 
  \raise30pt 
  \hbox{$ \displaystyle  0
  $ } } \nn\\[-20pt] 
 && \hspace{-1.5cm}
  \begin{picture}(20,83)(20,3)\mbox{\epsfxsize=3.6truecm
  \hbox{\epsfbox{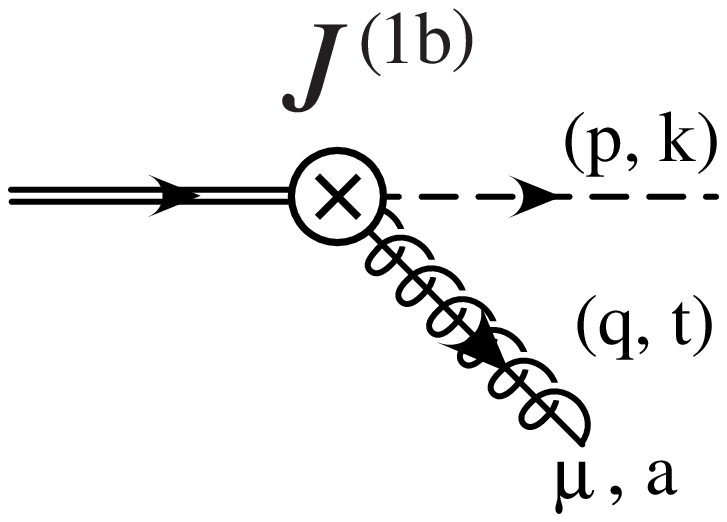}} } 
   \end{picture} \hspace{2.8cm}
  \parbox{9cm}{\mbox{\large \raisebox{3.4cm}{=} \hspace{0.1cm} 
  \raise95pt
  \hbox{ \large $\displaystyle 
  i\, B_i^{(d)}\big(\bn\mcdot \hat p,\bn\mcdot\hat q\big)\:
  \frac{g\, T^a}{m_b}  \:
  \bigg[ \Theta_i^{(d)\mu}
   -  \frac{\bn^\mu\: q^\perp_\alpha \Theta_i^{(d)\alpha} }{\bn\mcdot q} 
  \bigg] $ }}} \nn 
\end{eqnarray}
\vspace{-2.4cm} {\caption[1]{\setlength\baselineskip{10pt} Feynman rules for the
    $O(\lambda)$ currents $J^{(1b)}$ in Eq.~(\ref{J1aJ1b}) with zero and one
    gluon.  For the collinear particles we show their (label,residual) momenta,
    where label momenta are $p,q,q_i\sim \lambda^{0,1}$ and residual momenta are
    $k,t\sim \lambda^2$.  Momenta with a hat are normalized to $m_b$, $\hat p =
    p/m_b$ etc.  }
\label{figJ1b} }
\end{figure}   

In this subsection Feynman rules are given for the ${\cal O}(\lambda)$
heavy-to-light currents $J^{(1a)}$ and $J^{(1b)}$ in Eq.~(\ref{J1aJ1b}) which are
valid in a frame where $v_\perp=0$ and $v\mcdot n=1$. We also give the Feynman
rules that follow from the final form of the ${\cal L}_{\xi q}^{(1,2a,2b)}$
Lagrangians in Eq.~(\ref{Lxiq_final}).

\begin{figure}[t!]
\begin{eqnarray}
  && \begin{picture}(20,10)(20,0)
     \mbox{\epsfxsize=3.6truecm \hbox{\epsfbox{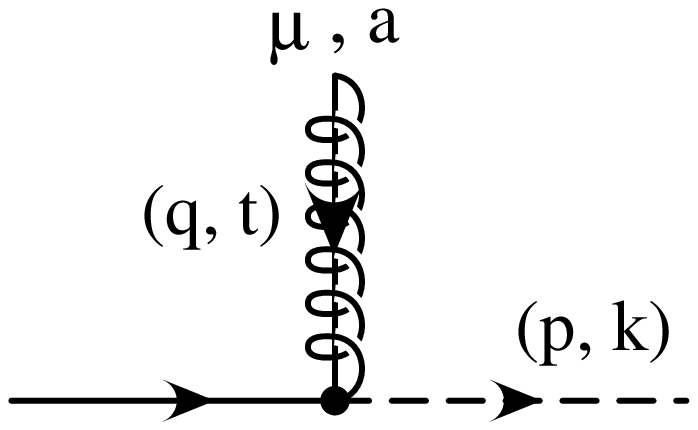}}  }
  \end{picture} \hspace{2.8cm} \mbox{\large 
  \raisebox{1.cm}{=} \hspace{1cm} 
  \raise30pt \hbox{ $   ig\,T^a\, 
 \Big[ \displaystyle \gamma_\mu^\perp 
   - \bn_\mu\, \frac{\displaystyle \qslash_\perp}{\displaystyle \bn\mcdot q}\: 
  \Big] $}  }
  \nn\\[-5pt]
  && \begin{picture}(20,80)(20,0)
     \mbox{\epsfxsize=3.6truecm \hbox{\epsfbox{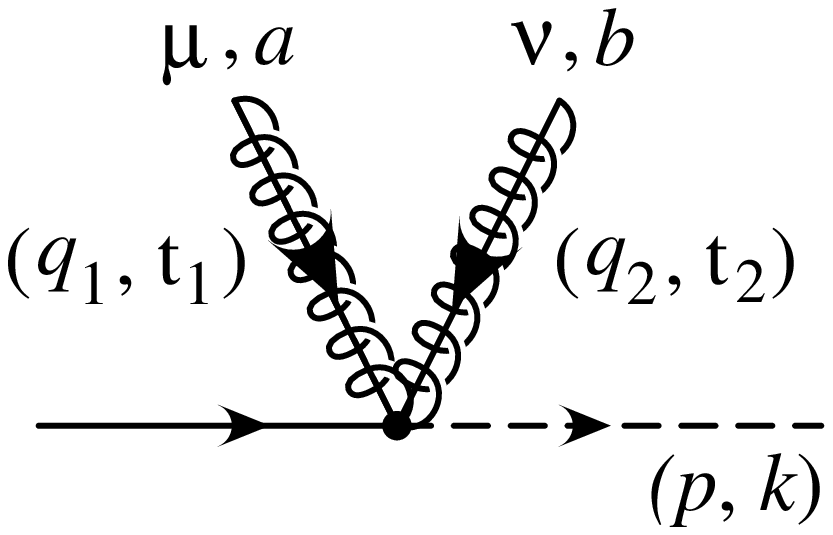}}  }
  \end{picture} \hspace{2.8cm} \mbox{ \large 
  \raisebox{1.3cm}{=}  \hspace{1cm}
  \raise50pt \hbox{$\displaystyle    ig^2\,\frac{T^b T^a}{\bn\mcdot q_1 } \bigg[
  \frac{\bn_\mu \bn_\nu\,\pslash^\perp }{ \bn\mcdot p}  - 
   {\gamma_\nu^\perp\, \bn_\mu} \bigg] 
    $ } }
 \nn\\[-35pt]
&& \hspace{2.in} \mbox{\large  
 \hbox{$\displaystyle  +\, ig^2\, \frac{T^a T^b}{\bn\mcdot q_2 }
  \bigg[ \frac{ \bn_\mu \bn_\nu\, \pslash^\perp} {\bn\mcdot p}
 - { \gamma_\mu^\perp\, \bn_\nu} \bigg] $ } }
 \nn\\[-25pt] \nn
\end{eqnarray}
{\caption[1]{\setlength\baselineskip{10pt} 
Feynman rules for the subleading usoft-collinear 
Lagrangian ${\cal L}_{\xi q}^{(1)}$ with one and two collinear
gluons (springs with lines through them). The solid lines are usoft quarks
while dashed lines are collinear quarks. For the collinear particles we
show their (label,residual) momenta. (The fermion spinors are suppressed.)
}
\label{figLuc1} }
\end{figure}
\begin{figure}[t!]
\begin{eqnarray}
 && \nn \\[-40pt]
 && 
   \begin{picture}(20,80)(20,0)
     \mbox{\epsfxsize=3.6truecm \hbox{\epsfbox{feyn1_ai.eps}}  }
  \end{picture} 
   \hspace{2.8cm} \mbox{\large 
  \raisebox{1.cm}{=} \hspace{0.5cm} 
  \raise30pt
  \hbox{$  \displaystyle i g\,T^a  
  \bigg[  \frac{\bnslash}{2} \Big( n_\mu -\frac{\bn_\mu\, n\mcdot q}{\bn\mcdot q}
  \Big) - \frac{\bn_\mu\, \tslash_\perp }{\bn\mcdot q} \bigg]$ }
  } 
 \nn\\[-15pt]
  && \begin{picture}(20,80)(20,0)
     \mbox{\epsfxsize=3.6truecm \hbox{\epsfbox{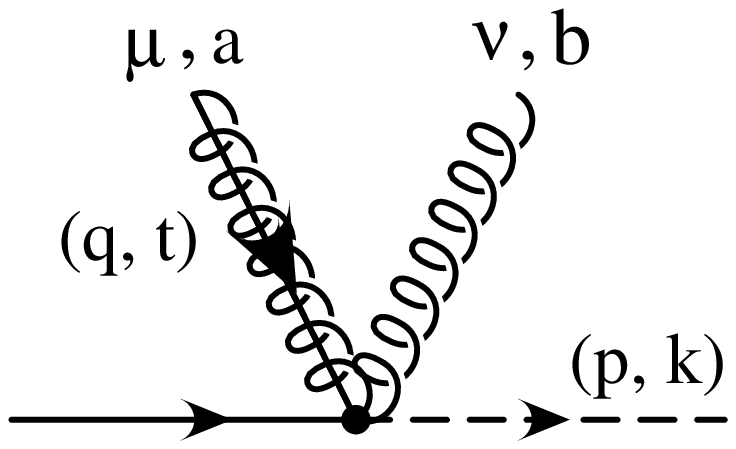}}  }
  \end{picture} \hspace{2.8cm} \mbox{\large 
  \raisebox{1.cm}{=} \hspace{0.5cm}
  \raise30pt \hbox{$ \displaystyle  \frac{-g^2 f^{abc}T^c}{\bn\mcdot q}\, 
    \bn_\mu \Big(\frac{\bnslash}{2}\, n_\nu + \gamma^\perp_\nu \Big)  $ } }
 \nn\\[-10pt] \nn
  && \begin{picture}(20,80)(20,0)
     \mbox{\epsfxsize=3.6truecm \hbox{\epsfbox{feyn2_ai.eps}}  }
  \end{picture} \hspace{2.8cm} \mbox{\large 
  \raisebox{1.cm}{=} \hspace{0.5cm}
  \raise50pt \hbox{\normalsize $\displaystyle  ig^2\,
 \frac{T^a T^b}{\bn\mcdot q_2} 
 \bigg[  -{n_\mu \bn_\nu}  \frac{\bnslash}{2}  
  + \bn_\mu \bn_\nu\, \frac{(\tslash_{1}\!+\!\tslash_{2})}
  {\bn\mcdot p} 
  \bigg] 
  $ } }
 \nn\\[-30pt]
&& \hspace{1.8in} \mbox{\large
 \hbox{\normalsize $\displaystyle +\, ig^2\, 
 \frac{T^b T^a}{\bn\mcdot q_1}
 \bigg[  -{n_\nu \bn_\mu}  \frac{\bnslash}{2}  
  + \bn_\mu \bn_\nu\, \frac{(\tslash_{1}\!+\!\tslash_{2})}{\bn\mcdot p} 
  \bigg] 
  $ } }
 \nn\\[-25pt] \nn
\end{eqnarray}
{\caption[1]{\setlength\baselineskip{10pt} Feynman rules for the $O(\lambda^2)$ 
usoft-collinear Lagrangian ${\cal L}_{\xi q}^{(2a)}$ with one and two gluons.
The spring without a line through it is an usoft gluon. For the collinear 
particles we show their (label,residual) momenta, where label momenta are 
$p,q,q_i\sim \lambda^{0,1}$ and residual momenta are $k,t,t_i\sim \lambda^2$.}
\label{figLuc2a} }
\end{figure}  

For the subleading currents the zero and one gluon Feynman rules for $J^{(1a)}$
and $J^{(1b)}$ are shown in Figs.~\ref{figJ1a} and \ref{figJ1b} respectively.
(From the results in the previous sections the Feynman rules for the currents
with $v_\perp\ne 0$ and $v\mcdot n\ne 1$ can also be easily derived.)  For
$J^{(1a)}$ the Wilson coefficients depend only on the total $\lambda^0$ collinear
momentum, while for $J^{(1a)}$ the coefficients depend on how the momentum is
divided between the quark and gluons. The $J^{(1a)}$ current has non-vanishing
Feynman rules with zero or one $A_n^\perp$ gluon and any number of $\bn\mcdot
A_n$ gluons. The possible gluons that appear in the $J^{(1b)}$ currents are
similar, but the current vanishes unless it has one or more collinear gluons
present.

For the mixed usoft-collinear Lagrangians from Eq.~(\ref{Lxiq_final}),
\begin{eqnarray}  \label{LucFeyn}
 {\cal L}^{(1)}_{\xi q}  &=&  
    \bar\xi_n \frac{1}{i\bn\mcdot D_c} ig \Bslash_c^\perp W q_{us}
    \mbox{ + h.c.  }\,, \nn\\
 {\cal L}^{(2a )}_{\xi q} &=& \bar\xi_n  \frac{1}{i\bn\mcdot D_c}\: 
     ig \Mslash\ \: W \, q_{us} \mbox{ + h.c.} \,, \nn\\
 {\cal L}^{(2b)}_{\xi q} &=&  \bar\xi_n \frac{\bnslash}{2} 
    i\Dslash_\perp^{\,c} \frac{1}{(i\bn\mcdot D_c)^2}\:  
    ig \Bslash_\perp^{\, c} W \: q_{us} \mbox{ + h.c.\hspace{0.1cm}} \,,
\end{eqnarray}
all Feynman rules involve at least one collinear gluon.  From ${\cal L}_{\xi
q}^{(1)}$ we obtain Feynman rules with zero or one $A_n^\perp$ gluons and any
number of $\bn\mcdot A_n$ gluons.  The one and two-gluon results are shown in
Fig.~\ref{figLuc1}.

\begin{figure}[t!]
\begin{eqnarray}
  && \begin{picture}(20,60)(20,0) \mbox{\epsfxsize=3.6truecm
  \hbox{\epsfbox{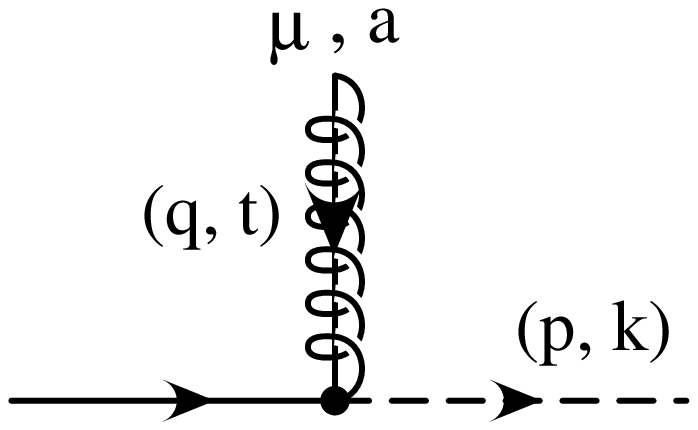}} } \end{picture} \hspace{2.7cm} \mbox{\large
  \raisebox{1.cm}{=} \hspace{0.2cm} \raise30pt \hbox{$ \displaystyle i g\,
  \frac{T^a}{\bn\mcdot q}\: \frac{\bnslash}{2} \Big[
  \qslash_\perp\gamma^\perp_\mu - \bn_\mu\: \frac{q_\perp^2}{\bn\mcdot q} \Big]
  $ } } \nn\\[-10pt] 
 && \begin{picture}(20,83)(20,3)\mbox{\epsfxsize=3.6truecm
  \hbox{\epsfbox{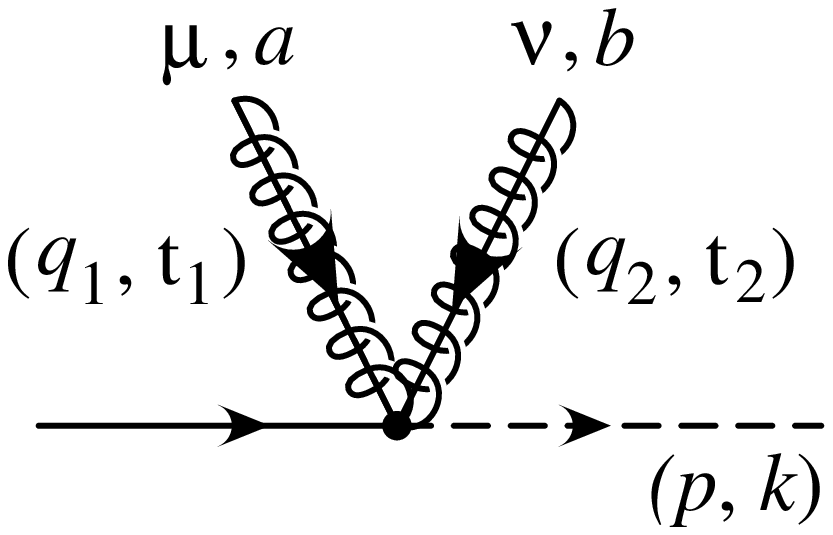}} } 
   \end{picture} \hspace{2.7cm}
  \parbox{9cm}{\mbox{\large \raisebox{2.5cm}{=} \hspace{0.2cm} 
  \raise90pt
  \hbox{\normalsize $\displaystyle ig^2\, \frac{ T^a T^b}{\bn\mcdot q_2}\:
  \frac{\bnslash}{2} \bigg[ {\gamma_\mu^\perp \gamma_\nu^\perp } -\frac{
  \pslash_\perp}{\bn\mcdot p} (\gamma^\perp_\mu \bn_\nu \!+\!\gamma^\perp_\nu
  \bn_\mu) - \frac{\gamma^\perp_\mu\: \bn_\nu\: \qslash_{2\perp} }{\bn\mcdot
  q_2} $ }}} \nn \\[-70pt] 
  && \hspace{1.8in} \mbox{\large \hbox{\normalsize
  $\displaystyle +\bn_\mu \bn_\nu \bigg( \frac{p_\perp^2}{(\bn\mcdot p)^2} +
  \frac{ \pslash_{\perp}\: \qslash_{2\perp} } { \bn\mcdot p\: \bn\mcdot
  q_2}\bigg) \bigg] 
  + \big[ (a,\mu,q_1,t_1)\leftrightarrow (b,\nu,q_2,t_2) \big]$ } }
  \nn
\OMIT{
&& \hspace{1.7in} \mbox{\large
 \hbox{\normalsize $\displaystyle +\, ig^2\,\frac{T^a T^b}{\bn\mcdot q_2} 
 \bigg[ \frac{ -\pslash_\perp\bn_\mu 
  \gamma_\nu^\perp \bnslash}{2\: \bn\mcdot (q_1\!+\!q_2)}
  + \frac{\gamma_\mu^\perp \gamma_\nu^\perp \bnslash}{2}  
  + \frac{\bn_\mu \bn_\nu \kslash_\perp}{ 
  \bn\mcdot (q_1\!+\!q_2)}\bigg] $ } }
 \nn\\[-25pt] \nn
}
\end{eqnarray}
{\caption[1]{\setlength\baselineskip{10pt}  
Feynman rules for the $O(\lambda^2)$ usoft-collinear Lagrangian 
${\cal L}_{\xi q}^{(2b)}$ with one and two gluons. For the collinear 
particles we show their (label,residual) momenta, where label momenta are 
$p,q,q_i\sim \lambda^{0,1}$ and residual momenta are $k,t,t_i\sim \lambda^2$.}
\label{figLuc2b} }
\end{figure}

For ${\cal L}_{\xi q}^{(2a)}$ we have Feynman rules with zero or one $\{n\mcdot
A_n,A_{us}^\perp\}$ gluon and any number of $\bn\mcdot A_n$ gluons. The one and
two-gluon results are shown in Fig.~\ref{figLuc2a}. Finally, for ${\cal L}_{\xi
q}^{(2b)}$ one finds Feynman rules with zero, one, or two $A_n^\perp$ gluons
and any number of $\bn\mcdot A_n$ gluons. In this case the one and two gluon
Feynman rules are shown in Fig.~\ref{figLuc2b}.
Note that it is important to treat the contributions from ${\cal L}_{\xi
  q}^{(2a)}$ and ${\cal L}_{\xi q}^{(2b)}$ separately since they show up in
different parts of the heavy-to-light factorization formulae derived in
Ref.~\cite{bps4} and shown in Eq.~(\ref{fFintro}).

For ${\cal L}_{\xi q}^{(2a)}+{\cal L}_{\xi q}^{(2b)}$ the Feynman rules are
different than one would derive using the intermediate form Eq.~(\ref{Lbcdf}),
since in transforming to the final form the equations of motion were
applied. However, observable predictions that are consistently made with either
set of Feynman rules will agree.

\section{Leading power predictions for $B$ to pseudoscalar mesons.}

\label{Bpisection}

As a phenomenological example, we consider the form factors for $B\to
\pi\ell\nu$, or more generally the form factors for $B\to P$ where $P$ is a
pseudoscalar meson (calculations using the factorization theorem in
Eq.~(\ref{fFintro}) for the vector meson are just as straightforward).  For
pseudoscalars there are three form factors in QCD, which are
conventionally defined by
\begin{eqnarray} \label{fullff}
\langle P(p)|\bar q \, \gamma^\mu b |\bar{B}(p_b)\rangle &=&
f_+(q^2)\left[p_b^\mu+p^{\mu}-\frac{m_B^2-m_P^2}{q^2}\,q^\mu\right]
+f_0(q^2)\,\frac{m_B^2-m_P^2}{q^2}\,q^\mu, \nn \\
\langle P(p)|\bar q \, i\sigma^{\mu\nu} q_\nu b|\bar{B}(p_b) \rangle &=&
-\frac{f_T(q^2)}{m_B+m_P}\left[q^2(p_b^\mu+p^{\mu})-
(m_B^2-m_P^2)\,q^\mu\right], 
\end{eqnarray}
where $q=p_b-p$.

For the region where $Q=\{E,m_b\}\gg \Lambda_{\rm QCD}$ (ie. small $q^2$) one
can use large energy factorization to study the form factors. For pions our
expansion parameter $\Lambda_{\rm QCD}/\bn\mcdot p \sim 0.5\,{\rm GeV}/(2E)$
becomes $1/4$ for $E\simeq 1\,{\rm GeV}$. This makes the region of $q^2$ where
the expansion is valid roughly $0 \lesssim q^2 \lesssim 10\,{\rm GeV^2}$.  In
SCET the form factors $f_+$, $f_0$, $f_T$ split themselves up into contributions
associated with three momentum regions: Wilson coefficients for $p^2\sim Q^2$,
two jet functions $J_{a,b}$ for $p^2\sim Q\Lambda_{\rm QCD}$, universal
light-cone wavefunctions for $p^2\sim \Lambda_{\rm QCD}^2$, and a single
non-factorizable form factor $\zeta_P$ (containing both $p^2\sim Q\Lambda_{\rm
  QCD}$ and $p^2\sim \Lambda_{\rm QCD}^2$). The leading contributions therefore
split into factorizable (F) and non-factorizable (NF)
contributions.\footnote{Here the phrase non-factorizable simply refers to the
  fact that the matrix elements can not be expressed in terms of convolutions
  with the standard light-cone wavefunctions.} This decomposition was defined by
the proof of a factorization formula for these form factors in Ref.~\cite{bps4}
\begin{eqnarray}\label{fF}
 f^{QCD}(q^2) &=& f^{\rm F}(Q) + f^{\rm NF}(Q) + \ldots \nn\\
f^{F}(Q) &=& N_0 \int_0^1\!\!\!\! dz\! 
    \int_0^1\!\!\!\! dx\! \int_0^\infty\!\!\!\!\! dr_+ \,
    T(z,Q,\mu_{\rm 0})\nn \\
 && \quad \times J(z,x,r_+,Q,\mu_{\rm 0},\mu)\: \phi_P(x,\mu) \:
  \phi_B(r_+,\mu) \,, \nn\\
f^{\rm NF}(Q) &=&C_k(Q,\mu)\: \zeta_k^P(Q,\mu) \,.
\label{fNF}
\end{eqnarray}
where $f^{\rm F}(Q)\sim f^{\rm NF}(Q)\sim Q^{-3/2}$ and the ellipses denote
terms that are suppressed by more powers of $1/Q$. Here $\phi_B=\phi_B^\pm$.

To separate the scales $Q^2$ and $Q\Lambda$ we match QCD onto an SCET$_{\rm I}$.
The scales $Q\Lambda$ and $\Lambda^2$ are then separated by matching SCET$_{\rm
  I}$ onto an SCET$_{\rm II}$~\cite{bps4}. Operators in SCET$_{\rm I}$ are
divided into F and NF categories depending on the form of the result of factoring
usoft gluons from collinear fields. In SCET$_{\rm I}$ the F contributions are
from the time-ordered products
\begin{eqnarray} \label{Tproducts} 
  T_1^F &=& \int\!\! \mbox{d}^4 x\: T\Big\{J^{(1a)}(0)\,,\,
   i{\cal L}^{(1)}_{\xi q}(x)\Big\}\,,\qquad 
  T_2^F = \int\!\! \mbox{d}^4 x\: T\Big\{J^{(1b)}(0)\,,\, i{\cal
  L}^{(1)}_{\xi q}(x)\Big\}\,,\nn\\ 
  T_3^F &=& \int\!\! \mbox{d}^4 x\: T\Big\{J^{(0)}(0)\,,\, i{\cal
  L}^{(2b)}_{\xi q}(x)\Big\}\,,
\end{eqnarray}
where the currents are taken from Eq.~(\ref{J1aJ1b}) and the usoft-collinear
Lagrangians from Eq.~(\ref{LucFeyn}).  After factorization of usoft and
collinear fields, the $T_i^F$'s are matched onto soft-collinear SCET$_{\rm II}$
operators. The collinear matrix elements are given in terms of jet functions
$J(z,x,r_+,Q)$, and the soft operators are given in terms of $B$ light-cone wave
functions $\phi_\pm(r_+)$ defined as~\cite{GN,bf}
\begin{eqnarray} \label{Bwf} 
  && \langle 0|\bar q(x^-) S_n(x^-,0)\,\Gamma h_v(0)|\bar B(v)\rangle 
   \\
  && \qquad =
  -\frac{i}{2} f_B m_B \int \mbox{d} r^+\: e^{-\frac{i}{2} r^+ x^-} {\rm Tr}
  \bigg\{ \frac{1\!+\!\vslash}{2} \Big[ \frac{\nslash\bnslash}{4} 
  \, \phi^+_B(r_+) + \frac{\bnslash\nslash}{4}\, \phi^-_B(r_+) \Big] 
  \gamma_5 \: \Gamma \bigg\}
  \nn\,.
\end{eqnarray} 

A few general properties of the factorizable term $f^F(Q)$ can be given without
an explicit computation.  First, the matrix elements of $T_{1,2}^F$ can only
depend on $\phi_B^+(r_+)$. This follows from the explicit form of the subleading
Lagrangian ${\cal L}^{(1)}_{\xi q}$, where $\bar\xi_n =\bar\xi_n
\frac{\nslash\bnslash}{4}$, so the usoft field $q_{us}$ appears only in the
combination $\bar q_{us} \frac{\nslash\bnslash}{4}$.  Using (\ref{Bwf}) this
implies that only the $\phi_B^+(r_+)$ term gives a nonvanishing contribution.
On the other hand, the factorizable operator $T_3^F$ depends on the combination
$\bar q_{us} \frac{\bnslash}{2}$, so its matrix element can only contain
$\phi_B^-(r_+)$. At tree-level the jet function $J$ from the matrix element of
$T_3^F$ vanishes, but a nonzero result could appear at one-loop order.  However,
the matrix element of $T_3^F$ contains the leading order current $J^{(0)}$, so
it obeys the same symmetry relations as those derived for the nonfactorizable
part $f^{NF}(Q)$~\cite{bps4}. Therefore, although this matrix element is
factorizable it does not increase the number of unknown non-perturbative
functions since for phenomenological analyses $\phi^-_B$ can be absorbed in
$\zeta_k^M$.  With this choice, all remaining factorizable contributions are
expressible in terms of just $\phi_B^+(r_+)$.
 
Using the approach explained in \cite{bps4} we can obtain the results for the
form factors. After factorization of usoft and collinear fields the T-products
of collinear fields coming from $T_{1,2}^F$ are given by (using
Eqs.~(\ref{J1aJ1b}) and (\ref{LucFeyn}))
\begin{eqnarray}
 {\cal J}^{1a}_{\omega}(x) &\equiv & 
  T \Big[ \bar \xi_n  \, i\!\!\DgppPld W\Big]_\omega^{i A}(0)\,
  \Big[W^\dagger ig \Bslash_\perp^{\, c} W \frac{1}{\bnP^\dagger} W^\dagger
   \: \xi_n\Big]_0^{j B}(x) \,, \\
 {\cal J}^{1b}_{\omega_1,\omega_2}(x) &\equiv & 
  T \bigg[ \big[ \bar \xi_n W \big]_{\omega_1} 
  \Big[ \frac{1}{\bnP} W^\dagger ig B_{\perp\alpha}^{\, c}  W \Big]_{\omega_2}
  \bigg]^{i A}(0)\,
  \Big[W^\dagger ig \Bslash_\perp^{\, c} W \frac{1}{\bnP^\dagger} W^\dagger
  \: \xi_n\Big]_0^{j B}(x)\,, \nn
\end{eqnarray}
where $i,j$ are Dirac indices and here $A,B$ are color indices in the
fundamental representation.  The functions ${\cal J}_{1a,1b}$ are collinear
gauge invariant and satisfy the spin structure constraints $\nslash {\cal
  J}_{1a,1b} = {\cal J}_{1a,1b}\nslash = 0$, and ${\rm tr} [ {\cal
  J}_{1a,1b}]=0$.

Taking into account constraints from the Dirac structure of the effective theory
fields one can easily find the most general form of the operators appearing in
the matching of ${\cal J}^{1a,1b}$ onto operators in ${\rm SCET}_{\rm II}$.  The
jet functions $J_{a,b}$ are defined by the terms which contribute on a
pseudoscalar state
\begin{eqnarray} \label{jets}
 {\cal J}_{\omega}^{1a}(x) &=& i \delta(x_+) \delta^2(x_\perp)
 \left[ \gamma_\perp^\alpha
  {\nslash}\gamma_5 \right]^{ji} \delta^{AB} \: \frac{1}{\omega}
\int\!\! d\bar\eta \int\!\! \frac{dr^+}{2\pi} \: e^{-\frac{i}{2} r^+ x^-}   \\
& &\quad \times 
 J_{a}(\bar\eta,r^+) \left[(\bar\xi_n W)\, \delta(\bar\eta-\bnP_+)\,
 \frac{\bnslash}{2}\gamma_5 (W^\dagger \xi_n)\right]_{\rm II} 
 + \cdots\nn\\
 {\cal J}^{1b}_{\omega_1,\omega_2}(x) &=&
 i \delta(x_+) \delta^2(x_\perp) \left[ \gamma_\perp^\alpha
 {\nslash}\gamma_5 \right]^{ji}\delta^{AB} \:  
 \frac{1}{\omega_1 \!+\! \omega_2} \int\!\! d\bar\eta \int\!\! 
  \frac{dr^+}{2\pi}\: e^{-\frac{i}{2} r^+ x^-} \\
& &\quad \times J_{b}(\bar\omega,\bar\eta,r^+) 
 \left[(\bar\xi_n W)\, \delta(\bar\eta-\bnP_+)\, \frac{\bnslash}{2}\gamma_5 
 (W^\dagger \xi_n)\right]_{\rm II} + \cdots\,,\nn
\end{eqnarray}
where $\bar\omega=\omega_1-\omega_2$.  We have suppressed the dependence of
$J_{1a,1b}$ on the $\mu$'s and on $\omega$, $\omega_1+\omega_2$ (the latter
combinations would be simply set to $\bn\mcdot p$ in the pseudoscalar matrix
element by momentum conservation~\cite{cbis}).  The ellipses in Eq.~(\ref{jets})
denote color octet terms and other operators which do not contribute for a
pseudoscalar meson $P$.

Using Eq.~(\ref{jets}) the operators in $\langle P_n(p)| T_{1,2}^F |\bar
B_v\rangle$ factor into a product of matrix elements that can be evaluated with
Eq.~(\ref{Bwf}) and Eq.(12,13) of Ref.~\cite{bps}. Switching variables to $x,z$
by using $\bar\omega=(2x-1)\bn\mcdot p$ and $\bar\eta=(2z-1)\bn\mcdot p$ we find
the following factorization theorems which are valid at leading
order\footnote{We kept a kinematic factor of $m_P$ in the prefactor of $f_T$
  even though it is formally power suppressed.}  in $1/Q$ and all orders in
$\alpha_s$
\begin{eqnarray} \label{fresults}
 f_+(q^2) \!&=&\! N_0 \int_0^1 \!\!\mbox{d}x \!  
  \int_0^\infty\!\!\!\!\! \mbox{d}r_+ \:
 \bigg[\frac{2E\!-\!m_B}{m_B}\: T^{(+)}_a(E,\mu_{\rm 0})\,
   \, J_a(x,r_+,Q,\mu_0,\mu)\: \nn\\
&& +\frac{2E}{m_b} \int_0^1 \!\!\mbox{d}z\:  T_b^{(+)}(E,z,\mu_{\rm 0})\:
   J_b(z,x,r_+,Q,\mu_0,\mu)\:\bigg] \phi_P(x,\mu)\: \phi_B^+(r_+,\mu) 
   \nn\\[3pt]
&& 
  + \Big\{C_1^{(v)}(2\hat E,\mu_{\rm 0})\!
     +\!\frac{E}{m_B} C_2^{(v)}(2\hat E,\mu_{\rm 0})
   \!+\! C_3^{(v)}(2\hat E,\mu_{\rm 0})\Big\}
   \zeta^P(Q,\mu_{\rm 0})\,, \\[6pt]
 f_0(q^2) \!&=&\!  N_0 \int_0^1 \!\! \mbox{d}x \! 
  \int_0^\infty\!\!\!\!\! \mbox{d}r_+
 \!\bigg[\frac{2E(m_B\!-\!2E)}{m_B^2}\: T^{(0)}_a(E,\mu_{\rm 0}) 
  \,  J_a(x,r_+,Q,\mu_0,\mu)\:
 \nn\\
 &&  + \frac{4E^2}{m_b m_B} \int_0^1 \!\!\mbox{d}z\: T_b^{(0)}(E,z,\mu_{\rm 0})
  J_b(z,x,r_+,Q,\mu_0,\mu) \bigg] \phi_P(x,\mu)\,\phi_B^+(r_+,\mu) 
    \nn\\[3pt]
&&  + \frac{2E}{m_B} \Big\{C_1^{(v)}(2\hat E,\mu_{\rm 0}) 
  \!+\! \frac{m_B\!-\!E}{m_B} C_2^{(v)}(2\hat E,\mu_{\rm 0}) 
  \!+\! C_3^{(v)}(2\hat E,\mu_{\rm 0})\Big\} \zeta^P(E,\mu_{\rm 0})\,, \nn \\
 f_T(q^2) \!&=&\!  N_0\frac{m_B\!+\!m_P}{m_B} \int_0^1 \!\! \mbox{d}x \! 
    \int_0^\infty\!\!\!\!\! \mbox{d}r_+
  \bigg[ -T_a^{(T)}(E,\mu_{\rm 0}) \:
  J_a(x,r_+,Q,\mu_0,\mu)\nn\\
&&  -\frac{2E}{m_b} \int_0^1 \!\!\mbox{d}z\: T_b^{(T)}(E,z,\mu_{\rm 0})
  J_b(z,x,r_+,Q,\mu_0,\mu)
  \bigg] \phi_P(x,\mu)\: \phi_B^+(r_+,\mu)
  \nn\\[3pt]
&&   + \frac{m_B\!+\!m_P}{m_B}
  \left\{C_1^{(t)}(2\hat E,\mu_{\rm 0}) - C_2^{(t)}(2\hat E,\mu_{\rm 0}) -
  C_4^{(t)}(2\hat E,\mu_{\rm 0})\right\}
  \zeta^P(E,\mu_{\rm 0})\,,\nn
\end{eqnarray}
where $\hat E =E/m_b$, $Q=\{E,m_b\}$, and the normalization coefficient is given
by $N_0=f_B f_P\, m_B/(4 E^2)$. The matrix element involving non-factorizable
operators gives $\zeta^P(Q,\mu)$ which is the reduced form factor describing
decays to a pseudoscalar meson $P$. The quantities in square brackets and curly
brackets are calculable, that is the $T_{a,b}$'s and $J_{a,b}$'s have expansions
in $\alpha_s(Q)$ and $\alpha_s(\sqrt{Q\Lambda})$ respectively. Note that the
$J_{a,b}$ are universal, meaning that at any order in $\alpha_s$ it is these
same jet functions which appear for any pseudoscalar meson and independent of
which form factor $f_{+,0,T}$ we consider. Therefore, the factorization theorem
still gives information even in the case where we assume that
$\alpha_s(\sqrt{Q\Lambda})$ is non-perturbative.

Working at ${\cal O}(\alpha_s(\mu_0))$ (ie. tree level) for the jet functions
gives
\begin{eqnarray} \label{Jtree}
 J_{1a}(x,r^+) &=& \frac{\pi C_F}{N_c} \: 
  \frac{\alpha_s(\mu_0)}{x r^+}\\
  J_{1b}(z,x,r^+) &=& \frac{\pi C_F}{N_c} \:
  \frac{\alpha_s(\mu_0)}{x r^+} \: \delta(z-x)\,.
\end{eqnarray}
At this level the $z$ integrals in Eq.~(\ref{fresults}) disappear because the
tree level jet gives a $\delta(z-x)$, and this causes the $z$ variable in the
$T_b$'s to be replaced by $x$.  The $T_{a,b}^{(j)}$ are combinations of Wilson
coefficients appearing in the $J^{(1a,1b)}$ currents given in Eq.~(\ref{J1aJ1b})
and should be evaluated at a scale $\mu_0^2\sim Q\Lambda$.  Expressed in terms
of the Wilson coefficients defined in Sec.~IV, they are given by
\begin{eqnarray}
 T^{(+)}_a(E,\mu) &=&  B_1^{(v)}(2\hat E,\mu) + \frac{\big[
   E B_2^{(v)}(2\hat E,\mu) + {m_B} B_3^{(v)}(2\hat E,\mu) \big]}{(2E-m_B)}\,, \\[5pt]
 T^{(0)}_a(E,\mu) &=& 
   B_1^{(v)}(2\hat E,\mu) + \frac{\big[(m_B\!-\!E) B_2^{(v)}(2\hat E,\mu) + 
   m_B B_3^{(v)}(2\hat E,\mu) \big]}{(m_B-2E)}  \nn \,, \\[5pt]
 T^{(T)}_a(E,\mu) &=& B_1^{(t)}(2\hat E,\mu) - B_2^{(t)}(2\hat E,\mu) 
   - 2B_3^{(t)}(2\hat E,\mu) + B_4^{(t)}(2\hat E,\mu)
   \,, \nn
\end{eqnarray}
where the dependence on $\hat\omega=2\hat E$ is shown, and
\begin{eqnarray}
  T^{(+)}_b(E,z,\mu) &=& B_{11}^{(v)}(2\hat E,z,\mu) 
    - \frac{E}{m_B} B_{12}^{(v)}(2\hat E,x,\mu)
    - B_{13}^{(v)}(2\hat E,z,\mu) \,, \\
  T^{(0)}_b(E,z,\mu) &=& B_{11}^{(v)}(2\hat E,z,\mu) 
    - \frac{(m_B-E)}{m_B} B_{12}^{(v)}(2\hat E,z,\mu)
    - B_{13}^{(v)}(2\hat E,z,\mu) \,,\nn \\
  T^{(T)}_b(E,z,\mu) &=&  B_{15}^{(t)}(2\hat E,z,\mu) 
    + B_{16}^{(t)}(2\hat E,z,\mu) 
    - B_{18}^{(t)}(2\hat E,z,\mu)
    \,,\nn
\end{eqnarray}
where $\hat\omega_1+\hat\omega_2=2\hat E$ and the dependence on $z$ is induced
from the $\hat\omega_1-\hat\omega_2$ dependence of the coefficients.

If we work at tree level in $J_{a,b}$ using Eq.~(\ref{Jtree}) and also in
$T_{a,b}$ then these coefficients are scale independent and satisfy
$T_a^{(+,0,T)} = T_b^{(+,0)} = 1$ and $T_b^{(T)}=0$. In this case if we take the
ratios $f_0/f_+$ and $f_T/f_+$ and expand assuming that the $f^F$ terms are
smaller than the $f^{NF}$ terms then our results agree with Ref.~\cite{bf}. We
note that using just the information in our factorization theorem that it is not
clear that one wants to expand in this way since the F and NF terms could
actually be similar in size as discussed in the introduction.
The expectation from QCD sum rules is that the ``soft'' NF
part of the form factors is larger than the ``hard'' F part~\cite{QCDsumrules}.

\section{Conclusion}

The soft-collinear effective theory (SCET) allows a rich structure of allowed
operators at higher orders in the expansion parameter $\lambda$.  In contrast to
simpler effective theories, the presence of fields ($\bn\cdot A_n$) and
derivatives ($\bn\cdot iD_c$) scaling like $\lambda^0$, allows a continuum set
of operators at any given order in $\lambda$. A similar situation is encountered
in deep inelastic scattering, where an infinite number of operators of
increasing dimension can contribute to the same order in $1/Q$.  In a generic
process with energetic hadrons it is therefore important to have a well-defined
procedure for organizing the structure of the soft-collinear operators at a
given order in $\lambda$. This organization is provided by SCET.

In this paper we formulated a general prescription for constructing the most
general ultrasoft-collinear operators appearing in the Lagrangian or in the
matching of an external current at a given order in $\lambda$ but to all orders
in $\alpha_s$. This was done by including constraints from collinear gauge
invariance, the Dirac structure of the effective theory fields and
reparameterization invariance. These conditions prove to be surprisingly
predictive, and constrain not only the number of allowed operators, but also
their functional dependence on label momenta.

For the case of the heavy-light currents, the constraints from the Dirac
structure of the effective theory fields have been included at leading order in
\cite{bfps}, and allow only 3 structures in the $v_\perp = 0$ frame. Here we
consider the more general case of an arbitrary heavy quark velocity $v$, which
is necessary in order to have a set of operators which closes under
reparameterization transformations.

At subleading order $O(\lambda)$ the Dirac constraints alone allow many more
operators. In particular, in addition to 2-body operators $(\bar \xi_n
W)_\omega\cdots h_v$, one has to include also 3-body currents of the form $(\bar
\xi_n W)_{\omega_1}\cdots (W^\dagger iD W)_{\omega_2} \cdots h_v$.  RPI
constraints on a subset of the two body operators were previously considered in
\cite{chay,bcdf}, and it was shown their coefficients are fixed in terms of the
coefficients of leading order currents. Here we extended the constraints to the
full set of allowed two-body and three-body operators, and showed that type (II)
RPI imposed severe constraints on the $(\omega_1,\omega_2)$ dependence of the
latter. For example, the scalar current $\bar q b$ is matched at $O(\lambda)$
onto 8 general operators in the effective theory. After imposing all
constraints, only one of these has a free Wilson coefficient, which has to be
determined from a matching calculation. A similar reduction is obtained for the
more complicated case of the vector/axial and tensor currents, for which one can
write (28,44) structures but only (4,7) Wilson coefficients are not fixed by
the symmetries of the effective theory.

In this paper we have focused on mixed usoft-collinear interactions, however for
many exclusive heavy-to-light processes the final operators that are needed are
of soft-collinear type as was the case for heavy-to-light form factors. In
practice it appears simplest to derive collinear-soft interactions {\em from}
the collinear-usoft ones using the two-stage matching technique, QCD $\to$ ${\rm
SCET}_{\rm I}$ $\to$ ${\rm SCET}_{\rm II}$, discussed in the proof of
factorization for heavy-to-light decays Ref.~\cite{bps4}. The operators in this
paper describe interactions in the intermediate ${\rm SCET}_{\rm I}$ theory. For
exclusive processes such as $B\to D\pi$~\cite{bps} where the intermediate
$p^2\sim Q\Lambda$ fluctuations in ${\rm SCET}_{\rm I}$ are responsible for
inducing simple operators in ${\rm SCET}_{\rm II}$ the procedure used in
Ref.~\cite{bps4} reduces to the one discussed in Ref.~\cite{bpssoft}.

Note Added: In the final stages of this work, Ref.~\cite{NH} appeared where a
direct study of light-light soft-collinear operators was performed. The Wilson
coefficients of these operators were determined by matching from QCD up to
one-loop order, and both 2-body and 3-body operators were found to contribute.
An intermediate theory with modes of momentum $p^\mu\sim Q(\lambda',1,\lambda')$
as dynamical degrees of freedom was also considered. This appears similar in
spirit to the QCD $\to$ ${\rm SCET}_{\rm I}$ $\to$ ${\rm SCET}_{\rm II}$
construction used in Ref.~\cite{bps4}, however in the intermediate ${\rm
SCET}_{\rm I}$ theory we found that the dynamical collinear modes should have
momentum scaling as $p^\mu\sim Q(\lambda',1,\sqrt{\lambda'})$. Finally,
reparameterization invariance constraints on soft-collinear operators were also
discussed in Ref.~\cite{NH}, and were shown to constrain the form of certain
Wilson coefficients.


\vspace{-0.2cm}
\begin{acknowledgments}
  We thank C.~Bauer and T.~Mehen for discussions.  This
  work was supported in part by the Department of Energy under the grants
  NSF PHY-9970781 and DE-FG03-00-ER-41132.
\end{acknowledgments}

\newpage

\end{document}